\documentclass[a4paper,12pt]{article}
\pdfoutput=1

\usepackage[utf8x]{inputenc}
\usepackage[T1]{fontenc}
\usepackage[lmargin=2.5cm,rmargin=2.5cm,tmargin=2.5cm,bmargin=3.5cm]{geometry}
\usepackage[font=small,labelfont=bf,width=0.95\textwidth]{caption}
\usepackage{amsmath,amssymb}
\usepackage{graphicx}
\usepackage[absolute]{textpos}
\usepackage{color}
\usepackage{booktabs}
\usepackage{fixltx2e}
\usepackage{xspace}
\usepackage{xcolor}
\usepackage{slashed}              
\usepackage[normalem]{ulem}
\usepackage[bookmarks=true,linktocpage]{hyperref}
\usepackage{cite}
\hypersetup{%
  pdftitle={Precise Higgs mass calculations in (non-)minimal
    supersymmetry at both high and low scales},
  pdfauthor={Peter Athron,Jae-hyeon Park,Tom Steudtner,Dominik
    Stöckinger,Alexander Voigt},
  pdfkeywords={FlexibleSUSY,supersymmetry,spectrum,generator,MSSM,NMSSM,E6SSM,MRSSM,effective,field,theory,EFT},
  colorlinks=true,urlcolor=blue,linkcolor=blue,citecolor=blue,filecolor=blue}

\newcommand{\unit}[1]{\,\text{#1}}

\newcommand{\ESSM}{\ensuremath{\text{E\textsubscript{6}SSM}}\xspace}
\newcommand{\FS}{\ensuremath{\text{FlexibleSUSY}}\xspace}
\newcommand{\FSv}{\ensuremath{\text{FlexibleSUSY 1.5.1}}\xspace}
\newcommand{\SARAH}{\ensuremath{\text{SARAH}}\xspace}
\newcommand{\SARAHv}{\ensuremath{\text{SARAH 4.9.0}}\xspace}
\newcommand{\Softsusy}{\ensuremath{\text{SOFTSUSY}}\xspace}
\newcommand{\Softsusyv}{\ensuremath{\text{SOFTSUSY 3.6.2}}\xspace}
\newcommand{\SPheno}{\ensuremath{\text{SPheno}}\xspace}
\newcommand{\SPhenov}{\ensuremath{\text{SPheno 3.3.8}}\xspace}
\newcommand{\SUSYHD}{\ensuremath{\text{\textsc{SusyHD}}}\xspace}
\newcommand{\SUSYHDv}{\ensuremath{\text{\textsc{SusyHD} 1.0.2}}\xspace}
\newcommand{\MhEFT}{\ensuremath{\text{MhEFT}}\xspace}
\newcommand{\FSTower}{\ensuremath{\text{FlexibleEFTHiggs}}\xspace}
\newcommand{\FeynHiggs}{\ensuremath{\text{FeynHiggs}}\xspace}
\newcommand{\FeynHiggsv}{\ensuremath{\text{FeynHiggs 2.12.0}}\xspace}
\newcommand{\Mma}{\ensuremath{\text{Mathematica}}\xspace}
\newcommand{\HSSUSY}{\ensuremath{\text{HSSUSY}}\xspace}
\newcommand{\nmssmtools}{\ensuremath{\text{NMSSMTools}}\xspace}
\newcommand{\nmssmtoolsv}{\ensuremath{\text{NMSSMTools 4.8.2}}\xspace}
\newcommand{\nmspec}{\ensuremath{\text{NMSPEC}}\xspace}
\newcommand{\nmssmcalc}{\ensuremath{\text{NMSSMCALC}}\xspace}
\newcommand{\MSbar}{\ensuremath{\overline{\text{MS}}}\xspace}
\newcommand{\DRbar}{\ensuremath{\overline{\text{DR}}}\xspace}
\newcommand{\MS}{\ensuremath{M_\text{SUSY}}\xspace}
\newcommand{\SM}{\ensuremath{\text{SM}}\xspace}
\newcommand{\MSSM}{\ensuremath{\text{MSSM}}\xspace}
\newcommand{\NMSSM}{\ensuremath{\text{NMSSM}}\xspace}
\newcommand{\MRSSM}{\ensuremath{\text{MRSSM}}\xspace}

\newcommand{\EFT}{\ensuremath{\text{EFT}}\xspace}
\newcommand{\mstL}{\ensuremath{m_{\tilde{t}_1}}}
\newcommand{\mstR}{\ensuremath{m_{\tilde{t}_2}}}
\newcommand{\Qmatch}{\ensuremath{Q_{\text{match}}}}
\newcommand{\Qlow}{\ensuremath{Q}}
\newcommand{\QFOMS}{\ensuremath{Q}}
\newcommand{\QFOMZ}{\ensuremath{Q}}
\newcommand{\alphaSMZ}{\ensuremath{\alpha_s^{\MSbar,\text{SM(5)}}(M_Z)}}
\newcommand{\alphaEMMZ}{\ensuremath{\alpha_{\text{e.m.}}^{\MSbar,\text{SM(5)}}(M_Z)}}

\newcommand{\LModelSoft}[1]{\mathcal{L}_{#1}^\mathrm{soft}}
\newcommand{\LModelYukawa}[1]{\mathcal{L}_{#1}^\mathrm{Yukawa}}
\newcommand{\DeltaMhFStowerYt}{\Delta M_h^{(y_t\text{ 0L vs.\ 1L})}}
\newcommand{\DeltaMhFSYt}{\Delta M_h^{(4\times y_t)}}
\newcommand{\DeltaMhQ}{\Delta M_h^{(Q)}}
\newcommand{\DeltaMhQmatch}{\Delta M_h^{(\Qmatch)}}

\newcommand{\gs}{\hat{g}_3}
\newcommand{\gsMSSM}{\bar{g}_3}
\newcommand{\ytlow}{\hat{y}_t}
\newcommand{\ytMSSMlow}{\bar{y}_t}
\newcommand{\vlow}{\hat{v}}
\newcommand{\vMSSM}{\bar{v}}
\newcommand{\gshigh}{{g}_3}
\newcommand{\gsMSSMhigh}{\tilde{g}_3}
\newcommand{\ythigh}{y_t}
\newcommand{\ytMSSMhigh}{\tilde{y}_t}
\newcommand{\vhigh}{v}
\newcommand{\vMSSMhigh}{\tilde{v}}
\newcommand{\lambdalow}{\hat{\lambda}}
\newcommand{\lambdahigh}{{\lambda}}
\newcommand{\kappaL}{\kappa_L}

\DeclareMathOperator{\re}{Re}

\allowdisplaybreaks

\title{\bf Precise Higgs mass calculations in (non-)minimal
  supersymmetry at both high and low scales}

\author{Peter Athron$^a$, Jae-hyeon Park$^b$, Tom Steudtner$^c$,\\
  Dominik St\"ockinger$^c$, Alexander Voigt$^d$}

\date{}

\begin{document}
\maketitle
\thispagestyle{empty}
\begin{center}
  \it
  $^a$ARC Centre of Excellence for Particle Physics at the
  Terascale,\\ School of Physics and Astronomy, Monash University, Victoria 3800\\[0.5em]
  ${}^b$Quantum Universe Center,
  Korea Institute for Advanced Study,\\
  85 Hoegiro Dongdaemungu,
  Seoul 02455, Republic of Korea\\[0.5em]
  ${}^c$Institut für Kern- und Teilchenphysik,\\ TU Dresden,
  Zellescher Weg 19, 01069 Dresden, Germany\\[0.5em]
  \it ${}^d$Deutsches Elektronen-Synchrotron DESY,\\ Notkestraße 85,
  22607 Hamburg, Germany
\end{center}
\begin{abstract}
  We present \FSTower, a method for calculating the SM-like Higgs pole
  mass in SUSY (and even non-SUSY) models, which combines an effective
  field theory
  approach with a diagrammatic calculation.  It thus achieves an all
  order resummation of leading logarithms together with
  the inclusion of all non-logarithmic 1-loop contributions.  We
  implement this method into \FS and study its properties in the MSSM,
  NMSSM, \ESSM\ and \MRSSM.  In the \MSSM, it correctly interpolates
  between the known results of effective field theory calculations in
  the literature for a high SUSY scale and fixed-order calculations in
  the full theory for a sub-TeV SUSY scale.  We compare our MSSM
  results to those from public codes and identify the origin of the
  most significant deviations between the \DRbar programs.  We then
  perform a similar comparison in the remaining three non-minimal
  models.  For all four models we estimate the theoretical uncertainty
  of \FSTower and the fixed-order \DRbar programs thereby finding that
  the former becomes more precise than the latter for a SUSY scale
  above a few TeV\@.
  Even for sub-TeV SUSY scales, \FSTower maintains
  the uncertainty estimate around $2$--$3\unit{GeV}$,
  remaining a competitive alternative to existing fixed-order computations.
\end{abstract}

\begin{textblock*}{7em}(\textwidth,1cm)
\noindent\footnotesize
CoEPP-MN-16-20 \\
DESY 16-057 \\
KIAS--Q16008
\end{textblock*}

\clearpage

\section{Introduction}

A hallmark of renormalizable supersymmetric (SUSY) theories is that
quartic scalar couplings are not free parameters, but fixed in terms
of gauge and (in some models) Yukawa couplings. As a result,
predictions of the Standard Model (SM)-like Higgs mass are restricted
to a limited range and precise calculations are very important for
testing SUSY models.  Since the discovery of a $125$ GeV Higgs boson
at the LHC \cite{Aad:2012tfa,Chatrchyan:2012xdj}, the need for precise predictions within SUSY
models has increased in several ways. First, the measurement is
already far more precise than existing theory predictions, motivating
significant improvements in both theory predictions and their
associated uncertainty estimates. Second, the non-observation of new
physics at the LHC, may imply heavier masses of new particles, so
predictions should be reliable both for light or heavy SUSY
masses. Third, the heavy Higgs boson mass provokes naturalness
questions that motivate non-minimal SUSY models and improving precision Higgs
mass calculations there.

Here we present \FSTower, a new method for calculating the Higgs mass
that can improve the precision of the Higgs mass prediction in minimal
and non-minimal SUSY models. This method uses effective field theory
(EFT) techniques, which improve the precision when the SUSY masses are much
heavier than the electroweak (EW) scale.  However, \FSTower also includes
terms which are important at low SUSY scales, previously only
included in fixed-order calculations.  This hybrid approach combines
the virtues of both worlds to give precise
predictions at both high and low SUSY scales. We also present an
extensive analysis of the remaining theory uncertainty and discuss in
detail the differences to other calculations, shedding
light on the theory uncertainties of existing calculations. The method
and uncertainty estimates are applied to the MSSM and three
non-minimal models, the \NMSSM, the \ESSM, and the \MRSSM.

The fixed-order and EFT approaches have both been used extensively in
the literature, for a complete picture, see e.g.\ the recent
review~\cite{Draper:2016pys}. In a fixed-order, or Feynman
diagrammatic computation, a perturbative expansion is performed to a
specified order in the gauge or Yukawa couplings. In the
MSSM, the dominant 2-loop corrections were added long ago
\cite{Heinemeyer:1998jw, Heinemeyer:1998kz, Heinemeyer:1998np, Zhang:1998bm, Espinosa:1999zm, Degrassi:2001yf, Espinosa:2000df, Brignole:2001jy, Dedes:2002dy, Brignole:2002bz, Dedes:2003km, Heinemeyer:2004xw}. Recent progress for the MSSM includes
incorporating electroweak gauge couplings 
\cite{Martin:2002wn},
  a genuine calculation of leading 3-loop effects
  \cite{Harlander:2008ju,Kant:2010tf},
  and
  momentum-dependent 2-loop contributions
  \cite{Borowka:2014wla,Borowka:2015ura,Degrassi:2014pfa}.
Many public codes for MSSM Higgs mass calculations are available, 
see e.g.\
  \cite{
Heinemeyer:1998yj,
Allanach:2001kg,
Djouadi:2002ze,
Allanach:2013kza,
Athron:2014yba%
  }.
There are also dedicated calculations and public codes for the NMSSM, 
see e.g.\
  \cite{Degrassi:2009yq,
Ellwanger:2006rn,
Baglio:2013iia,
Allanach:2013kza,
Athron:2014yba,Drechsel:2016jdg}.
  For any user-defined model,
  \SARAH/\SPheno performs
  an automatic 2-loop calculation at zero momentum
  in the gauge-less limit
  \cite{Goodsell:2014bna,Goodsell:2015ira}.

Fixed-order calculations are particularly
  reliable when the new particle masses are around
  the EW scale.
  If the new physics scale
  is too high, large logarithms appear at each order in perturbation
  theory, and
  the result can suffer from a large truncation error.
Recently, therefore,
Refs.~\cite{Hahn:2013ria,Drechsel:2016jdg,Bahl:2016brp} combined
fixed-order calculations with
the resummation of the leading
  and next-to-leading logarithms without double counting, reducing the
  theory uncertainty at high SUSY masses.

  EFT calculations use a matching-and-running procedure.  In the
  simplest case, all non-SM particles are integrated out at some high
  SUSY scale. The running SM parameters at the high scale are then
  determined by matching, run down to the EW scale using
  renormalization group methods, and the Higgs mass is computed at the
  weak scale in the SM\@.  Since the early works in this approach
  \cite{Barbieri:1990ja,Okada:1990gg,Haber:1993an,Casas:1994us,Carena:1995bx,Carena:1995wu,Haber:1996fp},
  developments include the analytical evaluation of 3-loop terms
  \cite{Martin:2007pg}, next-to-next-to-leading logarithm (NNLL) accuracy
  \cite{Draper:2013oza,Bagnaschi:2014rsa,Vega:2015fna,Bahl:2016brp},
 non-SM EFTs potentially for
additional thresholds
\cite{
Binger:2004nn,
Giudice:2011cg,
Bagnaschi:2014rsa,
Lee:2015uza,
Carena:2015uoe,
Bagnaschi:2015pwa%
}.
The RGEs can be solved either numerically as in this work or
perturbatively as done to two loops
\cite{Casas:1994us,Carena:1995bx,Carena:1995wu,Haber:1996fp} and three
loops \cite{Martin:2007pg} (see also
Appendix~\ref{sec:appendix-leading-logs}). Public programs
implementing this EFT-type calculation (for MSSM only) are 
\SUSYHD \cite{Vega:2015fna}, \FS/HSSUSY \cite{FlexibleSUSY:Website}
and the \MhEFT \cite{MhEFT:Website}.

As discussed e.g.~in Ref.~\cite{Vega:2015fna}, the disadvantage of
pure EFT-type calculations is that they miss 
non-logarithmic contributions that are suppressed by powers of the
SUSY mass scale already at the tree-level and 1-loop level. Hence, the
theory uncertainty increases strongly if the SUSY masses are close to
the EW scale. 

\FSTower is an EFT calculation with specially chosen matching
conditions, such that the Higgs mass calculation is exact at the
tree-level and 1-loop level.  This ensures that the theory uncertainty
remains bounded at both high and low SUSY scales, as we will show.  We
have implemented \FSTower into \FS \cite{Athron:2014yba}, a spectrum
generator generator for BSM models based on \SARAH
\cite{Staub:2008uz,Staub:2009bi,Staub:2010jh,Staub:2011dp,Staub:2012pb,Staub:2013tta}
and \Softsusy \cite{Allanach:2001kg,Allanach:2013kza} so that this
method can be used in a huge variety of models.  The level of
precision currently implemented is 1-loop mass matching and 3-loop
running in the SM\@. Currently a limiting assumption is that the
SM is the correct low-energy EFT at the EW scale and all
non-SM particles are integrated out at a heavy scale.

This paper is structured as follows: In
Section~\ref{sec:procedure_of_the_calculation} we give an overview of
the pure EFT and the fixed-order approaches and describe the \FSTower
method in more detail.  In Section~\ref{sec:MSSMtower_results} we
apply \FSTower to the MSSM and compare the results with those from
publicly available MSSM spectrum generators.  In addition, we analyse
the origin of the most significant deviations between the \DRbar
fixed-order calculations in \FS, \Softsusy and \SPheno.  We then
present several possibilities to estimate the theoretical uncertainty
of the Higgs mass prediction in the \DRbar fixed-order approaches and
in \FSTower.  In Section~\ref{sec:MSSM_uncertainty_summary} we
summarize and combine the uncertainty estimates and give an order of
magnitude for the SUSY scale above which we expect \FSTower to lead to
a more precise prediction than the \DRbar fixed-order programs.  In
Sections~\ref{sec:NMSSM}--\ref{sec:MRSSM} we apply \FSTower to the
NMSSM, \ESSM\ and the MRSSM and perform an uncertainty estimation.  We
conclude in Section~\ref{sec:conculsion}.

\section{Procedure of the calculation}
\label{sec:procedure_of_the_calculation}
The new \FSTower approach presented here is an EFT-type calculation of
the SM-like Higgs mass in the MSSM or any other non-minimal SUSY or
BSM model, where we assume the Standard Model is a valid EFT\@. \FSTower
is implemented into \FS \cite{Athron:2014yba}, a C++ and \Mma
framework to create modular spectrum generators for SUSY and non-SUSY
models. Before introducing \FSTower, we describe the SM and MSSM to
fix our notation and then describe fixed-order and ``pure EFT''
calculations implemented in several public programs all of which use
the \DRbar\ scheme.  There are also very accurate calculations in the
on-shell renormalization scheme, for example \FeynHiggs
\cite{Hahn:2013ria,Frank:2006yh,Degrassi:2002fi,Heinemeyer:1998np,Heinemeyer:1998yj,Borowka:2014wla,Hollik:2014bua,Hollik:2015ema}
and \nmssmcalc \cite{Ender:2011qh,Graf:2012hh,Baglio:2013iia}, but we
will not go into the details of the on-shell calculations.
In the following we use the programs \FSv, \Softsusyv, \SARAHv,
\SPhenov, \FeynHiggsv, \SUSYHDv and \nmssmtoolsv, if not stated
otherwise.

\subsection{The Standard Model and its minimal supersymmetric extension}

The Standard Model is invariant under local gauge transformations of the group,
\begin{align}
  G_\text{SM} = SU(3)_C\times SU(2)_L \times U(1)_Y ,
\end{align}
where the gauge couplings associated with $SU(3)_C$, $SU(2)_L$ and
$U(1)_Y$ are $g_3$, $g_2$ and $g_1$, respectively, with $g_1$ under
the $SU(5)$ GUT normalization.  Sometimes it is more convenient to
write expressions in terms of the $U(1)_Y$ gauge coupling, which we
denote $g_Y = \sqrt{3/5} \, g_1$. As usual, we also use
$e^2=g_Y^2g_2^2/(g_Y^2+g_2^2)$, $\alpha_{\text{e.m.}}=e^2/4\pi$ and
$\alpha_s=g_3^2/4\pi$.

The spontaneous breakdown of electroweak symmetry $SU(2)_L \times U(1)_Y \rightarrow U(1)_e$ occurs when the coefficient of the bilinear term in the Higgs potential,
\begin{align}V(\phi) = \mu^2 |\Phi|^2 + \lambda |\Phi|^4,
\end{align}
is negative.  This causes the neutral component of the Higgs field,
$\Phi$, which is a $SU(2)_L$ doublet, to develop a vacuum
expectation value (VEV),
$v = \sqrt{- {\mu^2} / {(2 \lambda)}}$.
The Standard Model fermions are the left handed $SU(2)_L$  quark and
lepton doublets
$Q_i$ and $L_i$, and the right handed $SU(2)_L$ singlets for up-type
and down-type quarks and charged leptons, $u_{Ri}$,  $d_{Ri}$,
$e_{Ri}$. They obtain mass through their Yukawa interactions
with the Higgs field,
\begin{align}
  \LModelYukawa{\SM} = (Y_u)_{ij} \, \overline{Q}_i \cdot \Phi^\dagger \, u_{Rj} + (Y_d)_{ij} \, \overline{Q}_i \, \Phi \, d_{Rj}  + (Y_e)_{ij} \, \overline{L}_i \, \Phi \, e_{Rj} + \textrm{h.c.},
\end{align}
when the neutral Higgs field develops a VEV\@. Here $i,j$ denote
generation indices,
and we define the $SU(2)_L$
dot product, $A \cdot B := A^1 B^2 - A^2 B^1$.
To simplify the notation we denote the
third generation Yukawa couplings as
$y_t, y_b, y_\tau$ which are the largest singular values of
$Y_u, Y_d, Y_e$, respectively.

The minimal supersymmetric extension of the Standard Model (MSSM) has
the superpotential,
\begin{align}
\mathcal{W}_\MSSM = \mu \, \hat{H}_u\cdot\hat{H}_{d} + (Y_u)_{ij} \, \hat{Q}_i\cdot\hat{H}_u \, \hat{u}^c_j   + (Y_d)_{ij} \, \hat{Q}_i\cdot\hat{H}_{d} \, \hat{d}^c_j +  (Y_e)_{ij} \, \hat{L}_i\cdot\hat{H}_{d} \, \hat{e}^c_j,
\label{Eq:superpot}
\end{align} 
where all superfields appear with a hat.
The chiral superfields have the $G_\text{SM}$ quantum numbers
\begin{equation}
  \begin{aligned}
 \hat{Q}  &:\textstyle (\mathbf{     3} ,\mathbf{2}, \frac{1}{6}) ,
&\hat{u}^c&:\textstyle (\mathbf{\bar{3}},\mathbf{1},-\frac{2}{3}) ,
&\hat{d}^c&:\textstyle (\mathbf{\bar{3}},\mathbf{1}, \frac{1}{3}) ,
&\hat{L}  &:\textstyle (\mathbf{     1} ,\mathbf{2},-\frac{1}{2}) ,
&\hat{e}^c&:\textstyle (\mathbf{     1} ,\mathbf{1}, 1          ) ,
\\
 \hat{H}_d&:\textstyle (\mathbf{     1} ,\mathbf{2},-\frac{1}{2}) ,
&\hat{H}_u&:\textstyle (\mathbf{     1} ,\mathbf{2}, \frac{1}{2}) ,
  \end{aligned}
\end{equation}
where the first and second symbol in the parentheses denotes the
representation of the corresponding superfield with respect to
$SU(3)_C$ and $SU(2)_L$ and the third component is the hypercharge in
standard normalization. The neutral components of the up-type Higgs
field, $H_u$, and down-type Higgs field, $H_d$, develop the VEVs,
$v_u/\sqrt{2}$ and $v_d/\sqrt{2}$, respectively.
As usual we define,
\begin{align}
  v &= \sqrt{v_u^2 + v_d^2}, & \tan \beta &= \frac{v_u}{v_d} ,
\end{align}
where $v \approx 246\unit{GeV}$.
The soft breaking Lagrangian is given by
\begin{align}
\LModelSoft{\MSSM} &= -\frac12\left[M_1\bar{\tilde{B}}^0\tilde{B}^0 +
  M_2\bar{\tilde{W}}\tilde{W} +
  M_3\bar{\tilde{g}}\tilde{g}\right]  -
m^2_{H_u}|H_u|^2 - m^2_{H_d}|H_d|^2 - \left[B\mu \, {H}_u\cdot{H}_{d} +
  \textrm{h.c.}\right] \nonumber\\
&\phantom{={}}
-\left[\tilde{Q}_i^\dagger(m^2_Q)_{ij}\tilde{Q}_j +
  \tilde{d}_{Ri}^\dagger(m^2_d)_{ij}\tilde{d}_{Rj}
   +
  \tilde{u}_{Ri}^\dagger(m^2_u)_{ij}\tilde{u}_{Rj}
  + \tilde{L}_{i}^\dagger(m^2_L)_{ij}\tilde{L}_{j} +
  \tilde{e}_{Ri}^\dagger(m^2_e)_{ij}\tilde{e}_{Rj}
  \right] \nonumber \\
&\phantom{={}}
+\left[(T_u)_{ij}\tilde{Q}_i \cdot  H_u\tilde{u}^\dagger_{\textrm{R}j}
  +
  (T_d)_{ij}\tilde{Q}_i \cdot H_{d} \tilde{d}^\dagger_{\textrm{R}j}
  + (T_e)_{ij}\tilde{L}_i \cdot H_{d}\tilde{e}^\dagger_{\textrm{R}j}
  + \textrm{h.c.}\right] .
\label{Eq:Lsoft}
\end{align}
In the following, we trade the
soft-breaking trilinear couplings for the customary $X_f$ parameters as
\begin{align}
  \begin{split}
    (Y_f)_{ij} (X_f)_{ij} = (T_f)_{ij} -  (Y_f)_{ij}
    \begin{Bmatrix}
      \mu^*\tan\beta \\ \mu^*\cot\beta
    \end{Bmatrix}
,
\quad
\text{for}    \left\{
       \begin{array}{l}\!\!f = d, e, \\ \!\!f = u,
       \end{array}
    \right.
  \end{split}
  \label{eq:MSSM_Xt_definition}
\end{align}
with the appearing matrices given in the super-CKM basis
\cite{Dugan:1984qf,Allanach:2008qq}.
For the third generation fermions we define $X_t := (X_u)_{33}$, $X_b
:= (X_d)_{33}$, $X_\tau := (X_e)_{33}$.
The gauginos have the following $G_\text{SM}$ quantum numbers:
\begin{equation}
  \begin{aligned}
 \tilde{B}  &:\textstyle (\mathbf{1},\mathbf{1},0) ,
&\tilde{W}  &:\textstyle (\mathbf{1},\mathbf{3},0) ,
&\tilde{g}  &:\textstyle (\mathbf{8},\mathbf{1},0) .
  \end{aligned}
\end{equation}
In the scenarios studied in the following we set the dimensionful
running \DRbar superpotential and soft-breaking parameters to the
common value of the SUSY scale, $\MS$, if not stated otherwise:
\begin{align}
  \begin{split}
    &(m_f^2)_{ij}(\MS) = \delta_{ij} \MS^2, \quad (f=Q,u,d,L,e) \\
    &M_i(\MS) = \MS, \quad (i = 1,2,3) \\
    &\mu(\MS) = \MS , \\
    &m_A^2(\MS) = \frac{B\mu(\MS)}{\sin\beta(\MS) \cos\beta(\MS)} = \MS^2 , \\
    &(X_f)_{ij}(\MS) = 0.
  \end{split}
  \label{eq:MSSM_MSUSY_scenario}
\end{align}
Sometimes we will go beyond the last equation and keep $X_t$ as a free
parameter and set it to a non-zero value.

In our numerical evaluations we will choose the numerical values
    $\alphaEMMZ =
    {1}/{127.944}$ for the running fine structure constant,
$M_t = 173.34 \unit{GeV}$, $M_\tau =
    {1.777\unit{GeV}}$, $M_Z =
    {91.1876\unit{GeV}}$ for the top quark, $\tau$ lepton and $Z$ boson pole masses, 
and $m_b^{\MSbar,\text{SM(5)}}(m_b) =
    {4.18\unit{GeV}}$ for the running $b$-quark mass, if not stated
    otherwise.

\subsection{Fixed-order calculations in \FS, \Softsusy and \SARAH/\SPheno}
\label{sec:fixed_order_calculations}

We now discuss the fixed-order approach for calculating the Higgs
mass that is implemented in \FS, \Softsusy and \SARAH/\SPheno.
There are two major steps in this calculation:

\begin{enumerate}
\item Find all \DRbar parameters at the SUSY scale.
\item Calculate the Higgs pole mass from the \DRbar parameters.
\end{enumerate}

The first step is rather complicated and involves an iteration. One
complication is that some parameters may be set at a higher-scale and
the values at the SUSY scale only obtained through the RG running,
though here we will simply set all non-SM parameters at the SUSY scale.
Nonetheless this is still non-trivial because some of the \DRbar
parameters must be chosen to fulfill the EWSB equations or are
determined by experimental data. For the EWSB conditions we will fix
the soft Higgs masses in this work, which is an option available in
all of the codes we use.  The VEV, $v$ is fixed from the \DRbar
running $m_Z$, leaving $\tan\beta$ as a free parameter at the SUSY
scale. The gauge couplings, $g_1$, $g_2$ and $g_3$ and Yukawa
couplings, $Y_u$, $Y_d$ and $Y_e$ can be extracted from data. This can
be done using the measured values of the running
\MSbar electromagnetic and strong couplings, the
Weinberg angle or an equivalent quantity, and the quark and lepton
masses. Specifically \FS, \Softsusy and \SPheno all use the following,
\begin{align}
  \alpha_s^{\DRbar,\text{SUSY}}(M_Z) &=
  \frac{\alphaSMZ}{1 -
    \Delta\alpha_s^{\text{SM}}(M_Z) -
    \Delta\alpha_s^{\text{SUSY}}(M_Z)} , \\
  m_Z^{\DRbar,\text{SUSY}}(M_Z) &=
  \sqrt{M_Z^2 + \re \Pi_{ZZ}^{T,\text{SUSY}}(M_Z^2)} ,
\label{eq:fixedalphaMZ}
\end{align}
where $M_Z$ is the $Z$-boson pole mass, $\Pi^T_{ZZ}(p^2)$ is the
transverse part of the 1-loop $Z$ self energy and $\alphaSMZ$ is
the \MSbar strong coupling in the SM with 5-flavours.
Using these and further similar relations, all \DRbar gauge couplings
and EWSB parameters of the fundamental SUSY theory can be determined
at the low scale $\QFOMZ=M_Z$. The Yukawa couplings are determined
similarly from the running vacuum expectation values and fermion masses, but
specific corrections beyond the 1-loop level are taken into
account. Most importantly, the running top quark mass in \FS\ and
\Softsusy is given by
\begin{align}
m_t^{\DRbar} &=
M_t+\re\left[
\widetilde{\Sigma}_t^{(1),S}(M_t)\right] +
M_t \re\left[
  \widetilde{\Sigma}_t^{(1),L}(M_t) +
  \widetilde{\Sigma}_t^{(1),R}(M_t)
\right] \nonumber \\
&\phantom{={}} + M_t
\left[\widetilde{\Sigma}_t^{(1),\text{qcd}}(m_t^{\DRbar})
+ \left(\widetilde{\Sigma}_t^{(1),\text{qcd}}(m_t^{\DRbar})\right)^2
+ \widetilde{\Sigma}_t^{(2),\text{qcd}}(m_t^{\DRbar})\right],
\label{eq:fixedmt_FlexibleSUSY}
\end{align}
where $M_t$ denotes the top pole mass,
$\widetilde{\Sigma}_t^{(1),S}(p)$,
$\widetilde{\Sigma}_t^{(1),L}(p)$ and
$\widetilde{\Sigma}_t^{(1),R}(p)$ denote the scalar,
left-handed and right-handed part of
the 1-loop top self energy without SM-QCD contributions, evaluated at
$p=M_t$,
and $\widetilde{\Sigma}_t^{(1,2),\text{qcd}}(m_t^{\DRbar})$ denote SM-QCD self energy
contributions, with a factor  $\slashed{p}$ removed,
evaluated at $p=m_t^{\DRbar}$
\cite{Avdeev:1997sz,Bednyakov:2002sf}:
\begin{align}
  \widetilde{\Sigma}_t^{(1),\text{qcd}}(m_t^{\DRbar}) &=
  - \frac{g_3^2}{12\pi^2} \left[5 - 3 \ln\left(\frac{(m_t^{\DRbar})^2}{\QFOMZ^2}\right)\right],\\
  \widetilde{\Sigma}_t^{(2),\text{qcd}}(m_t^{\DRbar}) &=
  -\frac{g_3^4}{4608 \pi^4}
  \Bigg[396 \ln^2\left(\frac{(m_t^{\DRbar})^2}{\QFOMZ^2}\right)-1476 \ln
   \left(\frac{(m_t^{\DRbar})^2}{\QFOMZ^2}\right)-48 \zeta (3) \nonumber \\
   &\phantom{={}}\qquad\qquad\quad +2011+16 \pi^2 (1+\ln 4)\Bigg] .
\end{align}
Eq.~\eqref{eq:fixedmt_FlexibleSUSY} is evaluated at the scale $M_Z$
and yields the running top mass $m_t^{\DRbar}(M_Z)$.
 \SPheno treats the top quark mass
differently and requires
\begin{align}
m_t^{\DRbar} &=
M_t+\re\left[
\widetilde{\Sigma}_t^{(1),S}(m_t^{\DRbar})\right]+
m_t^{\DRbar} \re\left[
  \widetilde{\Sigma}_t^{(1),L}(m_t^{\DRbar}) +
  \widetilde{\Sigma}_t^{(1),R}(m_t^{\DRbar})
\right] \nonumber \\
&\phantom{={}} +
m_t^{\DRbar}
\left[\widetilde{\Sigma}_t^{(1),\text{qcd}}(m_t^{\DRbar}) +
\widetilde{\Sigma}_t^{(2),\text{qcd}}(m_t^{\DRbar})
\right].
\label{eq:fixedmt_SPheno}
\end{align}
We will later comment on this difference between
Eqs.~\eqref{eq:fixedmt_FlexibleSUSY} and \eqref{eq:fixedmt_SPheno}.
Both these equations determine the running top mass implicitly and are solved
by an iteration, resulting in slightly different solutions.

In the second step, the Higgs boson mass is computed numerically by
solving
\begin{align}
  0&= \det\left[p^2\delta_{ij} - (m_{\phi}^2)_{ij}
    + \re \Sigma_{\phi,ij}(p^2) - \frac{t_{\phi,i}}{v_i}\right] ,
  \label{eq:fixedmh}
\end{align}
where $m_\phi^2$ denotes the CP-even Higgs tree-level mass matrix,
$\Sigma_\phi$ and $t_\phi$ are the \DRbar-renormalized CP-even Higgs
self energy and tadpole, respectively, and $v_1 \equiv v_d$, $v_2
\equiv v_u$.
In the MSSM, \FS, \Softsusy and \SPheno use the full 1-loop self
energy and 2-loop corrections of the order ${\mathcal O}((\alpha_t +
\alpha_b)^2 + (\alpha_t + \alpha_b) \alpha_s + \alpha_\tau^2)$ from
\cite{Degrassi:2001yf,Brignole:2001jy,Dedes:2002dy,Brignole:2002bz,Dedes:2003km}.  For
non-minimal SUSY models, \FS\ uses the full 1-loop self
energy (optionally extended by the 2-loop MSSM or NMSSM contributions).
\SARAH/\SPheno uses the 2-loop self energy in the gauge-less limit
and at zero momentum in any given model
\cite{Goodsell:2014bna,Goodsell:2015ira}.

\subsection{Pure EFT calculation in \SUSYHD and \FS/HSSUSY}
\label{sec:eft-approach}

EFT calculations have the virtue of resumming potentially large
logarithms of the generic heavy SUSY mass scale beyond any finite loop level.
The calculation is based on the approximation that all
non-SM particles, i.e.\ all SUSY particles and the extra Higgs states,
have a common heavy mass of order $\MS$, and that the SM is the correct
low-energy EFT below $\MS$. 

The determination of 
\DRbar parameters and the computation of the Higgs
mass is then done in three steps, carried out
iteratively, until convergence is reached:

\begin{enumerate}
\item Integrate out all SUSY particles 
at the SUSY scale, and determine the SM parameter $\lambda$ 
at $\MS$ by a matching of the SUSY theory to the SM\@.
\item Use the SM renormalization group equations to
run the SM parameters down to the EW scale. 
\item Match the
SM parameters to experiment at the EW scale, and compute the 
Higgs pole mass.
\end{enumerate}

In the pure EFT approach, the threshold corrections at the
SUSY scale are expressed as perturbative functions of the SM
parameters at $\MS$, dimensionless (combinations of) SUSY
parameters and at most logarithms of SUSY masses. No terms suppressed
by powers of $\MS$ appear. The known 1- and 2-loop threshold correction
to $\lambda$ from the MSSM read \cite{Bagnaschi:2014rsa,Vega:2015fna}
\begin{align}
\lambda^{\text{pure \EFT}} &=
\frac{1}{4} \left(g_Y^2 + g_2^2\right) \cos^22\beta
+\Delta\lambda^{(1)}+\Delta\lambda^{(2)},
\label{eq:delta_lambda_sum} \\
\begin{split}
  (4\pi)^2\Delta\lambda^{(1)} &= 3 (y_t^\SM)^2 \left[(y_t^\SM)^2 +
    \frac{1}{2} \left(g_2^2-\frac{g_Y^2}{3} \right)
    \cos 2\beta \right] \ln \frac{m_{Q_3}^2}{\Qmatch^2} \\
  &\phantom{={}} +3 (y_t^\SM)^2 \left[(y_t^\SM)^2 + \frac{2}{3} g_Y^2
    \cos 2\beta \right]
  \ln \frac{m_{U_3}^2}{\Qmatch^2} \\
  &\phantom{={}} +6 (y_t^\SM)^4 \tilde{X}_t
  \left[\widetilde{F}_1\left(x_{QU}\right)
    -\frac{\tilde{X}_t}{12} \widetilde{F}_2\left(x_{QU}\right)\right] \\
  &\phantom{={}}+ \frac{3}{4} (y_t^\SM)^2 \tilde{X}_t \cos 2\beta
  \left[g_Y^2 \widetilde{F}_3 \left(x_{QU}\right)
    + g_2^2 \widetilde{F}_4 \left(x_{QU}\right) \right] \\
  &\phantom{={}}-\frac14 (y_t^\SM)^2 \tilde{X}_t \cos^2 2\beta
  \left( g_Y^2 +g_2^2 \right) \widetilde{F}_5\left(x_{QU} \right) \\
  &\phantom{={}} +{\mathcal O}(g_Y^4, g_2^4, g_Y^2 g_2^2) ,
\end{split}
\label{eq:delta_lambda_1L}
\end{align}
where $\tilde{X}_t = X_t^2/(m_{Q_3} m_{U_3})$ and $x_{QU} = m_{Q_3}/m_{U_3}$.
In Eqs.~\eqref{eq:delta_lambda_sum}--\eqref{eq:delta_lambda_1L} $g_Y$,
$g_2$ and $y_t^\SM$ denote the Standard Model electroweak gauge and
top Yukawa couplings at the SUSY scale, respectively, all defined
in the \MSbar scheme.  With $\Qmatch$ we denote the matching scale,
which we identify with \MS, if not stated otherwise.
The loop functions
$\widetilde{F}_i(x)$ as well as $\Delta\lambda^{(2)}$ can be found in
\cite{Bagnaschi:2014rsa,Vega:2015fna}.
Since $\lambda$ is directly expressed in terms of running SM
parameters and fundamental SUSY input parameters, no other threshold
corrections are needed.

This pure EFT approach to calculate the Higgs pole mass is implemented in \SUSYHD \cite{Vega:2015fna}
and \FS/HSSUSY \cite{FlexibleSUSY:Website}\@.\footnote{%
  The \FS/HSSUSY model file has been written by Emanuele Bagnaschi, Georg
  Weiglein and Alexander Voigt and will be presented and studied in
  more detail by these authors in an upcoming publication.}
HSSUSY is now part of the public \FS distribution
and has the same essential features and method of
\SUSYHD within a C++ framework. Both programs use the same definition\footnote{%
  In HSSUSY we used analytical \Mma expressions for the 2-loop threshold
  corrections $\Delta\lambda^{(2)}$ of
  $\mathcal{O}(\alpha_t\alpha_s)$ provided by the authors of
  \cite{Bagnaschi:2014rsa} and
  $\mathcal{O}(\alpha_t^2)$ provided by the authors of \SUSYHD.}
\eqref{eq:delta_lambda_sum} for
$\lambda$ and 3-loop
RGEs to evolve $\lambda$ to the $M_t$ scale
\cite{Buttazzo:2013uya,Bednyakov:2013eba}.  At the $M_t$ scale,
both programs determine the SM gauge and Yukawa couplings by matching
to experiment. HSSUSY extracts the SM gauge and Yukawa couplings
 at the 1-loop level from
$\alphaSMZ$, $\alphaEMMZ$ and $G_F$ and quark and lepton masses
using the approach described in
\cite{Athron:2014yba}, thereby taking into account 1-loop and leading
2-loop corrections.
For the extraction of the top Yukawa coupling also the known
2-loop and 3-loop QCD corrections are taken into account 
\cite{Fleischer:1998dw,Melnikov:2000qh}.
\SUSYHD includes several further subleading corrections, e.g. fit
formulas for 2-loop
threshold corrections to the EW gauge couplings \cite{Buttazzo:2013uya}.
Finally, the Higgs pole mass is calculated at the scale $M_t$. HSSUSY
employs full 1-loop and leading 2-loop corrections
of $\mathcal{O}(\alpha_t\alpha_s + \alpha_t^2)$; \SUSYHD
uses a numerical fit formula approximating the full 2-loop
corrections.

\subsection{EFT calculation in \FSTower}
\label{sec:tower-approach}

The calculation of \FSTower\ follows the same logic as the EFT
calculation of \SUSYHD and \HSSUSY. The difference lies in the choice of the
matching conditions. In \FSTower, $\lambda(\MS)$ is
determined implicitly by the condition
\begin{align}
  (M_h^\SM)^2 = (M_h^\MSSM)^2 \quad \text{at} \quad \Qmatch ,
  \label{eq:polemassmatching}
\end{align}
i.e.\ by the condition that the lightest CP-even Higgs pole masses,
computed in the effective and the full theory at the SUSY scale in
fixed-order perturbation theory in the \MSbar/\DRbar\ schemes, agree.
The Standard Model Higgs pole mass is calculated at the scale \MS as
\begin{align}
  (M_h^\SM)^2 = (m_h^{\MSbar,\SM})^2 - \re \Sigma_h^{\MSbar,\SM}((M_h^\SM)^2) + t_h^{\MSbar,\SM}/v ,
  \label{eq:SM polemass}
\end{align}
where $m_h^{\MSbar,\SM}$ is the running \MSbar Higgs mass in the Standard Model,
$\Sigma^{\MSbar,\SM}_h$ is the \MSbar-renormalized Standard
Model Higgs self energy and $t_h^{\MSbar,\SM}$ is the
corresponding tadpole.
Using this,
the quartic Higgs coupling in the SM reads
\begin{align}
  &\lambda(\MS) = \frac{1}{v^2} \left[
    (M_h^\MSSM)^2
    + \re \Sigma_h^{\MSbar,\SM}((M_h^\SM)^2) - \frac{t_h^{\MSbar,\SM}}{v}
  \right] .
  \label{eq:lambda_1L_tower}
\end{align}
In the current implementation, only 1-loop self energies and tadpoles are used in
this matching condition; in the future it is planned to take into
account 2-loop corrections.

Likewise, the gauge couplings and the $Z$-boson and top quark mass,
are implicitly fixed by the conditions
\begin{align}
  &\alpha_{x}^{\DRbar,\text{SUSY}}(\MS) =
  \frac{\alpha_{x}^{\MSbar,\SM}(\MS)}{1 -
    \Delta\alpha_{x}^{\text{SUSY}}(\MS)} ,
  \quad x = \text{e.m.}, \text{s},
  \label{eq:matching_condition_alpha_em} \\
  &(m_Z^{\MSbar,\SM})^2 - \re \Pi_{ZZ}^{T,\SM}(M_Z^2) =
  (m_Z^{\DRbar,\text{SUSY}})^2 - \re \Pi_{ZZ}^{T,\text{SUSY}}(M_Z^2),
  \label{eq:matching_condition_MZ} \\
\begin{split}
  m_t^{\MSbar,\SM} -\re\left[
    \widetilde{\Sigma}_t^{(1),\SM}(M_t)\right] -
  m_t^{\MSbar,\SM} \left[
    \widetilde{\Sigma}_t^{(1),\text{SM-qcd}}(m_t^{\MSbar,\SM}) +
    \widetilde{\Sigma}_t^{(2),\text{SM-qcd}}(m_t^{\MSbar,\SM})\right] \\
  = m_t^{\DRbar} -\re\left[ \widetilde{\Sigma}_t^{(1)}(M_t) \right] -
  m_t^{\DRbar} \left[
    \widetilde{\Sigma}_t^{(1),\text{qcd}}(m_t^{\DRbar}) +
    \widetilde{\Sigma}_t^{(2),\text{qcd}}(m_t^{\DRbar}) \right],
\end{split}
\label{eq:matching_condition_Mt}
\end{align}
at the SUSY scale, where $\widetilde{\Sigma}_t^{(1)}$ again denote the
1-loop top self-energy contributions without the SM QCD part, and
$\widetilde{\Sigma}_t^{(1,2),\text{SM-qcd}}(m_t^{\MSbar,\SM})$
denote SM QCD self energy contributions in the \MSbar scheme, with a
factor $\slashed{p}$ removed, evaluated at $p=m_t^{\MSbar,\SM}$
\cite{Fleischer:1998dw}:
\begin{align}
  \widetilde{\Sigma}_t^{(1),\text{SM-qcd}}(m_t^{\MSbar,\SM})
  &= -\frac{g_3^2}{12 \pi^2} \left[4 - 3 \ln\left(\frac{(m_t^{\MSbar,\SM})^2}{\Qmatch^2}\right)\right],
  \label{eq:Sigma_mt_1L_SMQCD} \\
\begin{split}
  \widetilde{\Sigma}_t^{(2),\text{SM-qcd}}(m_t^{\MSbar,\SM}) &=
    - \frac{g_3^4}{4608 \pi^4}
    \Bigg[396 \ln^2\left(\frac{(m_t^{\MSbar,\SM})^2}{\Qmatch^2}\right)
          - 1452 \ln\left(\frac{(m_t^{\MSbar,\SM})^2}{\Qmatch^2}\right) \\
          &\qquad\qquad\qquad - 48 \zeta(3) + 2053 + 16 \pi^2 (1+\ln 4)\Bigg] \,.
  \end{split}
  \label{eq:Sigma_mt_2L_SMQCD}
\end{align}
Here quantities with the superscript $^{\SM}$ are SM
quantities, renormalized in the \MSbar scheme.
Three-loop RGEs are used to run the SM parameters to the EW
scale, as is done in \SUSYHD and \FS/\HSSUSY. The matching to
experimental quantities is done at $Q=M_Z$ in exactly the same way as
for \FS/\HSSUSY described in the previous subsection, except that only
2-loop SM-QCD corrections are used to extract $y_t^\SM(M_Z)$.
Finally, the Higgs pole mass is calculated in the Standard Model at
the scale $M_t$ using the full \MSbar-renormalized 1-loop self energy.
The crucial advantage of this choice of matching conditions is that
the resulting Higgs boson mass is exact at the 1-loop level and
contains resummed leading logarithms to all
orders.  In particular, \FSTower does not neglect terms of
$\mathcal{O}(v^2/\MS^2)$.  This is in contrast to the pure EFT
approach, where already at the tree-level terms suppressed by powers
of $\MS$ originating from the mixing of the light with the heavy Higgs
are missing.  Thus, \FSTower has no ``EFT uncertainty'' \cite{Vega:2015fna}, which is
present in \SUSYHD and \HSSUSY.

For completeness and illustration, the equivalence of the two choices
of matching conditions, Eqs.~\eqref{eq:delta_lambda_1L} and
\eqref{eq:lambda_1L_tower}, up to power-suppressed terms is proven
analytically at the 1-loop level in
Appendix~\ref{sec:appendix-equivalence-proof}.

\section{Numerical results in the MSSM and differences between
  calculations}
\label{sec:MSSMtower_results}

In the present section we discuss numerical results for the lightest,
SM-like Higgs boson in the \MSSM. The results of \FSTower\ are compared
to the results of  \Softsusy, \SARAH/\SPheno\
and \SUSYHD\ and variants of the original \FS. We focus mainly on
analysing the differences between the calculations and their origins
as well as on discussing theory uncertainties.

\subsection{MSSM for $\boldsymbol{X_t=0}$}
\subsubsection{Results of \FSTower\ and fixed-order calculations}

We begin with the special case of zero \DRbar stop mixing,
$X_t(\MS)=0$, and a common value $\MS$ for all \DRbar\ SUSY
mass parameters, as defined in Eqs.~\eqref{eq:MSSM_MSUSY_scenario}.
In this special case it is known that the
2-loop threshold corrections $\Delta\lambda^{(2)}$ are numerically very
small, and the leading 2-loop contributions of ${\cal O}(\alpha_s\alpha_t)$
even vanish \cite{Bagnaschi:2014rsa}. As a result, \FSTower\ happens to be essentially as
accurate as if 2-loop instead of 1-loop threshold corrections for
$\lambda$ were implemented. Our comparisons to other calculations will
therefore be sensitive to differences which do not originate from
missing 2-loop threshold corrections but from other, more subtle
effects. 
\begin{figure}[tbh]
  \centering
  \includegraphics[width=0.49\textwidth]{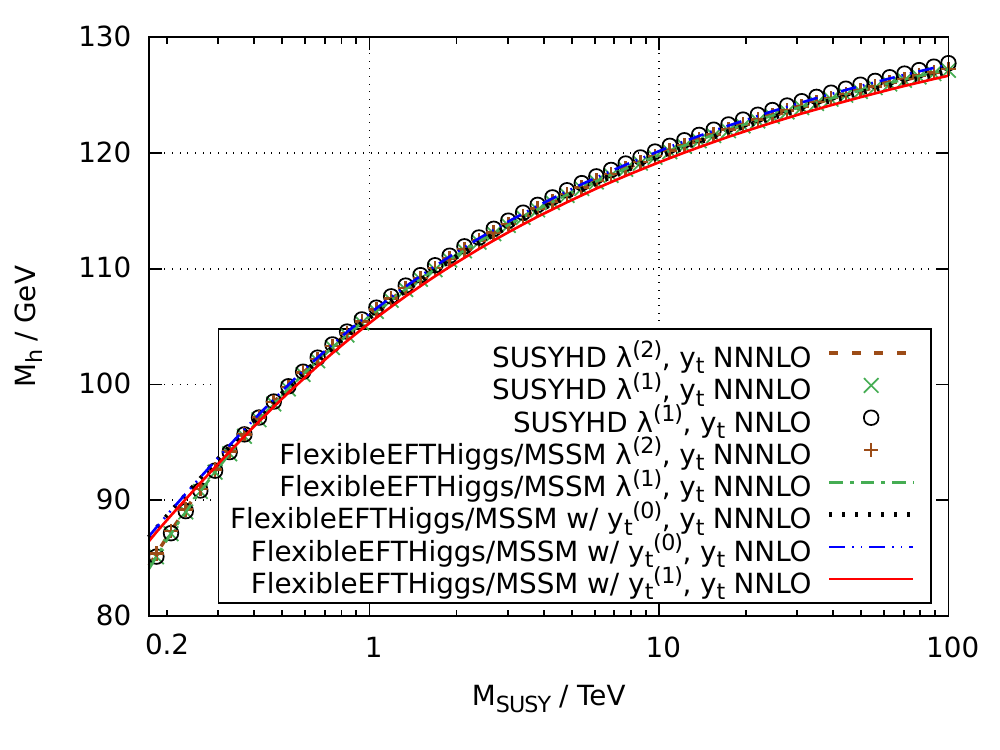}
  \includegraphics[width=0.49\textwidth]{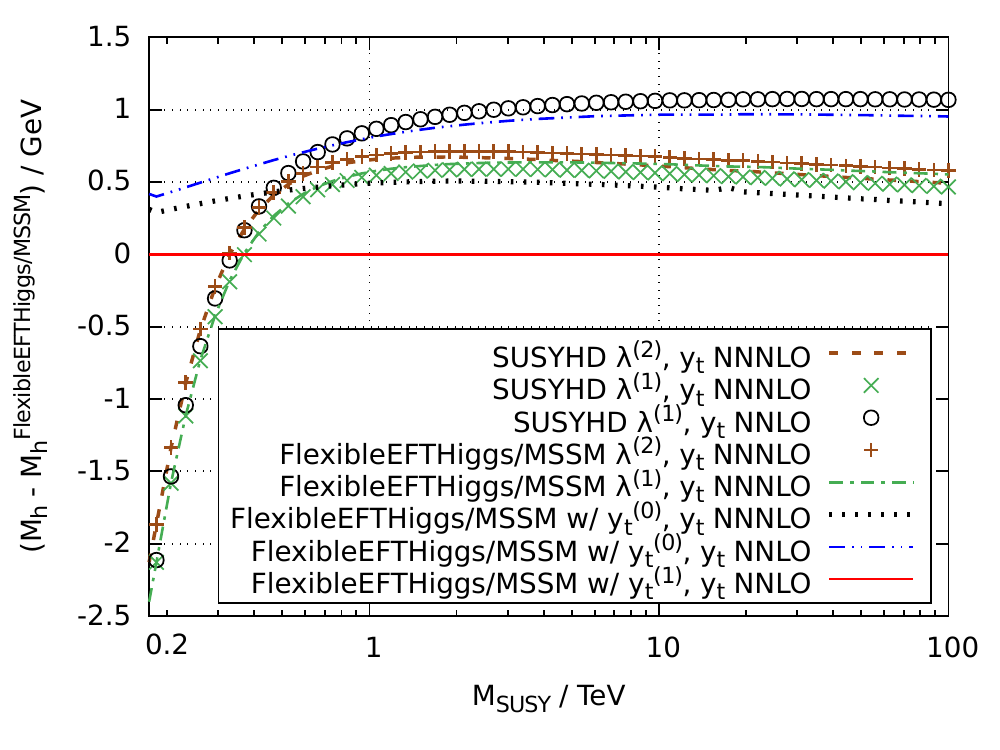}
  \caption{Influence of switching on or off different contributions to
    the lightest CP-even Higgs pole
    mass, $M_h$, in \SUSYHD
    and \FSTower for $\tan\beta = 5$, $X_t = 0$.}
  \label{fig:stepbystep}
\end{figure}
Figure~\ref{fig:stepbystep} compares \FSTower\ to \SUSYHD. It
demonstrates the validity of \FSTower\ and shows the numerical impact
of the various different design choices made in \FSTower\ and
\SUSYHD. The red solid line shows $M_h$ as a function of
$\MS$ for \FSTower\ with maximum precision,
i.e.\ with 1-loop mass matching conditions
Eqs.~\eqref{eq:polemassmatching}--\eqref{eq:matching_condition_Mt} at
the scale \MS, 3-loop running in the Standard Model, 1-loop matching
to the known low-energy parameters, including 2-loop QCD corrections
to $y_t^\SM$.
The other lines correspond to \SUSYHD\ and variants of \FSTower, where
the \SUSYHD-like calculation is transformed step by step into the
\FSTower-like one. We will now explain each step in detail.
\begin{itemize}
\item The brown dashed line corresponds to \SUSYHD with maximum
  precision.  The brown pluses correspond to \FSTower, where the
  calculation of all Standard Model parameters is performed using the
  same expressions as in \SUSYHD.  This means in particular that the
  fit formulas of Ref.~\cite{Buttazzo:2013uya} are used to obtain the
  running gauge and Yukawa couplings at the $M_t$ scale.  Thus, both
  programs use 2-loop threshold corrections to $\lambda(\MS)$ from
  Eq.~\eqref{eq:delta_lambda_sum}, 3-loop running in the Standard
  Model, calculation of $\alpha_s^{\MSbar,\SM}(M_t)$ using 4-loop QCD
  and 2-loop electroweak running from $M_Z$ to $M_t$ plus 3-loop
  matching, and calculation of $y_t^\SM(M_t)$ at NNNLO
  \cite{Buttazzo:2013uya}.  The two programs agree exactly with each
  other.\footnote{%
    For this reason the \FSTower version modified in this way might be
    viewed as a replica of \SUSYHD\ within the C++ framework of \FS.}

\item The green crosses and the green dash-dotted line correspond to
  \SUSYHD and the modified \FSTower respectively with only 1-loop matching of
  $\lambda$ at the high scale $\MS$.  The numerical difference from what would
  result from 2-loop matching is very small for large $\MS$, namely below
  $50\unit{MeV}$ for $\MS > 2\unit{TeV}$.  This confirms the statement
  that the 2-loop threshold correction is negligible for $X_t=0$ and a
  common SUSY mass scale.

\item The black dotted line corresponds to replacing the
  $\lambda$-matching, Eq.~\eqref{eq:delta_lambda_sum}, by the
  matching procedure of \FSTower,
  Eqs.~\eqref{eq:polemassmatching}--\eqref{eq:matching_condition_Mt},
  except that the equality of the top pole masses at \MS has been required at
  the tree-level only.  The Higgs pole mass matching is the essential
  design choice of \FSTower.
  The line converges to the $\lambda$-matching curves for large $\MS$,
  confirming that the two matching procedures become equivalent for
  $\MS\to\infty$.  For $\MS \lesssim 500\unit{GeV}$ the
  \SUSYHD-approach becomes unreliable.  The difference between the
  two matching procedures is formally of ${\cal
    O}((\text{tree-level, 1-loop})\times {v^2}/{\MS^2})$.  Terms of this order
  are ignored in \SUSYHD, but correctly taken into account in
  \FSTower, so the difference between the two matching procedures is a
  measure of part of the theory uncertainty of \SUSYHD.  In
  Ref.~\cite{Vega:2015fna} this theory uncertainty was labelled
  ``EFT uncertainty'', and the numerical result of
  Figure~\ref{fig:stepbystep} is compatible with the uncertainty estimate
  given in Ref.~\cite{Vega:2015fna}: For the scenario shown in
  Figure~\ref{fig:stepbystep} and $\MS > 1\unit{TeV}$ the difference is
  smaller than $200\unit{MeV}$.  For $\MS < 500\unit{GeV}$ the
  difference can reach up to $3\unit{GeV}$.

\item In the blue dashed-double-dotted line the low-scale
  computation of the running SM top Yukawa coupling has been changed,
  and the leading 3-loop QCD terms included so far have been switched
  off.  Even though the impact on the Higgs mass is formally of 4-loop
  order, the resulting numerical difference is rather
  sizeable, around $600\unit{MeV}$.  The importance of the 3-loop
  corrections to the top Yukawa coupling was already stressed in
  Refs.~\cite{Degrassi:2012ry,Buttazzo:2013uya,Vega:2015fna}.
  The black circles represent the
  equivalent change in \SUSYHD, where the 3-loop QCD corrections to
  the SM top Yukawa coupling are switched off.  In \SUSYHD the
  omission of this 3-loop correction leads to a change of the same size.

\item The red line shows the calculation in \FSTower.  It differs from
  the blue dashed-double-dotted line in the following ways: (i) $y_t^\MSSM(\MS)$ is calculated by
  matching the top pole mass at the 1-loop level (including 2-loop
  SM-QCD corrections) at \MS using
  Eq.~\eqref{eq:matching_condition_Mt}, (ii)
  $M_h$ is calculated at
  the scale $M_t$ by numerically solving Eq.~\eqref{eq:fixedmh} using
  the full momentum-dependent 1-loop Higgs self-energy, instead of
  setting the momentum to the \MSbar Higgs mass, $p^2 = m_h^2$, as
  done in \SUSYHD.  The inclusion of both changes leads to an
  approximately constant decrease of $M_h$ of about
  $1\unit{GeV}$.
\end{itemize}
\begin{figure}[htb]
  \centering
  \includegraphics[width=0.7\textwidth]{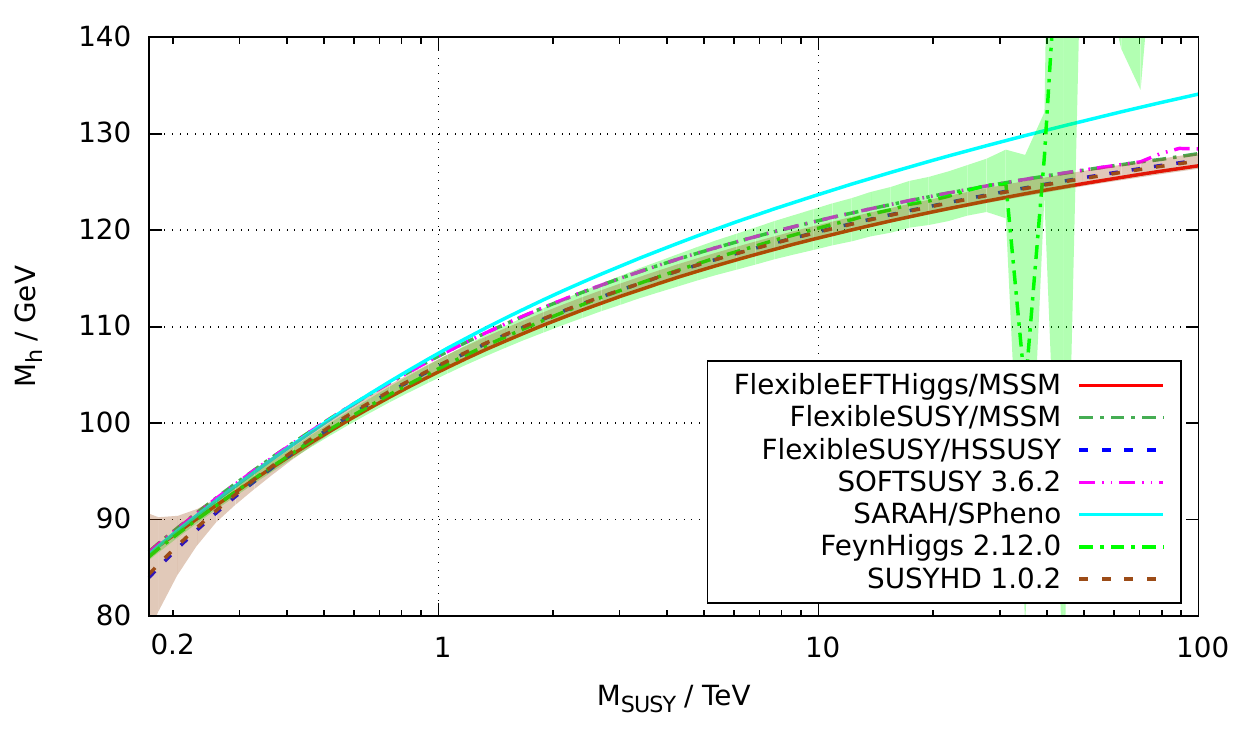}
  \caption{Comparison of predictions for $M_h$
     in the MSSM using the EFT with pole mass matching method
    (\FSTower/MSSM), the pure EFT calculation (\FS/HSSUSY and \SUSYHD) and
    the diagrammatic calculations (\FS/MSSM, \SARAH/\SPheno, \Softsusy and
    \FeynHiggs) for $\tan\beta = 5$ and $X_t = 0$.  The green and
    brown bands show the theory uncertainty as estimated by \FeynHiggs
    and \SUSYHD, respectively.}
  \label{fig:MSSM-scale}
\end{figure}
Figure~\ref{fig:MSSM-scale} compares the results of \FSTower\ and
\SUSYHD/\HSSUSY\ to the fixed-order results of \Softsusy, \SARAH/\SPheno, and
the original \FS. For comparison, also the results of \FeynHiggs\ are
shown; the differences between the recent versions of \FeynHiggs\ and
other calculations have been discussed e.g.~in
Refs.~\cite{Borowka:2015ura,Vega:2015fna,Draper:2016pys}.
In line with the discussion of
Figure~\ref{fig:stepbystep}, \SUSYHD\ and \FSTower\ agree up to $0.5\unit{GeV}$ at
high $\MS$, but \SUSYHD\ deviates more strongly at low $\MS$ due to the
missing terms of ${\cal O}((\text{tree-level, 1-loop})\times{v^2}/{\MS^2})$.

Figure~\ref{fig:MSSM-scale} shows in addition that \FSTower\ agrees at
low $\MS$ with all fixed-order calculations. This is the consequence
of the choice of the pole-mass matching condition Eq.~\eqref{eq:polemassmatching}, and it reflects the fact that \FSTower\ corresponds to an exact 1-loop calculation plus resummation of higher-loop logarithms.

\subsubsection{Theory uncertainty estimations}
\label{sec:FStower_Xt=0_uncertainty_estimation}

The comparisons shown in the previous figures allow us to make
several observations about various ways to estimate theory
uncertainties.
Ref.~\cite{Vega:2015fna} has divided the theory uncertainty of
\SUSYHD\ into three parts, one of which is the ``EFT uncertainty'' due
to truncating the low-energy EFT at the dimension-4 level
(i.e.\ taking the renormalizable SM as the EFT). This EFT uncertainty
arises from missing power-suppressed terms of ${\cal
    O}((\text{tree-level, 1-loop})\times {v^2}/{\MS^2})$; hence it becomes
large for $\MS\lesssim 500\unit{GeV}$. As mentioned in the context of
Figure~\ref{fig:stepbystep}, the choice of the Higgs pole mass 
matching condition in \FSTower\ avoids this uncertainty by construction.
As a consequence, the difference between \SUSYHD\ and
\FSTower\ at low $\MS$ can be regarded as a measure of the
EFT uncertainty of \SUSYHD. 

In \FSTower\ the Higgs mass prediction is exact at the 1-loop level
due to the 1-loop Higgs pole mass matching condition.  At the
2-loop level, power-suppressed as well as non-power-suppressed (but
non-logarithmic) terms are missing; these will be discussed in the
next subsection.

Now we turn to an extensive discussion of the differences between EFT
and fixed-order calculations at high
$\MS$, and on the resulting theory uncertainty of the fixed-order
calculations. Figure~\ref{fig:MSSM-scale} shows that at high $\MS$,
the two fixed-order calculations of \SPheno\ and
\FS/\Softsusy\ deviate significantly from each other, and that \FS/\Softsusy\ agrees
well with the EFT calculations.
These differences originate from $\ge3$-loop terms, which are
taken into account differently. For a deeper understanding and
illustration, we derive the leading 3-loop logarithms for all these
approaches:
\begin{itemize}
\item The all-order leading-log part of the EFT results of \FSTower\
  and \SUSYHD\ can be obtained analytically by integrating 1-loop RGEs
  and using tree-level matching at the high and low scales.
\item \SPheno, \Softsusy\ and \FS\ do a fixed-order 2-loop computation
  of $M_h$ in the \DRbar-scheme at the scale $\MS$. Once the running
  parameters at the scale $\MS$ are replaced by their low-energy
  counterparts via the definitions of Section
  \ref{sec:fixed_order_calculations}, implicit terms of $\ge3$-loop order are
  generated. These implicit higher-order terms are different in \FS/\Softsusy
  and \SPheno, because of the different definitions of the top Yukawa
  coupling in Eqs.~\eqref{eq:fixedmt_FlexibleSUSY},
  \eqref{eq:fixedmt_SPheno}, respectively.
\end{itemize}
The leading logarithms in $\alpha_s$ and $\alpha_t$ up to 3-loop level
obtained in these ways can be written as  
\begin{align}
  \begin{split}
    (M_h^2)^{X}& = m_h^2 + \vlow^2 \ytlow^4\Big[12 t_S \kappaL -12
    t_S^2 \kappaL^2 \left(16 \gs^2 -3 \ytlow^2 \right) 
+4 t_S^3
    \kappaL^3 \Delta^X_\text{3LLL}
+\cdots\Big],
\\
\Delta^X_\text{3LLL} &=\begin{cases}
736 \gs^4 -240 \gs^2 \ytlow^2 -99 \ytlow^4
&(X=\EFT), \\
\frac{736}{3} \gs^4 +144
   \gs^2 \ytlow^2 -\frac{351}{2} \ytlow^4
&(X=\text{\FS/\Softsusy}), \\
\frac{992}{3} \gs^4 +240
   \gs^2 \ytlow^2 -\frac{297}{2}
   \ytlow^4
&(X=\text{\SPheno}) ,
\end{cases}
  \end{split}
\label{eq:Mh_LL_3_approaches_low}
\end{align}
where $X$ denotes the calculational approach ($X=\EFT$ denotes
\FSTower\ or \SUSYHD) and $\kappaL=1/(16\pi^2)$, $t_S=\ln(\MS/M_t)$,
$\vlow=v^\SM(M_t)$, $\gs=g_3^\SM(M_t)$, $\ytlow=y_t^\SM(M_t)$.  We
have worked in the large-$\tan\beta$ limit, and the
details of this calculation are shown in
Appendix~\ref{sec:appendix-leading-logs}; for the $\EFT$-case similar
analytical results including subleading logarithms are presented in
Refs.~\cite{Martin:2007pg,Draper:2013oza}.

By construction, all codes agree at the 2-loop level, and the EFT
calculations contain the correct 3-loop leading log. However, the
implicit 3-loop leading logs of \SPheno\ and \FS/\Softsusy
in \eqref{eq:Mh_LL_3_approaches_low} are both incorrect, and different.\footnote{%
In Ref.~\cite{Draper:2013oza}, the EFT calculation was compared to
``fixed-order calculations''. In that reference, ``fixed-order'' was
simulated via perturbative truncation of the full EFT result. Hence,
even at the 3-loop  order, the ``fixed-order'' calculations of 
Ref.~\cite{Draper:2013oza} always agree with the EFT result. This is
different from the concrete fixed-order calculations implemented in
\SPheno, \Softsusy\ and \FS, which are 2-loop codes but nonetheless
include partial corrections at the $\ge3$-loop level.
}
\begin{figure}[thb]
  \centering
  \includegraphics[width=0.7\textwidth]{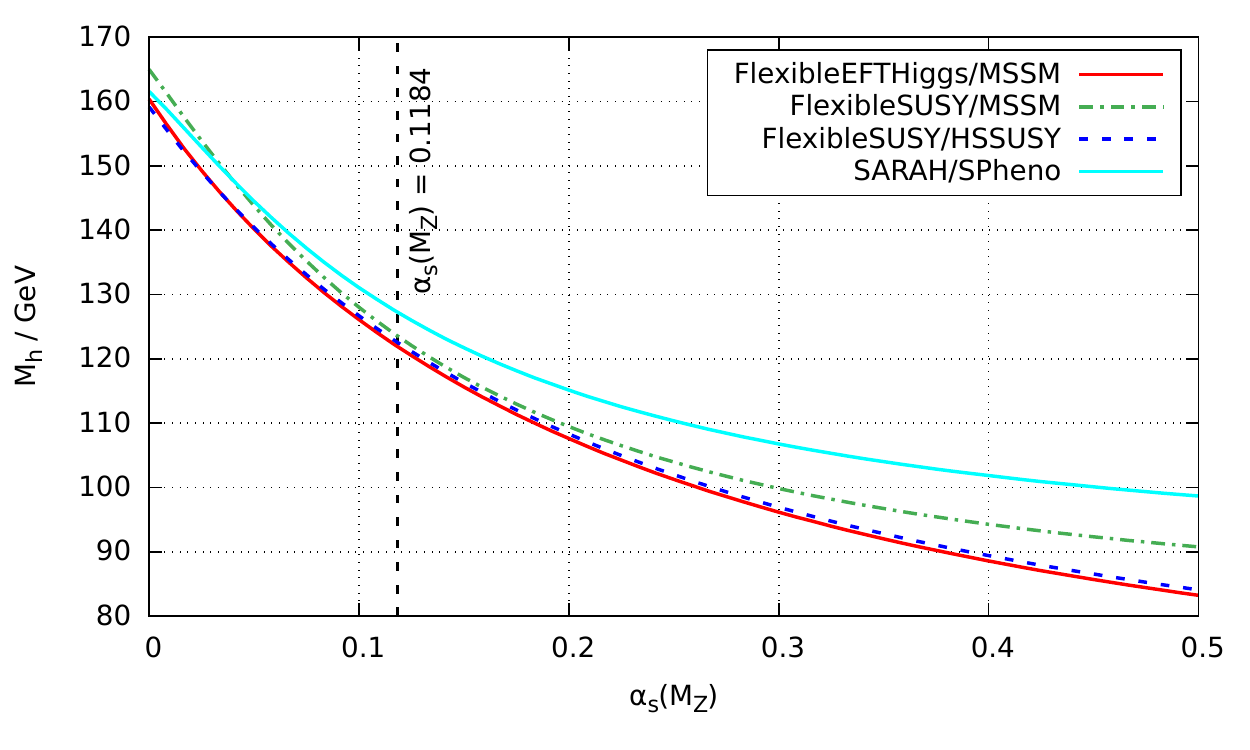}
  \caption{Accidentality of the agreement between EFT and (\DRbar) fixed-order 
    results:  $M_h$ in the MSSM as a function
    of $\alphaSMZ$ for $\MS = 20\unit{TeV}$,
    $\tan\beta = 5$, $X_t = 0$, $M_t = 173.34 \unit{GeV}$,
    $m_b^{\MSbar,\text{SM(5)}}(m_b) =
    \frac{4.18\unit{GeV}}{100}$, $M_\tau =
    \frac{1.777\unit{GeV}}{100}$.}
  \label{fig:accidentality}
\end{figure}
The analytical results show why \FS/\Softsusy and \SPheno deviate from
each other at high $\MS$. They also make it clear that the difference
between \FS/\Softsusy and \SPheno should be regarded as part of the
theory uncertainty of both programs. In fact, inspection of the
coefficients of the 3-loop leading logs in
Eqs.~\eqref{eq:Mh_LL_3_approaches_low} indicates that the theory
uncertainty of both \FS/\Softsusy and \SPheno could be significantly
larger than their difference. In this sense it is surprising that the
EFT results are actually close to \FS/\Softsusy\ but far away from
\SPheno in Figure~\ref{fig:MSSM-scale}.  The reason for this is an
accidental cancellation between the ${\cal O}(\alpha_s^2\alpha_t)$
terms in Eqs.~\eqref{eq:Mh_LL_3_approaches_low} and formally
subleading terms.  This cancellation can be made more obvious, if one
expresses $M_h$ in terms of the Standard Model \MSbar parameters at
$\MS$:
\begin{align}
(M_h^2)^{X}&= \nonumber
m_h^2
+ \vhigh^2 
\ythigh^4\Big[12 t_S \kappaL
+12 t_S^2 \kappaL^2 
\left(16 \gshigh^2 - 9\ythigh^2 \right)
+
4 t_S^3\kappaL^3 \bar{\Delta}^X_\text{3LLL}
+\ldots\Big],
\\
\bar{\Delta}^X_\text{3LLL} &=\begin{cases}
736 \gshigh^4-672 \gshigh^2 \ythigh^2+90
   \ythigh^4
&(X=\EFT),\\
\frac{736 \gshigh^4}{3}-288 \gshigh^2
   \ythigh^2+\frac{27 \ythigh^4}{2}
&(X=\text{\FS/\Softsusy}),\\
\frac{992 \gshigh^4}{3}-192 \gshigh^2
   \ythigh^2+\frac{81 \ythigh^4}{2}
&(X=\text{\SPheno}),
\end{cases}
\label{eq:Mh_LL_3_approaches}
\end{align}
where $\vhigh = v^\SM(\MS)$, $\gshigh = g_3^\SM(\MS)$, $\ythigh =
y_t^\SM(\MS)$.  In the EFT result there is an accidental, numerical
cancellation between the different 3-loop terms which has been observed and
discussed in Refs.~\cite{Martin:2007pg,Draper:2013oza}. In spite of different numerical
coefficients, a similar cancellation happens in \FS/\Softsusy and (to a smaller
extent) in \SPheno. As a consequence, the EFT results are
closer to the fixed-order ones than what could be expected. 

To
highlight this accidentality we show Figure~\ref{fig:accidentality},
which displays $M_h$ in the three approaches for different values of
$\alphaSMZ$.  $\MS$ is set to $20\unit{TeV}$ to
amplify the 3-loop leading logarithms. The plot shows that accidentally
the fixed-order \FS and the EFT calculations agree around the true value of
$\alphaSMZ \approx 0.1184$.

\begin{figure}[tbh]
  \centering
  \includegraphics[width=0.7\textwidth]{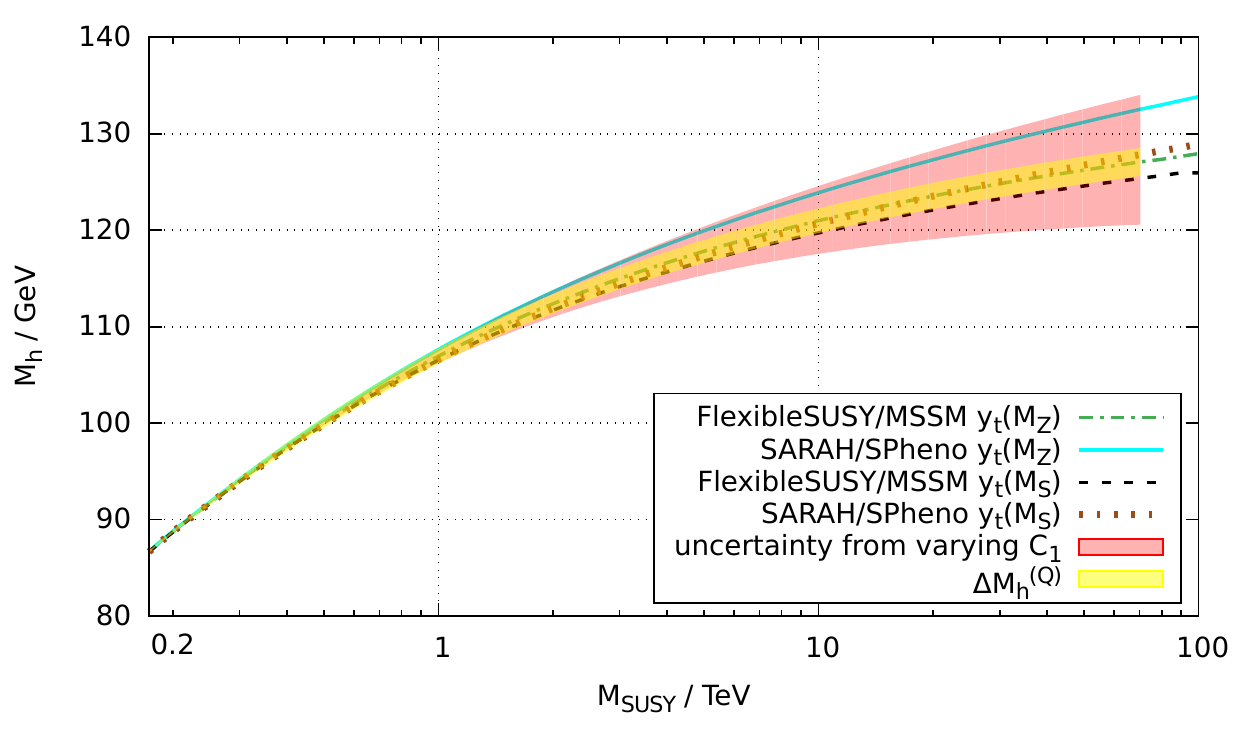}
  \caption{Illustration of the theory uncertainty estimate
    $\DeltaMhFSYt$ of the fixed-order calculations of \FS and \SARAH/\SPheno
    using four different ways to calculate $y_t^{\MSSM}$.
    We choose $\tan\beta = 5$ and $X_t = 0$.  The red
    band shows the variation of $M_h$ when the coefficient $C_1$ is
    varied within the interval $[-184/9,184/9]$.  The yellow uncertainty
    band shows $\DeltaMhQ$, defined as the variation of $M_h$ when the
    renormalization scale at which $M_h$ is calculated is varied
    within $[\MS/2, 2\MS]$.}
  \label{fig:MSSM-scale-yt-variants}
\end{figure}

After these considerations we turn to the question of estimating the
theory uncertainty of the \DRbar\ fixed-order calculations.
As noted above, the difference between the fixed-order MSSM
calculations in \FS/\Softsusy and \SPheno\ is due to a different
treatment of the 3-loop leading logarithms, so it can be regarded as
an estimate for part of the
theory uncertainty of the two calculations. On a more general level,
we therefore discuss two ways to estimate the theory uncertainty of
these fixed-order calculations:
\paragraph{1. Using known MSSM higher-order results:}
  In the MSSM, we know that the leading 1-loop contributions are
  governed by the running top mass $m_t$. On the other hand the full
  2-loop MSSM SUSY-QCD contributions to $m_t$ are known
  \cite{Bednyakov:2002sf}.  Evaluating
  Eq.~\eqref{eq:fixedmt_FlexibleSUSY} at the 2-loop leading log
  level taking into account the full 2-loop SUSY-QCD contributions of
  \cite{Bednyakov:2002sf} would shift the running top mass by
  \begin{align}
    \Delta m_t^{(2)}(M_Z) = -\frac{184}{9} \frac{g_3^4 M_t}{(4\pi)^4}
    \ln^2\frac{\MS}{M_Z} .
  \end{align}
Thus, to estimate the theory uncertainty, we can add the term
  \begin{align}
    C_1 \frac{g_3^4 M_t}{(4\pi)^4} \ln^2\frac{\MS}{M_Z}
  \end{align}
  to the r.h.s.\ of Eq.~\eqref{eq:fixedmt_FlexibleSUSY} and vary the
  coefficient $C_1$ within the interval $[-184/9,184/9]$. This
  changes the  3-loop leading
  logarithms in the Higgs boson mass prediction by a motivated amount. The
  resulting uncertainty band is shown in red in
  Figure~\ref{fig:MSSM-scale-yt-variants}.  We find that this
  uncertainty band contains both the \FS curve (green dash-dotted
  line) and the \SPheno curve (turquoise solid line).
\paragraph{2. Generating higher-order terms:}
  Another option is to change the calculation of $y_t^\MSSM$
  such that changes of higher-order are automatically induced.
  The different treatment of $y_t^\MSSM$ in \FS\ and \SPheno,
  i.e.\ using Eq.~\eqref{eq:fixedmt_FlexibleSUSY} or
  \eqref{eq:fixedmt_SPheno}, provides two examples. There are further
  motivated possibilities to define $y_t^\MSSM$, namely to employ
  either Eq.~\eqref{eq:fixedmt_FlexibleSUSY} or
  Eq.~\eqref{eq:fixedmt_SPheno} at the renormalization scale $\MS$
  instead of at $M_Z$.  All four variants to
  calculate $y_t^\MSSM$ are equal at the 1-loop level but different
  at the 2-loop level, so the resulting Higgs masses differ by
  3-loop terms. In
  Figure~\ref{fig:MSSM-scale-yt-variants} we show the four
  corresponding Higgs mass
  predictions. The two new ones are shown as the black dashed line and
  the brown dotted line, respectively.  We
  find that the four approaches to calculate $y_t^\MSSM$ are
  distributed within the red uncertainty band. Their differences thus
  represent an 
  alternative way to estimate the theory uncertainty from the
  missing 3-loop leading logarithms in the fixed-order
  calculations.  We therefore define
  \begin{align}
    \DeltaMhFSYt = \max_{
      y_t^{(i)},\,y_t^{(j)} \in \left\{
        y_t^{\eqref{eq:fixedmt_FlexibleSUSY}}(M_Z),
        y_t^{\eqref{eq:fixedmt_SPheno}}(M_Z),
        y_t^{\eqref{eq:fixedmt_FlexibleSUSY}}(\MS),
        y_t^{\eqref{eq:fixedmt_SPheno}}(\MS)
      \right\}
    } \left| M_h(y_t^{(i)}) - M_h(y_t^{(j)}) \right| ,
    \label{eq:def_DMh4yt}
  \end{align}
  where $y_t^{\eqref{eq:fixedmt_FlexibleSUSY}}(Q)$ refers to the
  definition of Eq.~\eqref{eq:fixedmt_FlexibleSUSY} and
  $y_t^{\eqref{eq:fixedmt_SPheno}}(Q)$ refers to
  Eq.~\eqref{eq:fixedmt_SPheno} evaluated at the scale $Q$.
  The advantage of this second way is that it can be
  applied also in non-minimal models, where the 2-loop contributions
  to $y_t^{\text{SUSY}}$ are unknown.

Another frequently used way to estimate the theory uncertainty is to
vary the renormalization scale $\QFOMS$ at which $M_h$ is calculated
and the loop-corrected EWSB conditions are solved.  The variation
interval is usually chosen to be $[Q_0/2,
2Q_0]$, where $Q_0$ is the default
renormalization scale to be used to calculate the Higgs pole mass in
the chosen approach.  In the fixed-order programs $Q_0 =
\MS$ is used, while in \FSTower we use $Q_0 = M_t$.
\begin{figure}[tbh]
  \centering
  \includegraphics[width=0.49\textwidth]{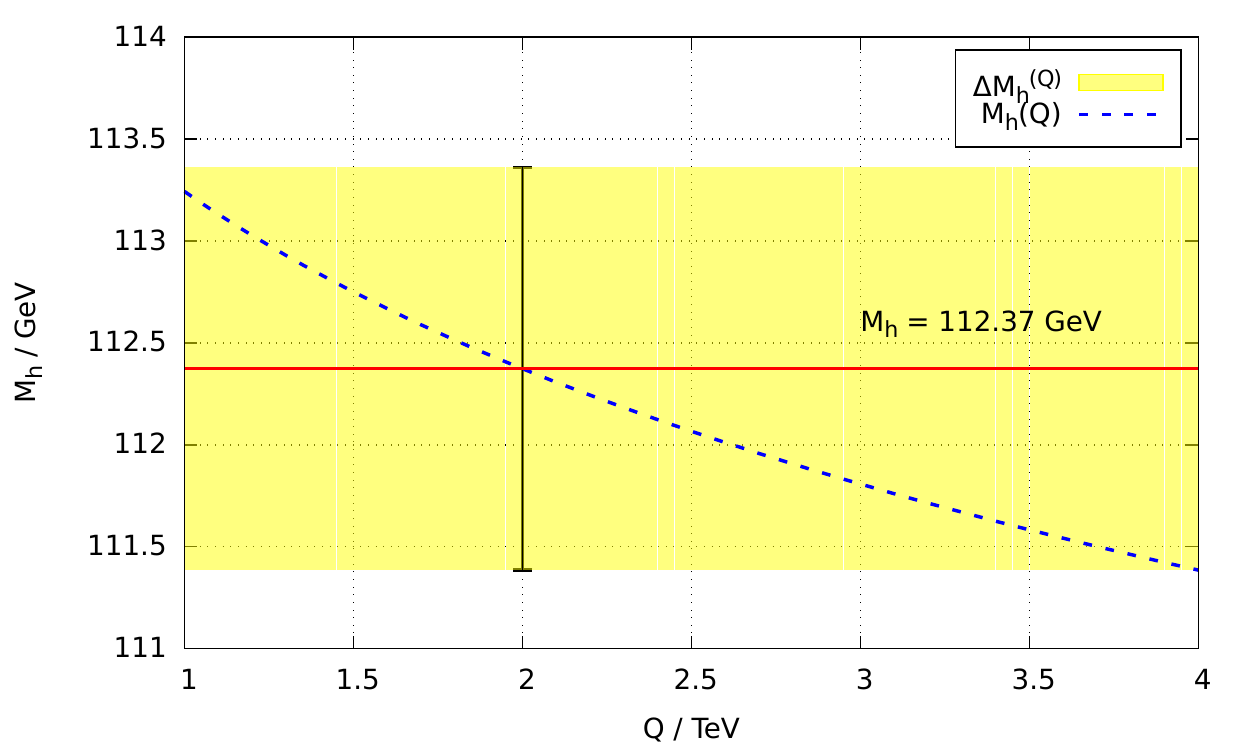}
  \includegraphics[width=0.49\textwidth]{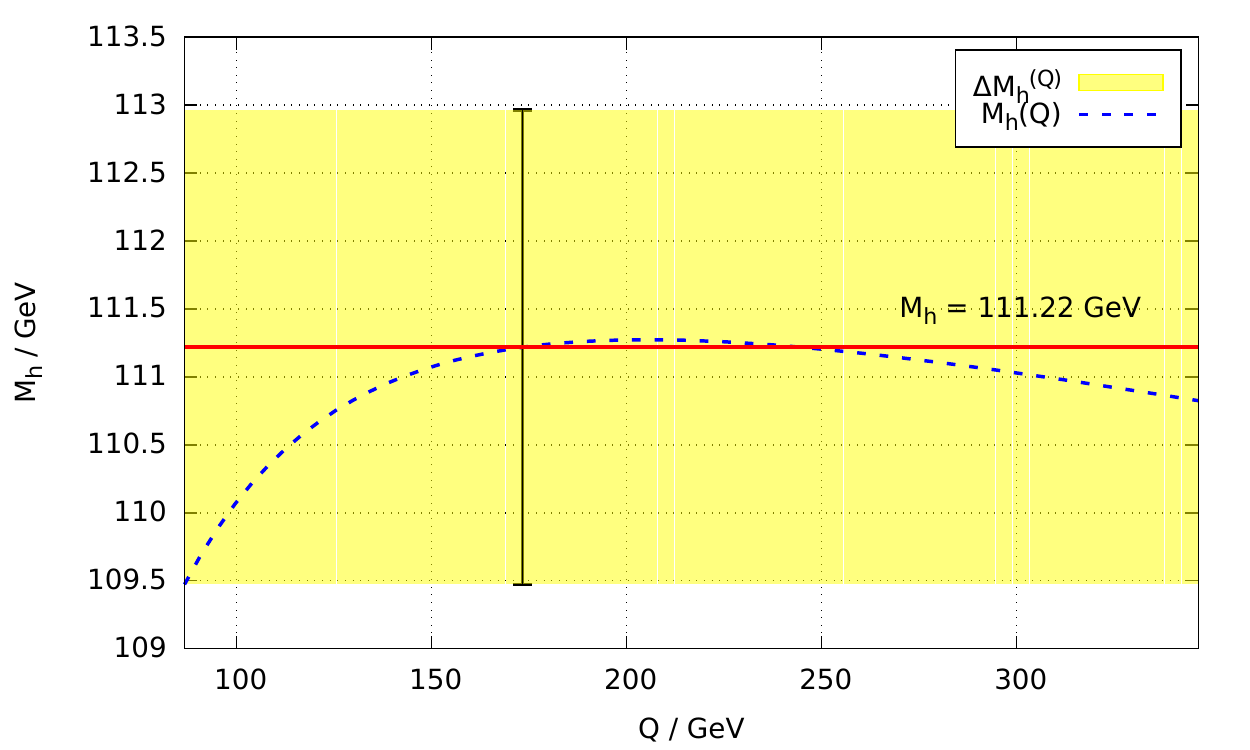}
  \caption{Illustration of the theory uncertainty estimate $\DeltaMhQ$
    for the fixed-order calculations of \FS and \SARAH/\SPheno (left panel)
    as well as \FSTower (right panel).  The blue dashed line shows
    $M_h$ in the MSSM for $\tan\beta = 5$,
    $X_t = 0$, $\MS = 2\unit{TeV}$ as a function of the
    renormalization scale $Q$.  The black vertical error bar is placed at the
    default value of the renormalization scale, $Q_0$, and the red
    horizontal line marks the corresponding Higgs pole mass,
    $M_h(Q_0)$.  The black error bar and the yellow band show the
    resulting uncertainty estimate $\DeltaMhQ$.}
  \label{fig:Mh_Q}
\end{figure}
Figure~\ref{fig:Mh_Q} shows $M_h$ as a function of $Q$ in the MSSM
calculated for $\tan\beta = 5$, $X_t = 0$, $\MS = 2\unit{TeV}$ with \FS
in the fixed-order approach (left panel) and with \FSTower (right
panel).  The renormalization scale has been varied within the interval
$[Q_0/2, 2Q_0]$.  In each approach one can see that the sizes of the
resulting upwards and downwards variations of the Higgs pole mass is
not equal and might even be highly non-linear.  Due to this effect, we
define the uncertainty $\DeltaMhQ$ to be
\begin{align}
  \DeltaMhQ = \max_{Q\in [Q_0/2, 2Q_0]}\left|M_h(Q_0) - M_h(Q)\right| .
  \label{eq:def_DMhQ}
\end{align}
Thus, in this scenario we obtain $\DeltaMhQ = 1.0\unit{GeV}$ for the
fixed-order approach, and $\DeltaMhQ = 1.7\unit{GeV}$ for \FSTower.
The yellow band in Figure~\ref{fig:MSSM-scale-yt-variants} shows the
variation of $M_h$ in the fixed-order approach when $\QFOMS$ is varied
within the interval $[\MS/2, 2\MS]$.

By construction, the width of this band, i.e.~the magnitude of
$2\times\DeltaMhQ$ is given by terms of ${\cal
  O}(\text{3-loop}\times\ln^2(\MS/M_Z)\times\ln(2))$ or  ${\cal
  O}(\text{2-loop}\times\ln(\MS/M_Z)\times\ln(2))$, where in the
latter case only 2-loop contributions {\em beyond} the ${\mathcal O}((\alpha_t
+ \alpha_b)^2 + (\alpha_t + \alpha_b) \alpha_s + \alpha_\tau^2)$ can
contribute. This should be contrasted with the magnitude of
$\DeltaMhFSYt$ from \eqref{eq:def_DMh4yt}, which is a measure of the 
leading missing/incorrect terms of ${\cal
  O}(\text{3-loop}\times\ln^3(\MS/M_Z))$. Hence the two uncertainty
estimations are sensitive to different contributions, but particularly
$\DeltaMhQ$ alone would underestimate the theory uncertainty at high
$\MS$. 

\subsection{MSSM for $\boldsymbol{X_t\ne 0}$}\label{sec:MSSM_for_Xt_ne_0}
\subsubsection{Results of \FSTower\ and fixed-order calculations}

Now we turn to the MSSM with $X_t \ne 0$. The main new aspect is that
the 2-loop threshold correction for $\lambda(\MS)$, which is not
implemented in \FSTower, is now non-negligible. Hence, we can in particular
discuss the theory uncertainty of \FSTower\ from these missing
non-logarithmic 2-loop contributions. However, our analysis is
intended to be more general. It aims to be applicable also to the case
of the non-minimal SUSY models discussed in the subsequent sections,
as well as in the future when the 2-loop threshold correction is implemented in
\FSTower. It might also shed light on the theory uncertainty of
existing programs such as \SUSYHD.

\begin{figure}[tbh]
  \centering
  \includegraphics[width=0.49\textwidth]{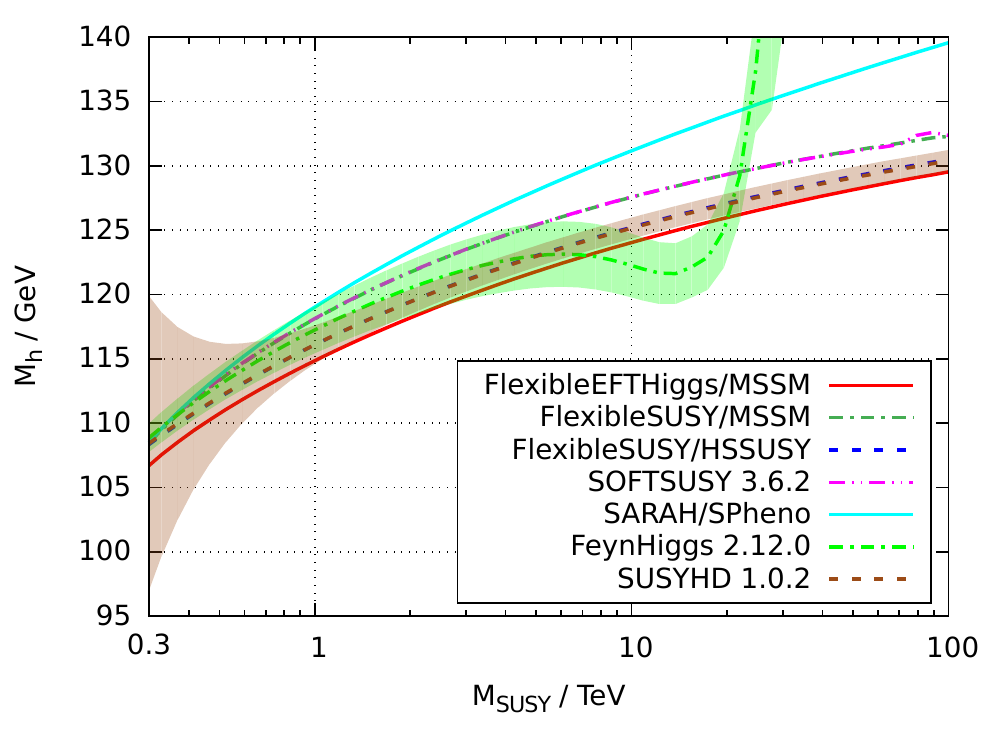}
  \includegraphics[width=0.49\textwidth]{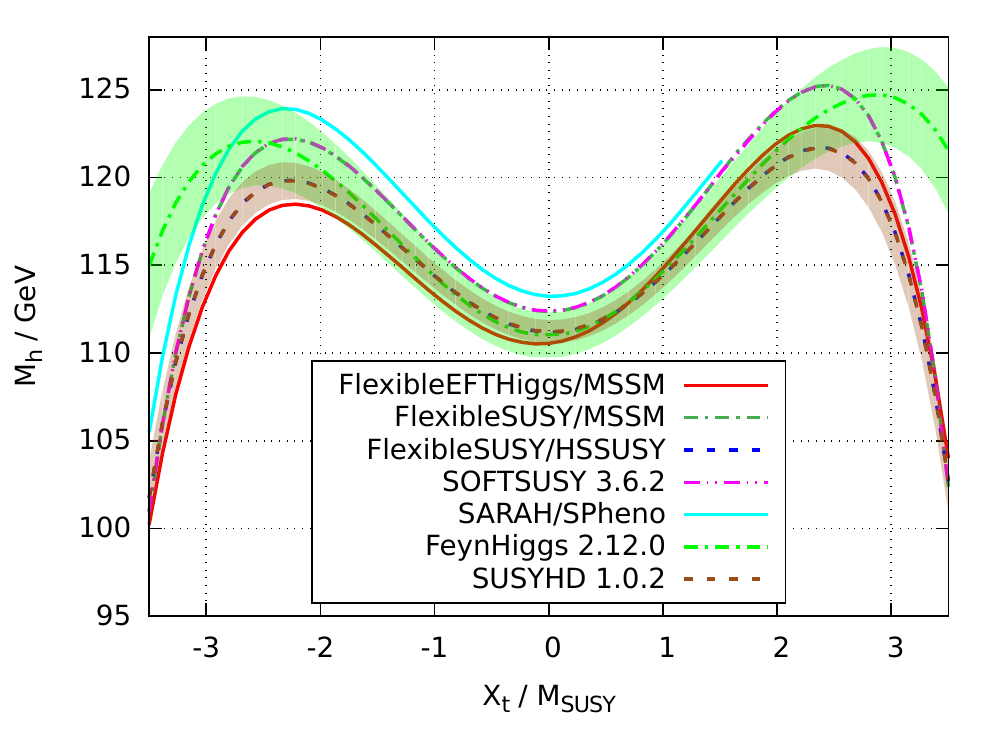}
  \caption{Comparison of predictions for $M_h$
    in the MSSM using the EFT with pole mass matching method
    (\FSTower/MSSM), the pure EFT calculation (\FS/HSSUSY and \SUSYHD) and
    the diagrammatic calculations (\FS/MSSM, \SARAH/\SPheno, \Softsusy and
    FeynHiggs) for $\tan\beta = 5$. The green and brown bands show the
    theory uncertainty as estimated by \FeynHiggs and \SUSYHD,
    respectively. In the left panel we fix $X_t = -2\MS$ and vary \MS.
    In the right panel we fix $\MS = 2\unit{TeV}$ and vary $X_t$.}
  \label{fig:MSSM-Xt}
\end{figure}

In Figure~\ref{fig:MSSM-Xt} we show $M_h$ in the MSSM as a function of
$\MS$ for $X_t=-2\MS$ in the left panel, and $M_h$ as a function of
$X_t$ for $\MS=2\unit{TeV}$ in the right panel, for some publicly
available spectrum generators.  The Higgs boson mass is calculated
using \FSTower (red solid line), \FS (green dash-dotted line),
\Softsusy (pink dashed-double-dotted line), \SARAH/\SPheno
(turquoise solid line), \FeynHiggs (light green dash-dotted
line) and \SUSYHD (brown dashed line).

The large difference between \SPheno\ and \FS/\Softsusy in these two
figures is again due to the different, incorrect $\ge 3$-loop leading
logs.  As Figure~\ref{fig:MSSM-Xt} shows, the difference is
increasing with $\MS$, and thus should be regarded as an estimate of
part of the theory uncertainty of \SPheno\ and \FS/\Softsusy.

For $X_t=0$, \SUSYHD\ and \FSTower\ differ by around $1\unit{GeV}$,
which corresponds mainly to the inclusion of higher-order terms in the
matching of $y_t$ at $\MS$, to 1-loop terms suppressed by ${v^2}/{\MS^2}$,
which are missing in \SUSYHD, and to the
different definition of the running top mass at the low scale.  As can
be seen in Figure~\ref{fig:MSSM-Xt}, for $X_t\ne 0$ the difference
between \SUSYHD and \FSTower can become larger, due to the missing
2-loop $M_h$ matching in \FSTower.  Still, in accordance with the
non-logarithmic nature of the 1-loop threshold corrections, the difference
between \SUSYHD\ and \FSTower\ does not significantly increase with $\MS$.

\subsubsection{Theory uncertainty estimations}
\label{sec:FStoweruncertainty}

Now we turn to estimating the theory uncertainty of \FSTower. As
explained in Section~\ref{sec:FStower_Xt=0_uncertainty_estimation}
it has no ``EFT uncertainty'', because power-suppressed terms are
automatically taken into account up to the 1-loop level. But
\FSTower is missing the 2-loop threshold corrections in its current
version, leading to a theory uncertainty.   We propose several methods to
estimate the theory uncertainty of
$M_h$ in \FSTower originating from these missing 2-loop threshold
corrections:
\begin{figure}[tbh]
  \centering
  \includegraphics[width=0.49\textwidth]{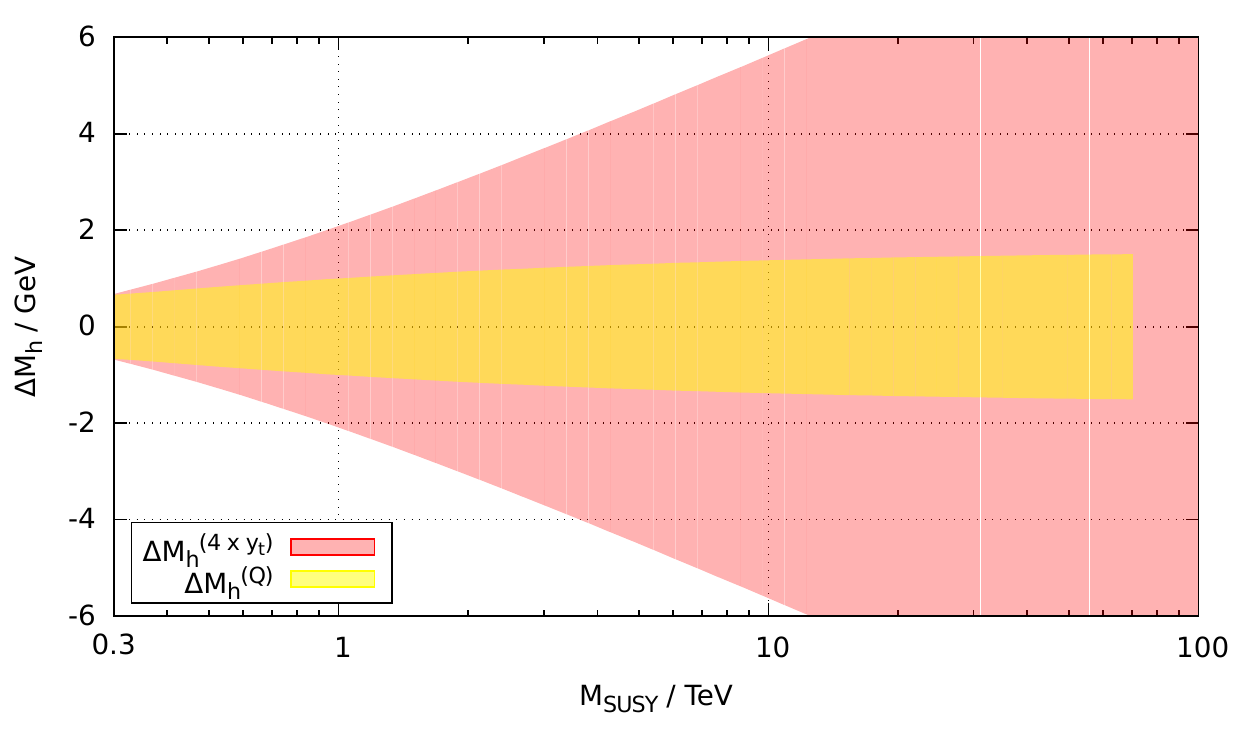}
  \includegraphics[width=0.49\textwidth]{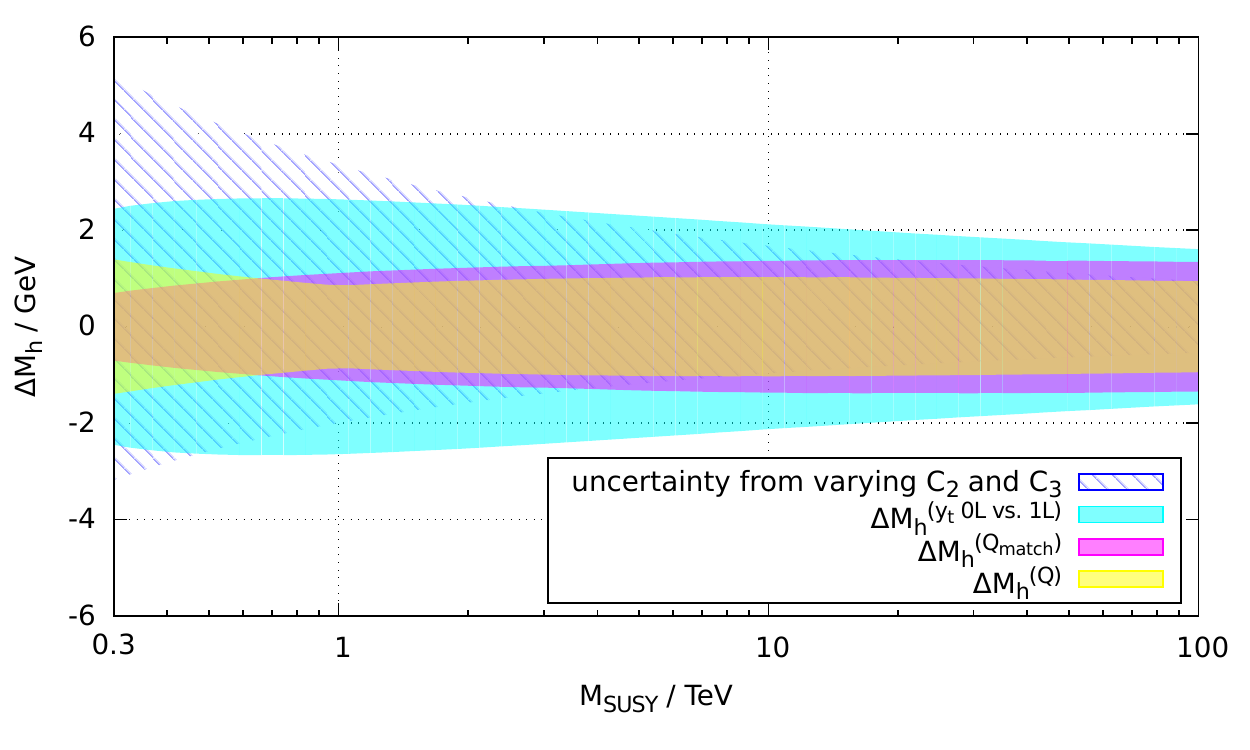}
  \includegraphics[width=0.49\textwidth]{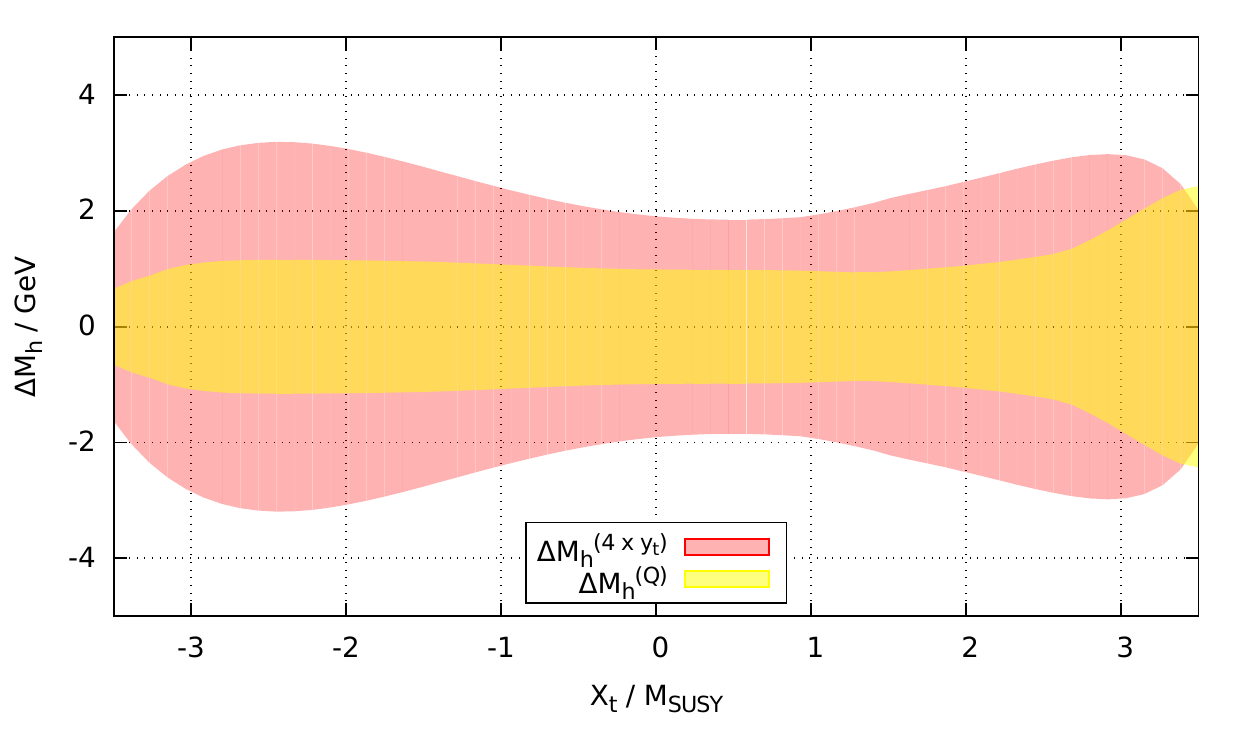}
  \includegraphics[width=0.49\textwidth]{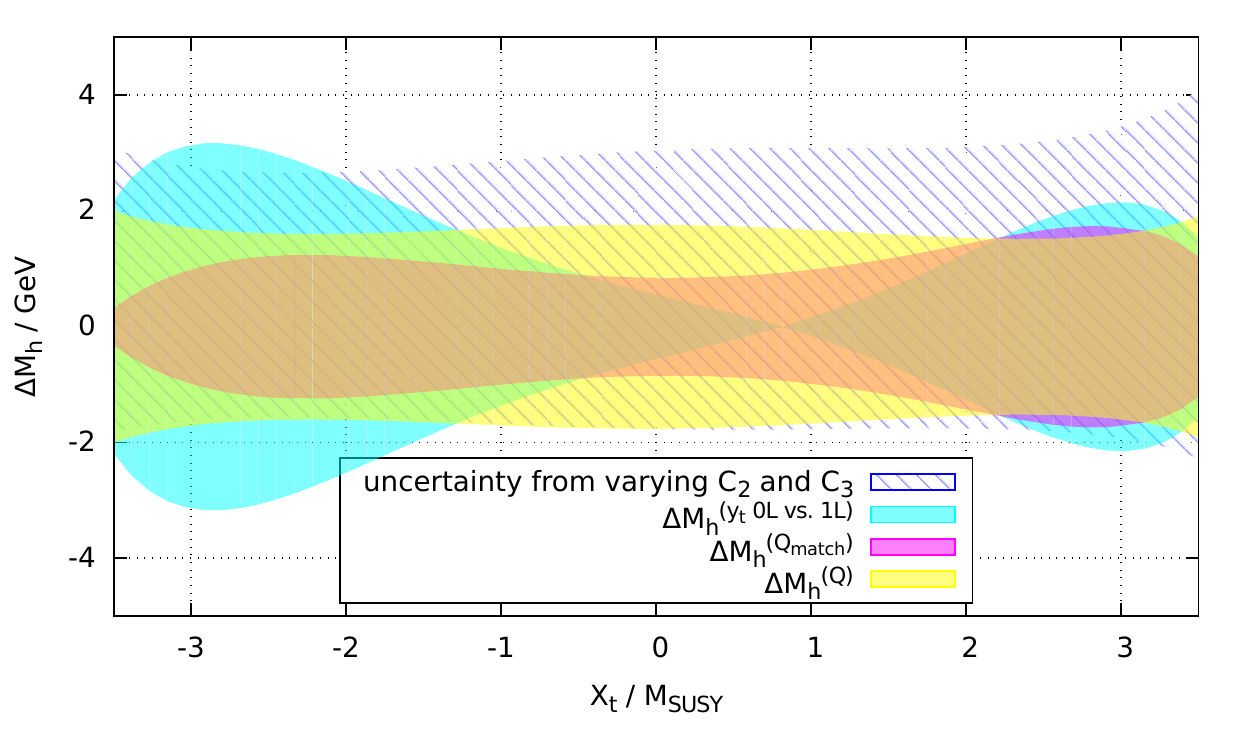}
  \caption{Uncertainty estimates for the fixed-order 2-loop
    calculations with \FS and \SARAH/\SPheno  and for \FSTower. Left panels: fixed-order uncertainty
    estimates of Sec.~\ref{sec:FStower_Xt=0_uncertainty_estimation}
    using renormalization scale variation, $\DeltaMhQ$ 
    (yellow) and from the different top Yukawa definitions,
    $\DeltaMhFSYt$ (red).  Right panels: \FSTower
    uncertainty estimations of Sec.~\ref{sec:FStoweruncertainty} using
    renormalization scale uncertainty, 
    $\DeltaMhQ$ (yellow), matching scale variation, $\DeltaMhQmatch$
    (pink) and the uncertainty from different loop orders for the
    top pole mass matching, $\DeltaMhFStowerYt$ (turquoise). 
    In the top row $\tan\beta = 5$ and $X_t = -2 \MS$
    is used.  In the bottom row we set $\tan\beta = 5$ and $\MS =
    2\unit{TeV}$.  }
  \label{fig:MSSM_tower_uncertainties}
\end{figure}
\paragraph{1. Using known MSSM higher-order results:}
Actually the leading MSSM 2-loop
threshold 
corrections for $\lambda(\MS)$ are known and are of ${\cal
  O}(\alpha_t\alpha_s)$ and $\mathcal{O}(\alpha_t^2)$
\cite{Bagnaschi:2014rsa,Vega:2015fna}. They
have the form
  \begin{align}
    \Delta\lambda^{(2)}_{(\alpha_t \alpha_s)}
    &= \frac{g_3^2 (y_t^\SM)^4}{(4\pi)^4} \times C_2, &
    \Delta\lambda^{(2)}_{(\alpha_t^2)}
    &= \frac{(y_t^\SM)^6}{(4\pi)^4} \times C_3 ,
    \label{eq:delta_lambda_coeffs}
  \end{align}
  where the coefficients $C_2$ and $C_3$ depend on $\Qmatch$,
  $X_t/\MS$ and $\tan\beta$ in our common SUSY mass scale scenario.
  If one sets $\Qmatch=\MS$ and varies $X_t$ within the reasonably
  large interval $[-3\MS,+3\MS]$ and $\tan\beta$ within $[1,\infty]$,
  the coefficients vary within $C_2\in[-314, 231]$ and $C_3\in[-6,
    489]$.  These minimal and maximal values for $C_2$ and $C_3$ can
  be used to estimate the maximal effect of the missing 2-loop
  threshold corrections to $\lambda(\MS)$ in \FSTower by adding the
  terms \eqref{eq:delta_lambda_coeffs} to the r.h.s.\ of
  Eq.~\eqref{eq:lambda_1L_tower}.  The uncertainty estimated in this
  way is shown as the dashed area in the panels on the r.h.s.\ of
  Figure~\ref{fig:MSSM_tower_uncertainties}.  As can also
  be seen from Figure~\ref{fig:MSSM-Xt}, the variation of $C_2$ and
  $C_3$ does not reflect the fact that the 2-loop
  threshold corrections are negligible for $X_t = 0$ and are large for $X_t \approx
  \sqrt{6}\MS$.  Therefore, the variation of $C_2$ and $C_3$ certainly
  leads to an overestimation of the theory uncertainty of \FSTower for
  small values of $X_t$.  However, one can expect that the theory
  uncertainty estimated in this way is reasonable for maximal mixing
  scenarios.

\paragraph{2. Generating higher-order terms:}
  Another option to estimate the uncertainty is to change the
  calculation of $\lambda(\MS)$ in \FSTower
  such that changes of higher-order are automatically induced.
  Here we have two quantities at our disposal, which we expect to have
  a sizable impact on the value of $\lambda(\MS)$: (i) the value of
  $y_t^\MSSM$,
  (ii) the choice of the renormalization
  scale, $\Qmatch$, at which $\lambda$ is calculated.
  \begin{itemize}
  \item[(i)]
    The dominant 1-loop threshold correction to $\lambda$ is governed
    by the top Yukawa coupling. Thus, changing $y_t$  by
    motivated 1-loop terms shifts $\lambda$ by 2-loop terms. Such
    motivated terms can be obtained by switching  the $y_t^\MSSM$
    definition at the SUSY scale, Eq.\
    \eqref{eq:matching_condition_Mt}, between the 1-loop level and
    the tree-level. The differences in $y_t^\MSSM$ are sensitive to
    $\alpha_s$, $\alpha_t$, $X_t$, and contain logarithmic
    as well as non-logarithmic terms. The resulting
    shift in $\lambda$
    should therefore provide a good estimate of the magnitude of the actual
    dominant 2-loop     threshold corrections to     $\lambda$. As an
    automatic way to 
    evaluate the theory uncertainty from the missing 2-loop threshold
    corrections we propose to define
    \begin{align}
      \DeltaMhFStowerYt = \left|M_h^{\text{\FSTower}}(y_t^{\MSSM,(1)}) -
        M_h^{\text{\FSTower}}(y_t^{\MSSM,(0)})\right|,
      \label{eq:delta_Mh_yt}
    \end{align}
    where the two terms on the r.h.s.\ correspond to the \FSTower
    prediction using the $y_t^\MSSM(\MS)$ definition
    \eqref{eq:matching_condition_Mt} either  at the 1-loop or at
    the tree-level. 

    The turquoise uncertainty band in the panels on the r.h.s.\ of
    Figure~\ref{fig:MSSM_tower_uncertainties} shows the
    variation of $M_h$ by $\pm \DeltaMhFStowerYt$, i.e.~the variation
    from using either the tree-level or 1-loop top Yukawa coupling.
    The figures show that this estimated uncertainty is of the same order as
    the uncertainty obtained from variation of $C_2$ and $C_3$ for
    most SUSY scales.  Furthermore, we find that the uncertainty
    estimate vanishes for $X_t \approx 0$ and is maximal for maximal
    mixing ($X_t \approx \pm\sqrt{6}\MS$).  Thus, this estimated
    uncertainty reflects the expectation that the missing 2-loop
    threshold corrections for $\lambda(\MS)$ are small for vanishing
    $X_t$ and can be sizable for maximal mixing.

  \item[(ii)] By variation of the matching scale $\Qmatch$ within the
    interval $[\MS/2, 2\MS]$, the size of logarithmic higher-order
    contributions to $\lambda(\MS)$ can be estimated.  Varying
    $\Qmatch$ involves (a) RG running of all Standard Model parameters
    to $\Qmatch$ using 3-loop RGEs, (b) RG running of all MSSM
    parameters to $\Qmatch$ using 2-loop RGEs and (c) calculation of
    $\lambda$, as well as the MSSM gauge and Yukawa couplings and
    $v^\MSSM$ at the scale $\Qmatch$ using
    Eqs.~\eqref{eq:polemassmatching}--\eqref{eq:matching_condition_Mt}.
    Thus, the matching scale variation is sensitive to missing 2-loop
    renormalization scale-dependent logarithmic contributions in the
    calculation of $\lambda$.

    The effect of the matching scale variation is shown by the red
    band on the r.h.s.\ of Figure~\ref{fig:MSSM_tower_uncertainties}.  We
    find that the uncertainty is nearly independent of $X_t$, which is
    in agreement with the expectation:
    As can be seen from Eq.~\eqref{eq:delta_lambda_nonzero_Xt}, the
    renormalization scale-dependent part of the 1-loop threshold
    correction, $\Delta\lambda^{(1)}$, is not $X_t$-dependent.
    Furthermore, the $\beta$ functions of the MSSM parameters
    $g_Y^\MSSM$, $g_2^\MSSM$, $\tan\beta$, $v_u$ and $v_d$, which
    determine $\lambda$ at the tree-level, do not depend on $X_t$
    either \cite{Sperling:2013eva,Sperling:2013xqa}.  For this reason,
    the variation of $\Qmatch$ is not directly sensitive to
    $X_t$-dependent terms.  Therefore, one can expect that the
    variation of $\Qmatch$ alone is not sufficient to estimate the
    theory uncertainty from missing 2-loop threshold
    corrections.\footnote{For $\Delta\lambda^{(2)}$ this is no longer
      the case: The renormalization scale-dependent part of
      $\Delta\lambda^{(2)}$ depends on $X_t$, see Eq.~(21) in
      Ref.~\cite{Vega:2015fna}.}
  \end{itemize}

Another source of uncertainty in \FSTower comes from the missing
2-loop contributions to $M_h$ in the SM\@.  As done above, one way to
estimate the leading logarithmic 2-loop $M_h$ contributions is to vary
the renormalization scale $\Qlow$, at which $M_h$ is calculated,
within the interval $[M_t/2, 2M_t]$.  This uncertainty estimate is
shown in form of the yellow band on the r.h.s.\ of
Figure~\ref{fig:MSSM_tower_uncertainties}.  Comparing all uncertainty
bands for \FSTower, we find that for this scenario the Higgs mass
theory uncertainty is dominated by the missing 2-loop contributions to
$\lambda(\MS)$.

\section{Combined MSSM uncertainty estimation}
\label{sec:MSSM_uncertainty_summary}

\begin{figure}[tbh]
  \centering
  \includegraphics[width=0.49\textwidth]{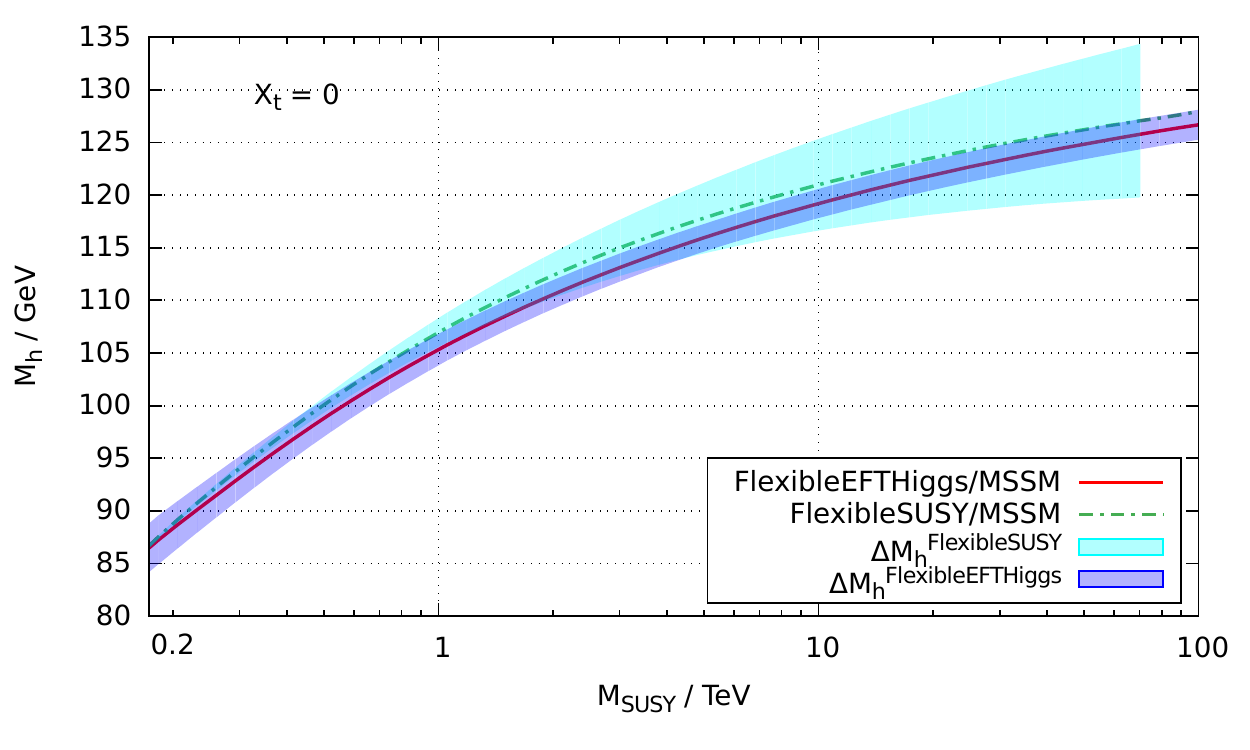}
  \includegraphics[width=0.49\textwidth]{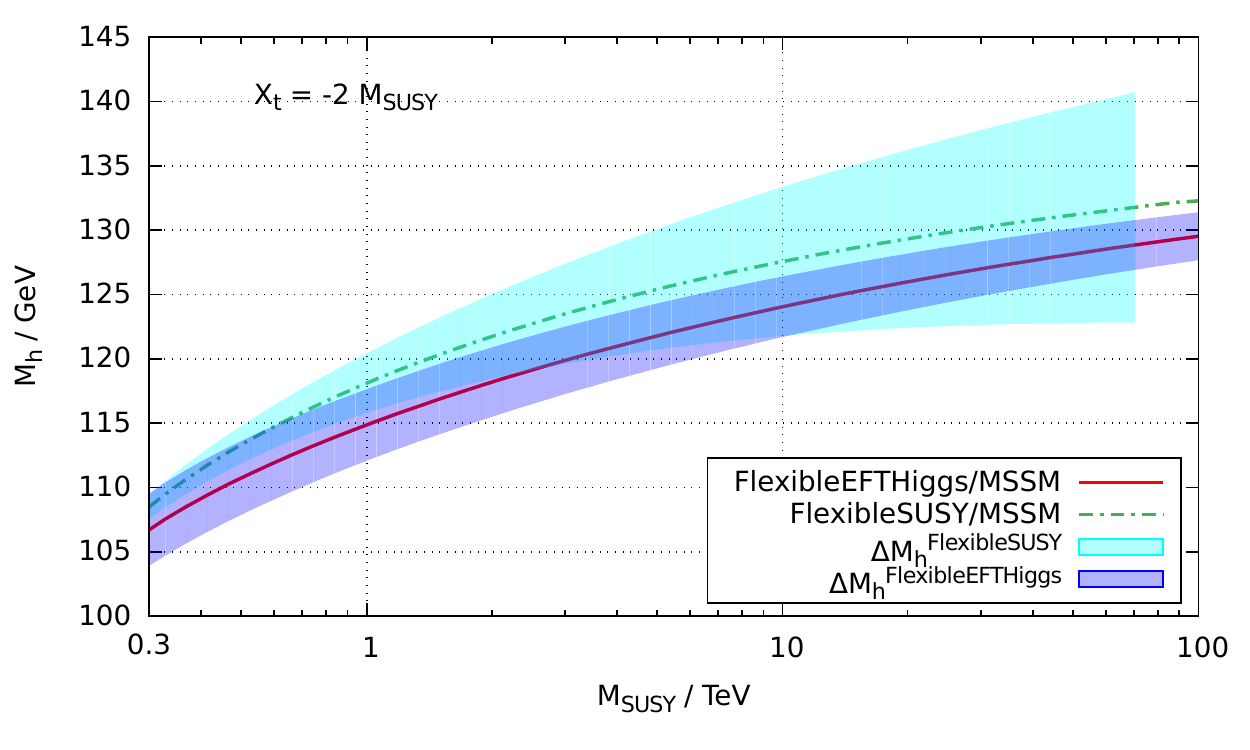}\\
  \includegraphics[width=0.49\textwidth]{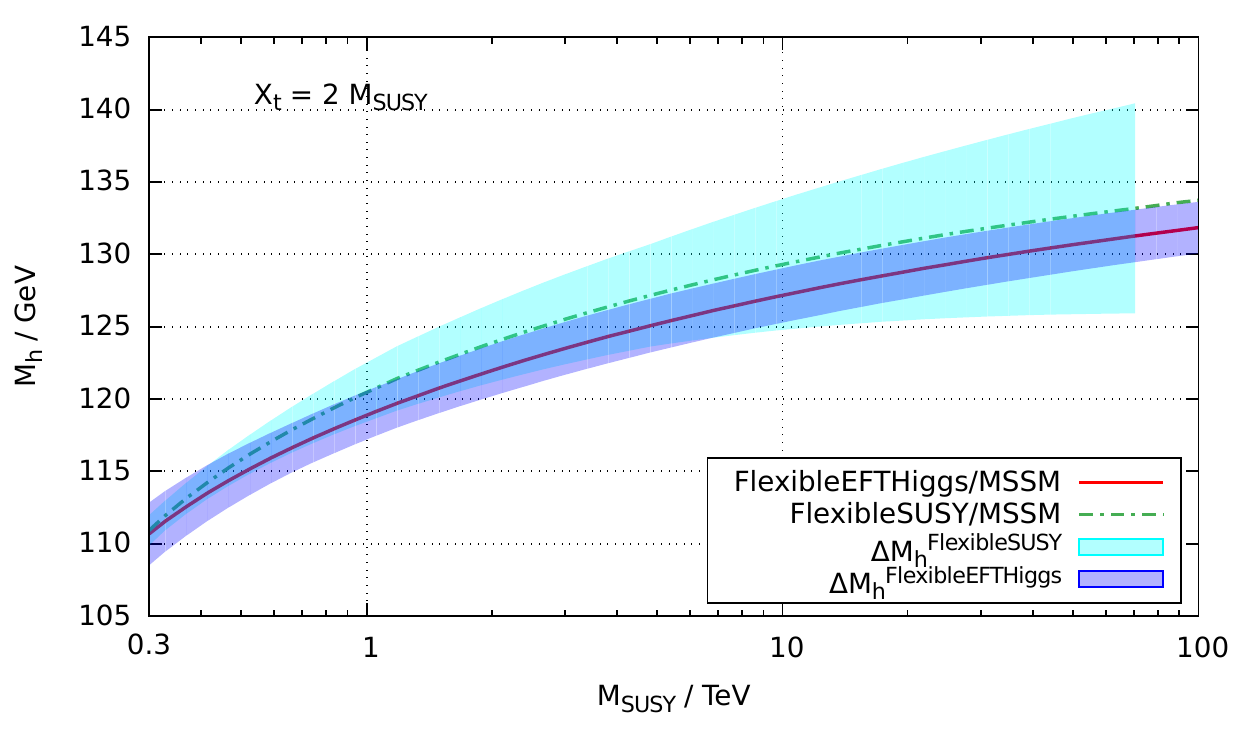}
  \includegraphics[width=0.49\textwidth]{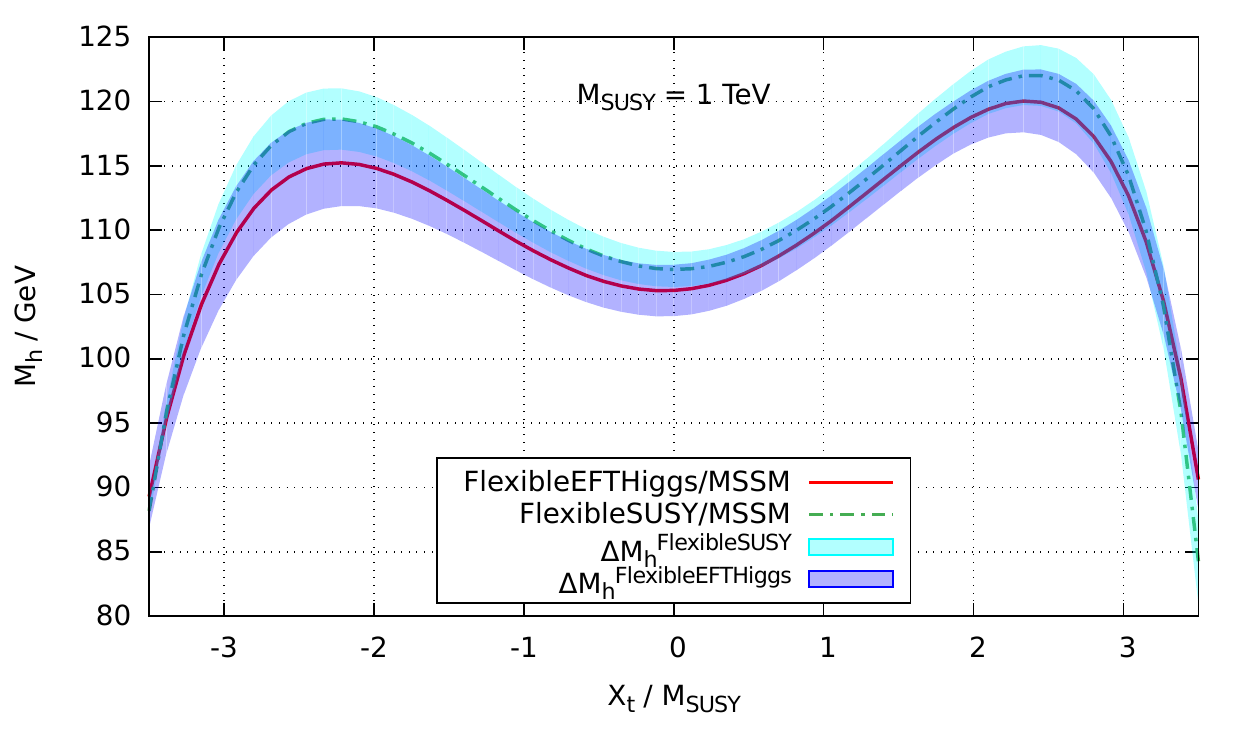}\\
  \includegraphics[width=0.49\textwidth]{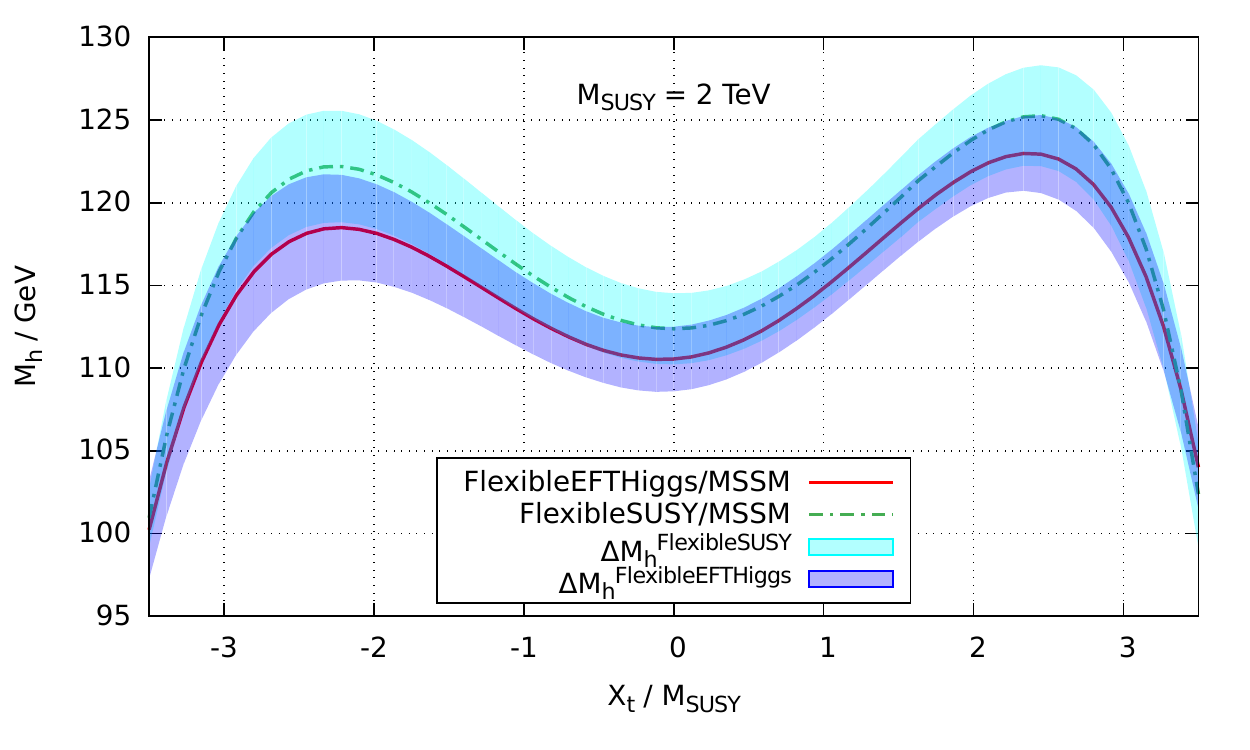}
  \includegraphics[width=0.49\textwidth]{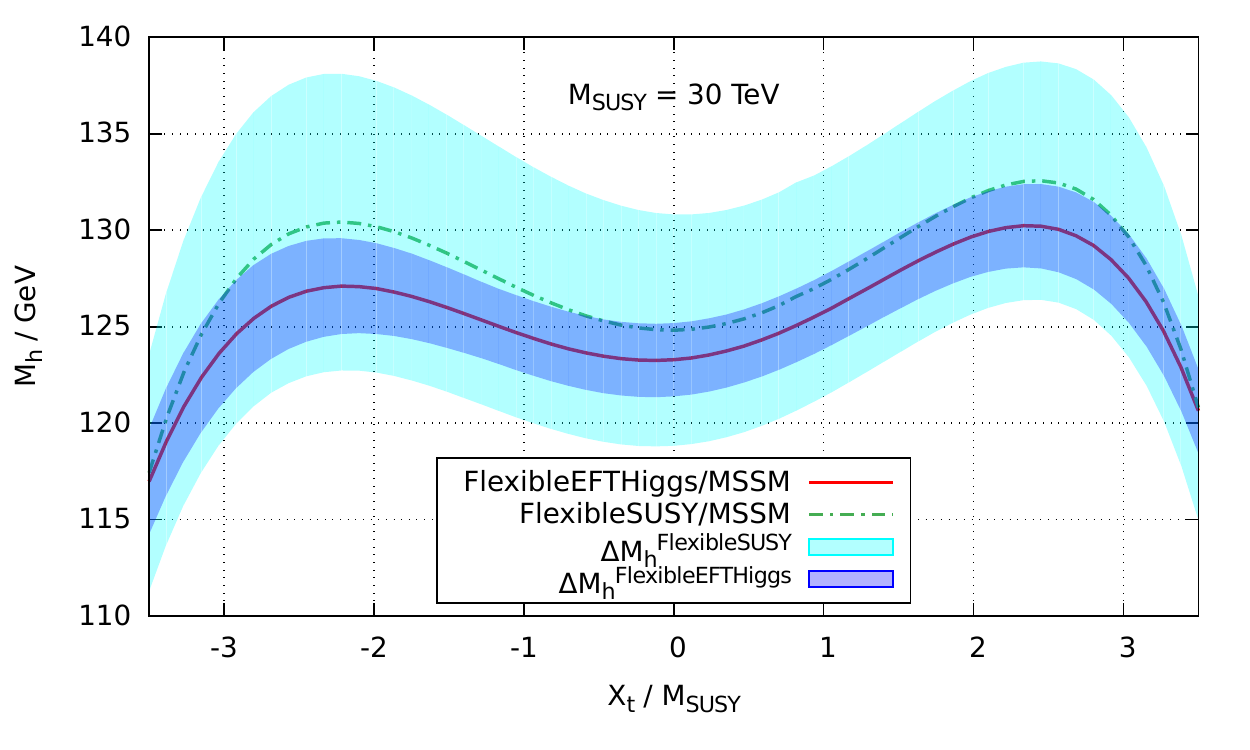}
  \caption{Predictions for $M_h$ and combined theoretical uncertainty estimates for \FS and
    \FSTower in the MSSM for $\tan\beta = 5$.  The first three panels
    show $M_h$ as a function of $\MS$ for $X_t / \MS = 0$, $-2$ and
    $2$, respectively.  The last three panels show $M_h$ as a function
    $X_t$ for $\MS = 1\unit{TeV}$, $2\unit{TeV}$ and $30\unit{TeV}$,
    respectively.}
  \label{fig:MSSM_combined_uncertainties}
\end{figure}

In the previous section we discussed many different ways to estimate
contributions to theory uncertainties, relevant for existing
fixed-order calculations as well as for \SUSYHD and \FSTower.
In this section we summarize and combine these various uncertainty
estimates, focusing on \FSTower and the fixed-order codes (the
discussion equally applies to \FS,
\Softsusy and \SPheno).
Figure~\ref{fig:MSSM_combined_uncertainties} shows the Higgs pole
mass calculated with \FSTower and the fixed-order \FS, including
estimates of theory uncertainties. The plots demonstrate that the
new approach always has an uncertainty of around 2--3$\unit{GeV}$ and
becomes more accurate for $\MS$ in the few-TeV range. We now provide
the details of the uncertainty estimates.

\paragraph{\FSTower calculation:}
Following the classification of the theory uncertainties in
Ref.~\cite{Vega:2015fna}, \FSTower has two
basic sources of theory uncertainty: from missing higher-order
corrections in the matching procedure at the high scale (``high-scale
uncertainty''), and from missing
higher-order corrections in the Higgs pole mass computation in the EFT
at the low scale (``low-scale uncertainty'').

An important property of \FSTower is the inclusion of all
non-logarithmic 1-loop contributions to $M_h$ due to the special
choice of the matching procedure. As discussed in
Section~\ref{sec:MSSMtower_results}, the
resulting ``EFT uncertainty'' discussed in Ref.~\cite{Vega:2015fna}
due to missing power-suppressed tree-level or 1-loop terms is 
therefore not present in \FSTower by construction.

The high-scale uncertainty of \FSTower is estimated in two ways,
introduced and discussed in detail in Section~\ref{sec:FStower_Xt=0_uncertainty_estimation}:
\begin{enumerate}
\item\label{en:Delta_Mh_tower_yt} Use of $y_t^\MSSM(\MS)$, which has been
  obtained from the top pole mass matching at the SUSY scale either at the tree-level or
  at the 1-loop level.  We denote the corresponding shift in the Higgs
  pole mass as $\DeltaMhFStowerYt$, as defined in
  Eq.~\eqref{eq:delta_Mh_yt}.

\item\label{en:Delta_Mh_tower_Qmatch} Variation of the matching scale
  $\Qmatch$ within the interval $[\MS/2, 2\MS]$.  We denote the corresponding
  Higgs pole mass uncertainty estimate by $\Delta
  M_h^{(\Qmatch)}$, see Eq.~\eqref{eq:def_DMhQ}.
\end{enumerate}
The low-scale uncertainty is estimated as follows:
\begin{enumerate}
\item[3.]\label{en:Delta_Mh_tower_Q} Variation of the renormalization
  scale $\Qlow$, at which the Higgs pole mass is calculated, in the
  interval $[M_t/2,2M_t]$.  We denote the corresponding Higgs pole mass
  uncertainty estimate by $\Delta M_h^{(\Qlow)}$, see
  Eq.~\eqref{eq:def_DMhQ}.
\end{enumerate}
Since the two high-scale uncertainty estimates
\ref{en:Delta_Mh_tower_yt} and \ref{en:Delta_Mh_tower_Qmatch} 
are partially sensitive to the same higher-order MSSM corrections, we
combine $\DeltaMhFStowerYt$ and $\DeltaMhQmatch$ by taking the
maximum of the two for each parameter point.  $\Delta M_h^{(\Qlow)}$ is
sensitive to logarithmic higher-order Standard Model corrections,
which is why we add it in quadrature to the former:
\begin{align}
  \Delta M_h^{\FSTower} = \sqrt{\left(\max\left\{\DeltaMhFStowerYt,
        \DeltaMhQmatch\right\}\right)^2 + \left(\Delta
      M_h^{(\Qlow)}\right)^2} .
  \label{eq:def_Delta_Mh_tower}
\end{align}%
In Figure~\ref{fig:MSSM_combined_uncertainties} we find that for small
values of $X_t$ this combined 
uncertainty estimate is of the order $2\unit{GeV}$ for \FSTower.  The
uncertainty grows up to around $3\unit{GeV}$ for maximal mixing.
Since \FSTower is an EFT calculation, its uncertainty does
not depend on the SUSY scale: Even for large
$\MS\approx 30\unit{TeV}$ the uncertainty is of the order or below
$3\unit{GeV}$. Likewise, because there is no ``EFT uncertainty'', the
uncertainty does not grow significantly for low $\MS$.

\paragraph{Fixed-order calculation:}

The theory uncertainty of the fixed-order calculations arises from
missing higher-order corrections. As described in
Section~\ref{sec:FStower_Xt=0_uncertainty_estimation} we propose two
measures of leading missing contributions:
\begin{enumerate}
\item Using the four different definitions of $y_t^\MSSM$, as
  described in Section~\ref{sec:FStower_Xt=0_uncertainty_estimation}.
  We denote the maximum difference between the Higgs masses obtained
  using these four definitions as $\DeltaMhFSYt$, see
  Eq.~\eqref{eq:def_DMh4yt}. This is particularly sensitive to the
  leading 3-loop logarithms governed by the top Yukawa coupling.

\item Variation of the renormalization scale $\QFOMS$, at which the
  Higgs pole mass is calculated, within the interval $[\MS/2,2\MS]$.
  We denote the corresponding Higgs pole mass uncertainty estimate 
  by $\Delta M_h^{(\QFOMS)}$, see Eq.~\eqref{eq:def_DMhQ}. This is
  particularly sensitive to subleading logarithms governed by all
  couplings of the \MSSM.
\end{enumerate}
We have combined these two uncertainty estimates as
\begin{align}
  \Delta M_h^{\FS} = \sqrt{\left(\DeltaMhFSYt\right)^2 + \left(\Delta
      M_h^{(\QFOMS)}\right)^2}.
  \label{eq:def_Delta_Mh_fixed_order}
\end{align}
In Figure~\ref{fig:MSSM_combined_uncertainties},
$\Delta M_h^{\FS}$ grows logarithmically with $\MS$ as expected.  For
small values of $X_t$ and SUSY scales below $1\unit{TeV}$, the
combined uncertainty estimate is below $1\unit{GeV}$.  For larger SUSY
scales and larger $X_t$ values, the uncertainty can grow up to $9\unit{GeV}$.

We remark that further subleading effects, such as finite,
non-logarithmic corrections arising e.g.~from going beyond the ${\mathcal O}((\alpha_t +
\alpha_b)^2 + (\alpha_t + \alpha_b) \alpha_s + \alpha_\tau^2)$
approximation at the 2-loop level, are not necessarily captured by the
estimate \eqref{eq:def_Delta_Mh_fixed_order}; hence particularly at
low $\MS$, the true uncertainty of the fixed-order calculations might
be larger than this estimate.

\section{Numerical results in the NMSSM}
\label{sec:NMSSM}

Here we consider the next-to-minimal supersymmetric standard model
(NMSSM) \cite{Ellwanger:2009dp,Maniatis:2009re}, where the MSSM
superfield content is extended by an extra gauge singlet superfield
$\hat{S}$. In early calculations of higher-order corrections to NMSSM
Higgs masses both effective field theory techniques
\cite{Espinosa:1991fc,Espinosa:1992hp,Elliott:1993ex,Yeghian:1999kr,Ellwanger:1999ji,Ellwanger:2005fh}
and fixed-order calculations in the effective potential approximation
\cite{Ellwanger:1993hn,Pandita:1993tg,Elliott:1993uc,Elliott:1993bs}
were employed.  More recently \DRbar calculations with full 1-loop
corrections ~\cite{Degrassi:2009yq,Staub:2010ty}, 2-loop corrections
of ${\cal O}(\alpha_s(\alpha_b+\alpha_t))$
\cite{Degrassi:2009yq}, and finally 2-loop corrections involving all
superpotential parameters were calculated \cite{Goodsell:2014pla}.
Recent progress in a mixed on-shell-\DRbar scheme has also been made,
with full 1-loop corrections \cite{Ender:2011qh,Graf:2012hh} and
2-loop corrections of ${\cal O}(\alpha_s \alpha_t)$
\cite{Muhlleitner:2014vsa}.

We assume that there is a $Z_3$
symmetry, which forbids the $\mu$-term so that when the new scalar
singlet, $S$, develops a VEV and generates an effective $\mu$-term, it
solves the $\mu$ problem of the \MSSM.  The superpotential is then,
\begin{align}
\mathcal{W}_{\textrm{NMSSM}} = \mathcal{W}_{\textrm{MSSM}}(\mu = 0) + \lambda \, \hat{S} \, \hat{H}_u \cdot\hat{H}_d + \frac13\kappa \, \hat{S}^3.
\end{align}
The soft breaking Lagrangian density is,
\begin{align}
\LModelSoft{\NMSSM} =  \LModelSoft{\MSSM} (B\mu=0) + \lambda A_\lambda S H_u H_d + \frac13\kappa A_\kappa S^3 + m_S^2 |S|^2.
\end{align}  
The Higgs fields develop VEVs,
\begin{align}
  \langle H_u \rangle &= \frac{1}{\sqrt{2}}\begin{pmatrix} 0 \\ v_u \end{pmatrix}, &
  \langle H_d \rangle &= \frac{1}{\sqrt{2}}\begin{pmatrix} v_d \\ 0 \end{pmatrix}, &
  \langle S \rangle &= \frac{1}{\sqrt{2}} v_s .
  \label{eq:NMSSM_Higgs_VEVs}
\end{align}

Here we implement our new method for predicting the Higgs mass, and
our uncertainty estimates for this and the \FS fixed-order calculation, to the
\NMSSM.  We then compare our \FSTower calculation to the predictions
using some of the publicly available software. To keep our analysis
simple we set the soft-breaking squared sfermion mass parameters and
gaugino masses to $\MS$, as defined in
Eqs.~\eqref{eq:MSSM_MSUSY_scenario}, $\tan\beta(\MS) = 5$, and require
that the two additional Yukawa couplings in the NMSSM are equal,
\begin{equation}
  \lambda(\MS) = \kappa(\MS) .
\end{equation}
We also require that $v_s$ is fixed so that $\mu_{\text{eff}} = \MS$, i.e.
\begin{equation}v_s(\MS) =  \frac{\sqrt{2} \MS}{\lambda(\MS)}.\end{equation}
The new trilinears are fixed to,
\begin{align} A_\lambda(\MS) &= \frac{1}{\lambda} \left(\frac{\sqrt{2}\tan\beta \MS^2}{v_s (\tan^2 \beta + 1)} - \kappa \lambda \frac{v_s}{\sqrt{2}}\right), & A_\kappa(\MS) &= -\frac{\sqrt{2}\MS^2}{v_s} ,
\end{align}
where all \DRbar quantities on the right hand side are evaluated at \MS. The complicated expression for $A_\lambda$ ensures the mass of the MSSM-like CP-odd state, which appears in the CP-odd mass matrix, is equal to \MS.  
The soft-breaking squared Higgs mass parameters $m_{H_u}^2, m_{H_d}^2,
m_S^2$ are fixed by the EWSB minimization conditions.
\begin{figure}[h!]
  \centering
  \includegraphics[width=0.49\textwidth]{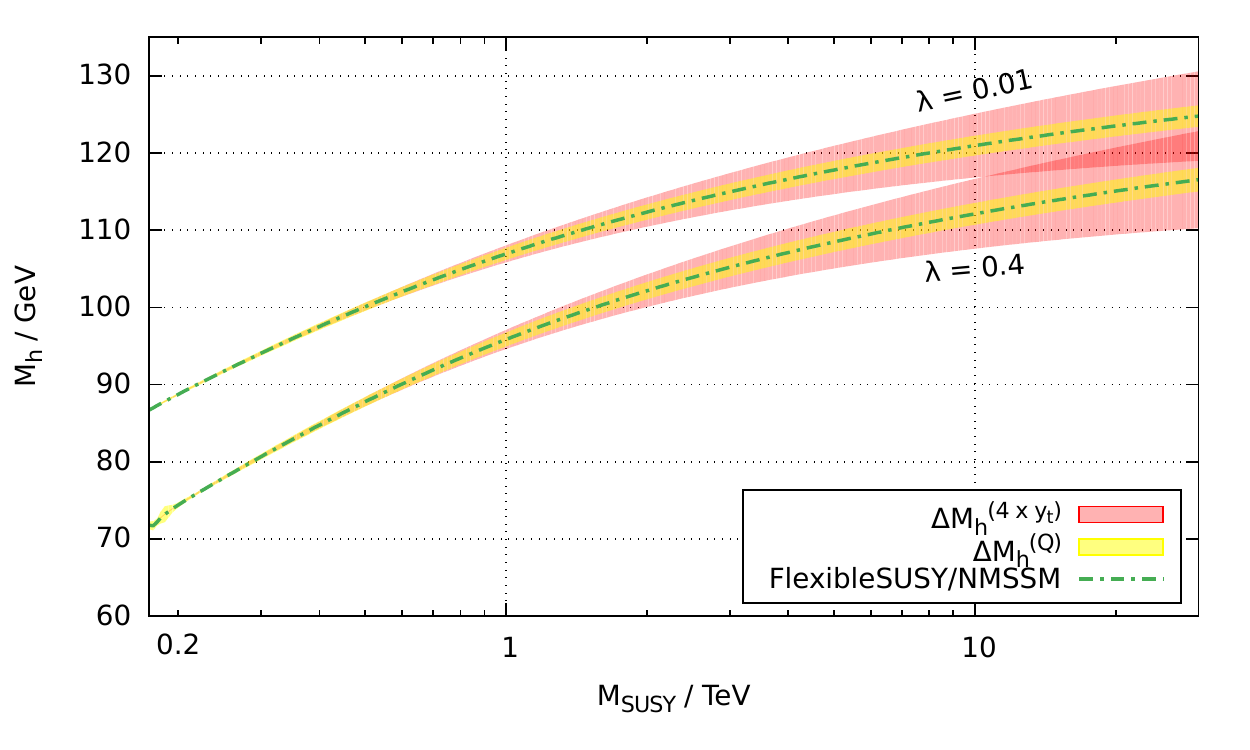}
  \includegraphics[width=0.49\textwidth]{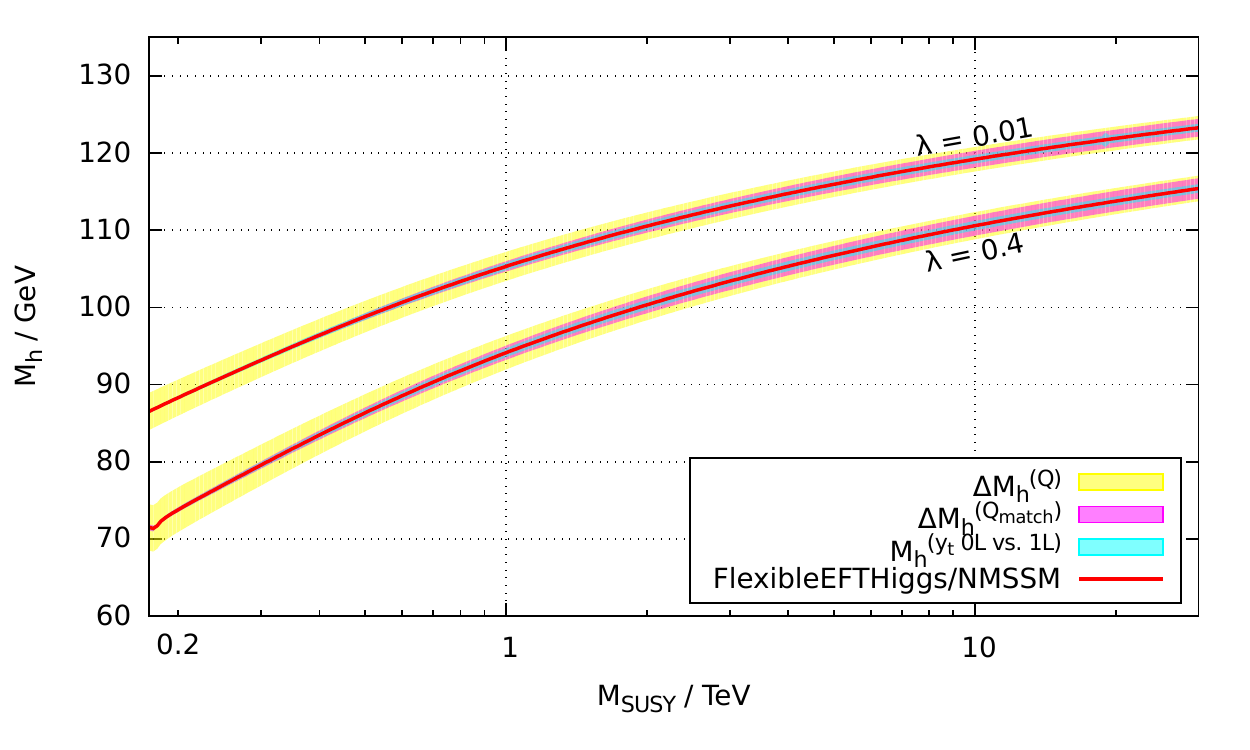} \\
  \includegraphics[width=0.49\textwidth]{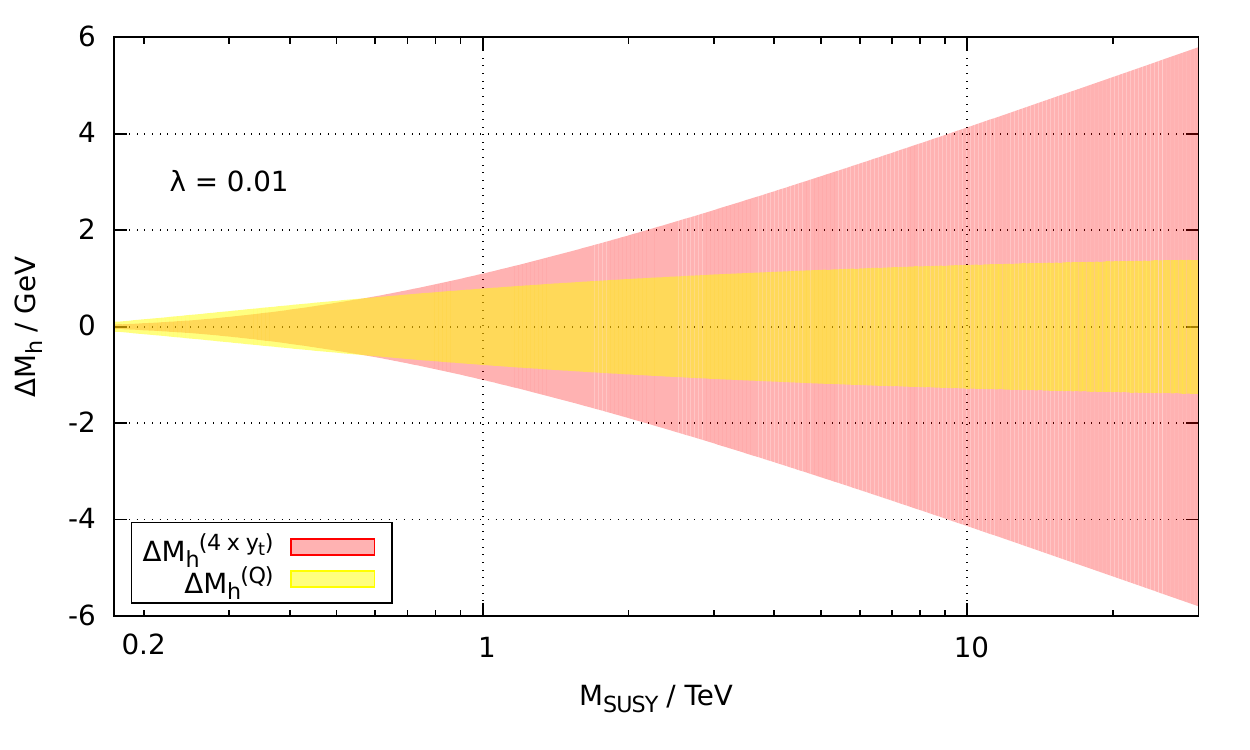}
  \includegraphics[width=0.49\textwidth]{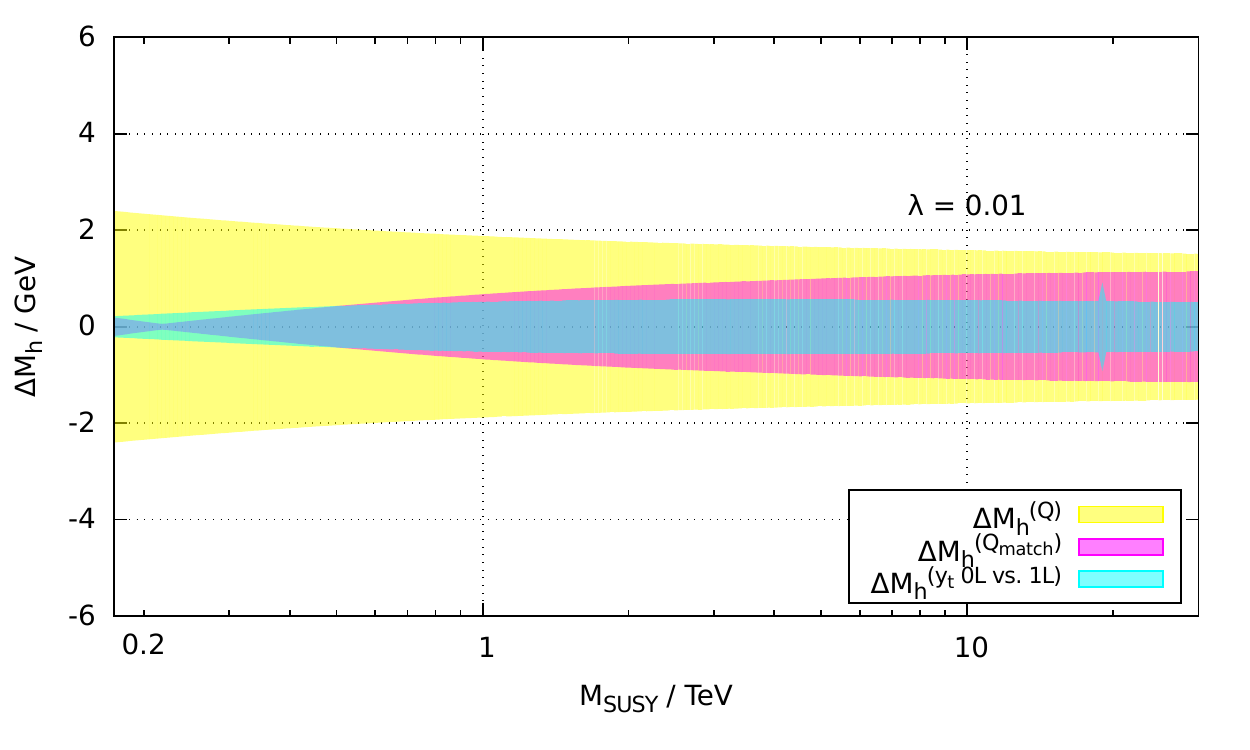}
  \\
  \includegraphics[width=0.49\textwidth]{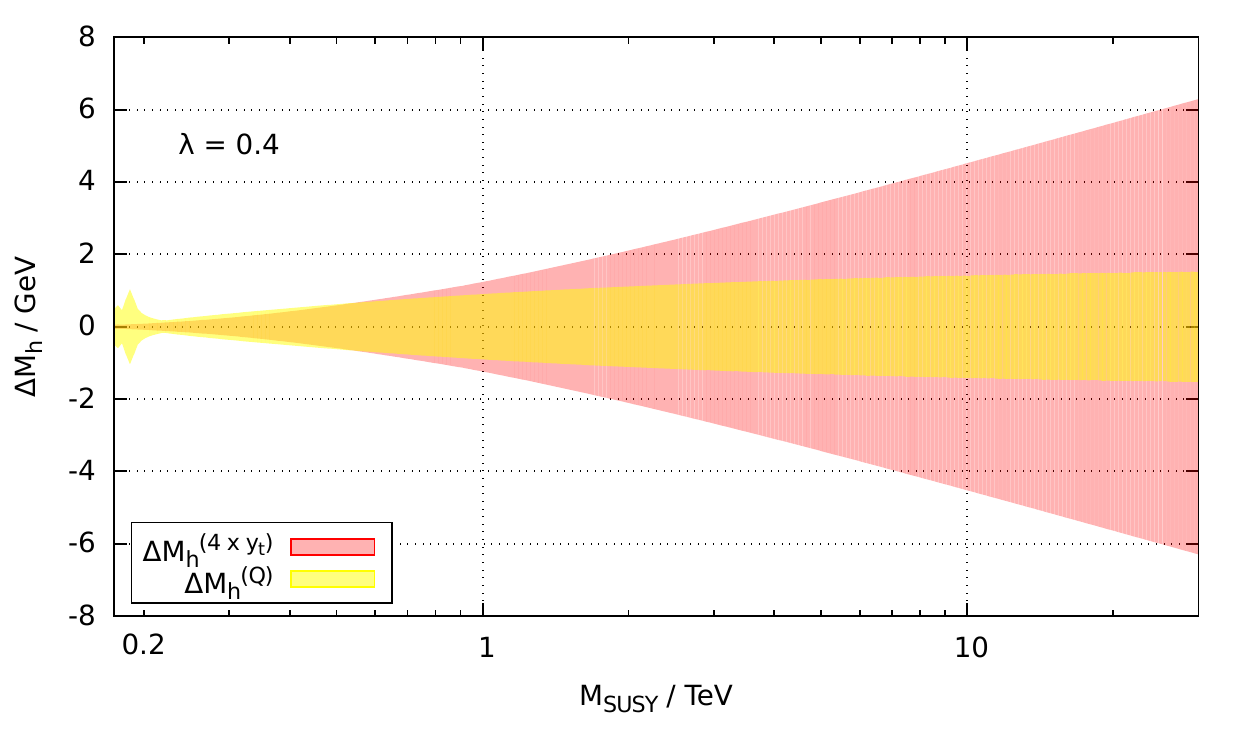}
  \includegraphics[width=0.49\textwidth]{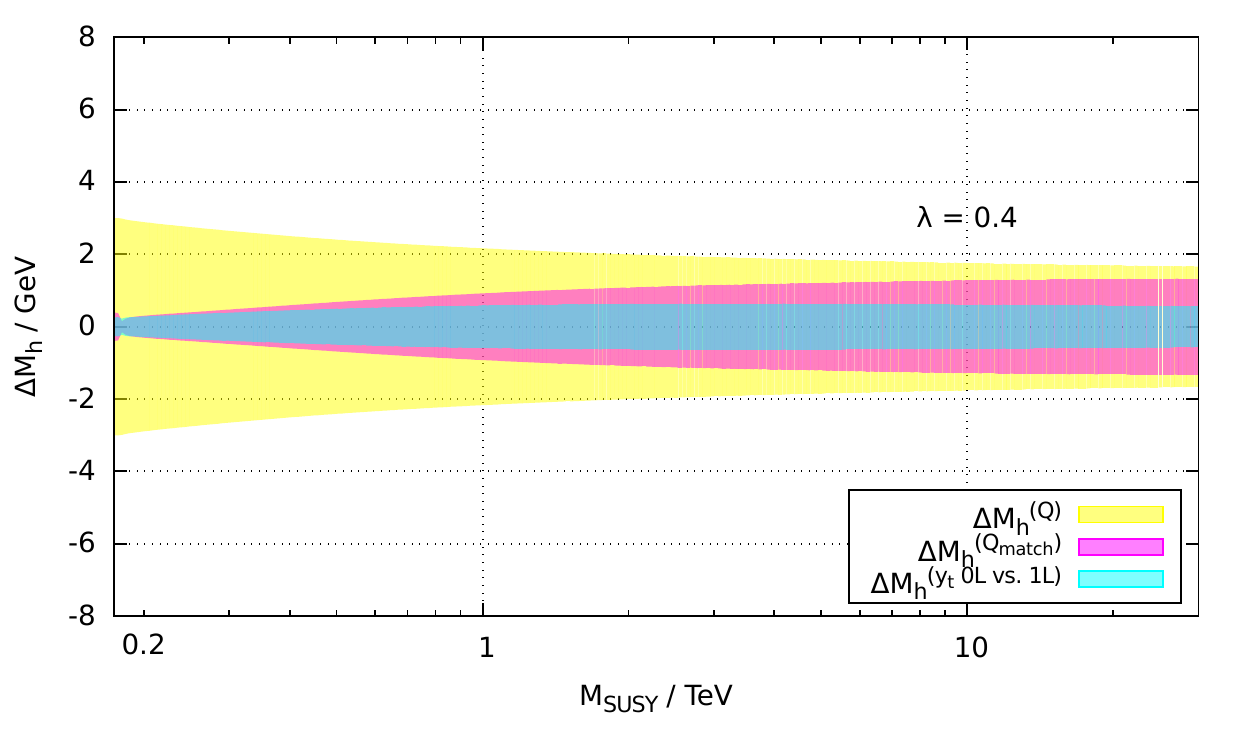}
  \caption{Higgs mass predictions and uncertainty estimates of
    Sec.~\ref{sec:MSSMtower_results} applied to the fixed-order
    calculations with \FS and to \FSTower in the \NMSSM. The top row shows
    the Higgs mass predictions and the uncertainty estimates, the
    lower two rows only the uncertainty estimates, as in
    Fig.~\ref{fig:MSSM_tower_uncertainties}.  The values of the
    singlet Yukawa couplings are $\kappa = \lambda = 0.01$ or $\kappa
    = \lambda = 0.4$, as indicated in the plots.  In all panels we
    have fixed $X_t = 0$ and $\tan\beta=5$. }
 \label{fig:MhFSNMSSM}
\end{figure}

In Figure \ref{fig:MhFSNMSSM} we compare the NMSSM predictions for the
Higgs mass using the \FS fixed-order calculation and the new \FSTower
calculation. In the top panels one can see that as in the MSSM the \FS
prediction is remarkably close to \FSTower.  However the uncertainty
bands for fixed-order calculation in the left panels of Figure
\ref{fig:MhFSNMSSM} show that nonetheless this is a coincidence and
the true uncertainty of the fixed-order calculation is much larger.
Two different values of the new singlet Yukawa couplings, $\lambda$
and $\kappa$, are shown and one can see in this case increasing these
couplings reduces the Higgs mass due to increased singlet mixing, but
has little impact on the comparison between the two approaches.

The panels on the right of Figure \ref{fig:MhFSNMSSM} show the
uncertainty estimation bands for the \FSTower calculation.  By comparing the
plots in the middle panel one can see that as \MS is increased, the
fixed-order uncertainty rises rapidly while the \FSTower uncertainties
have only a weak dependence on \MS, in line with our expectations and
in agreement with the results obtained in the \MSSM.

\begin{figure}[tbh]
  \centering
  \includegraphics[width=0.49\textwidth]{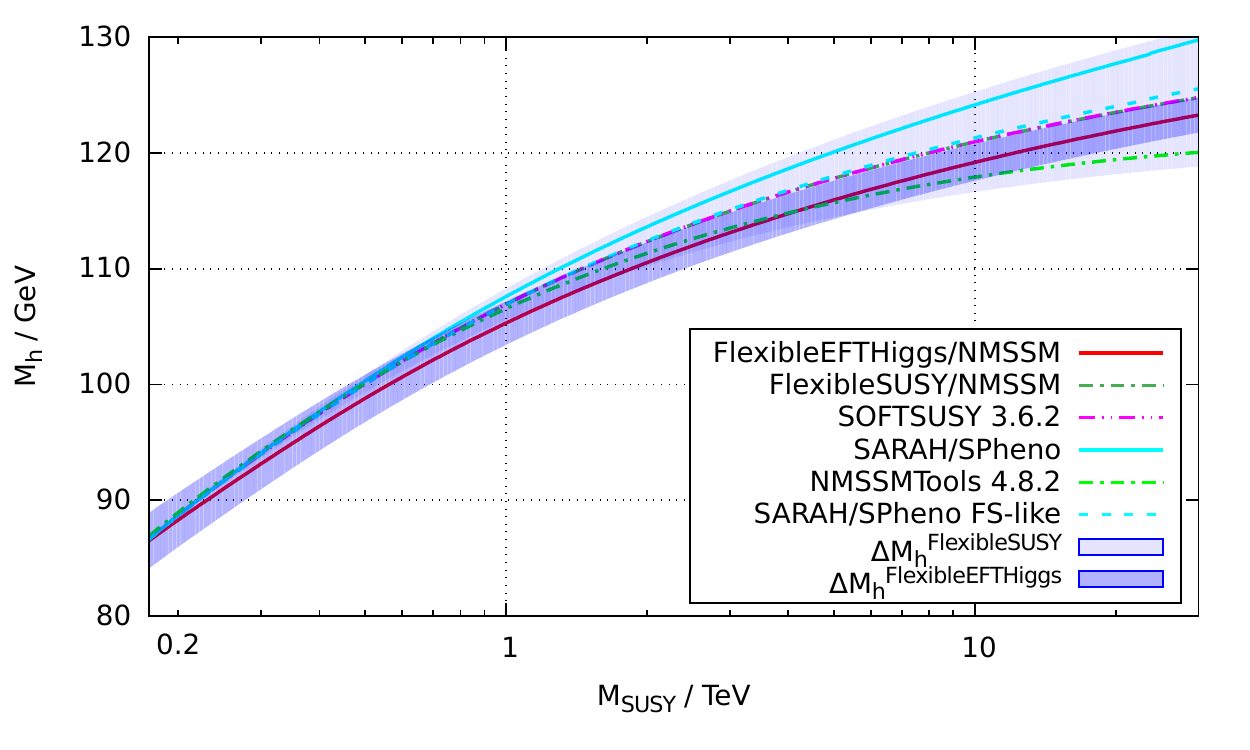}
  \includegraphics[width=0.49\textwidth]{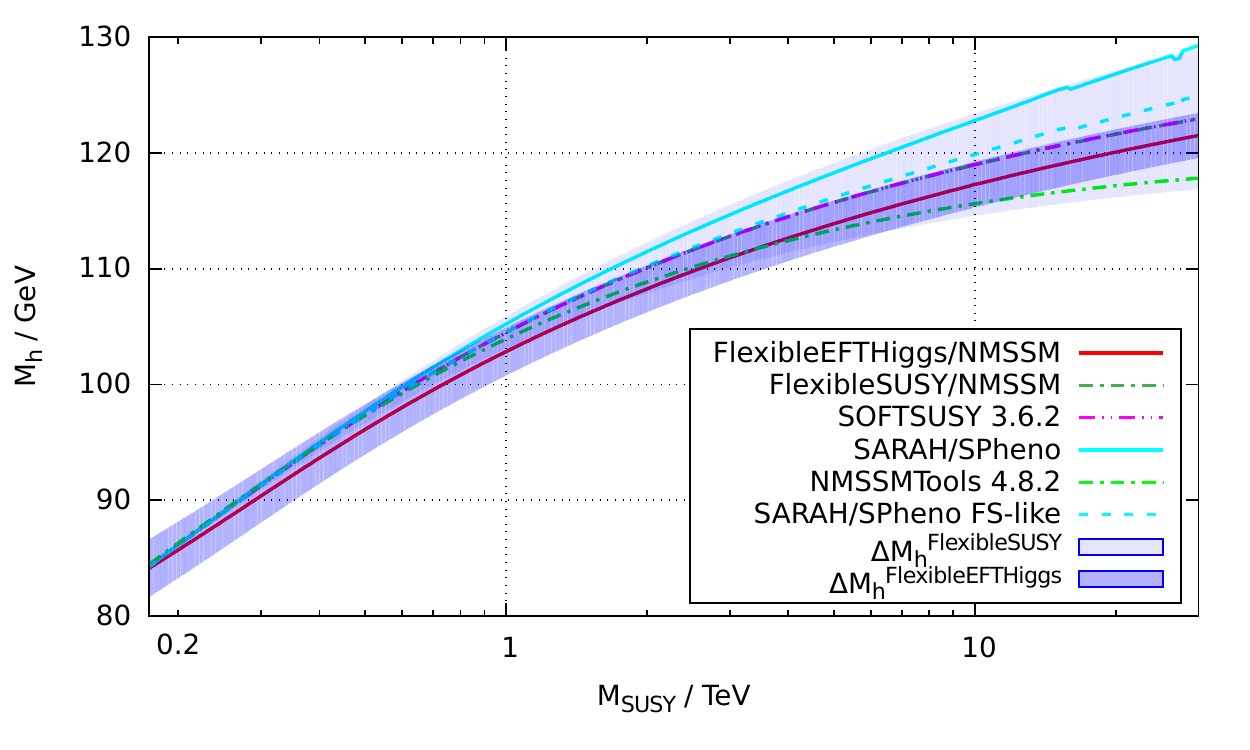}\\
  \includegraphics[width=0.49\textwidth]{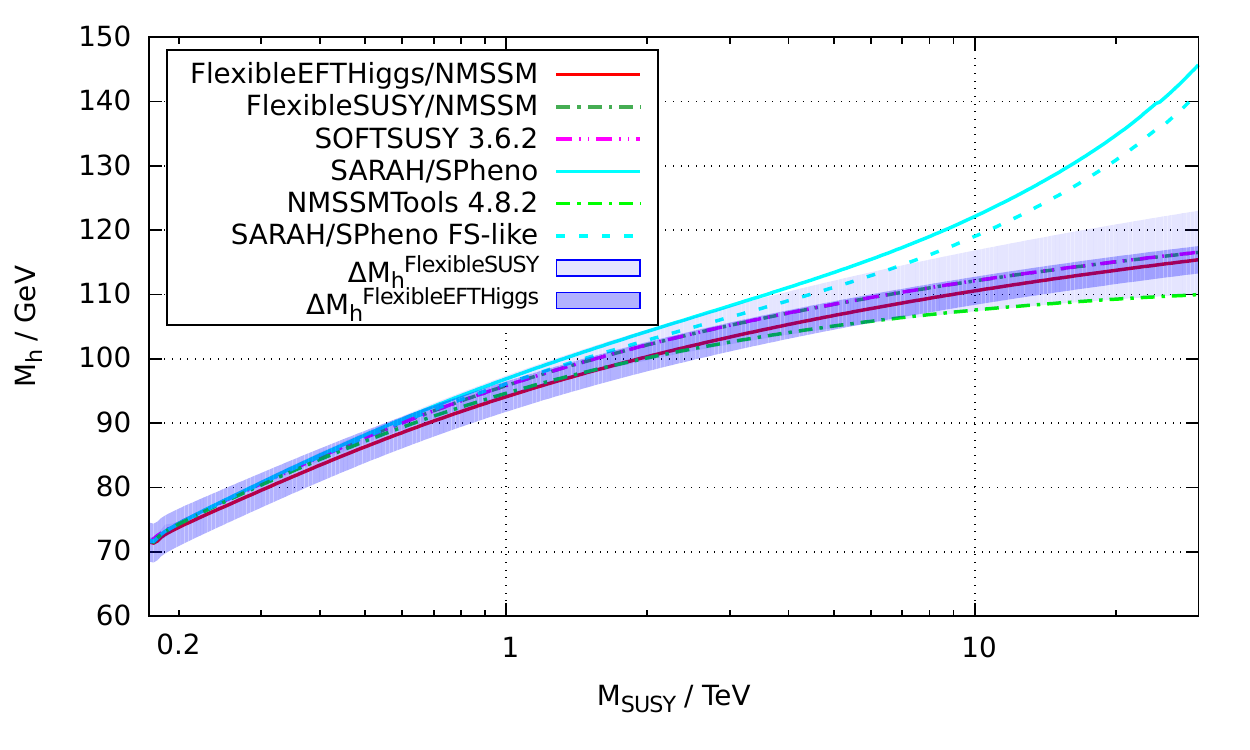}
  \includegraphics[width=0.49\textwidth]{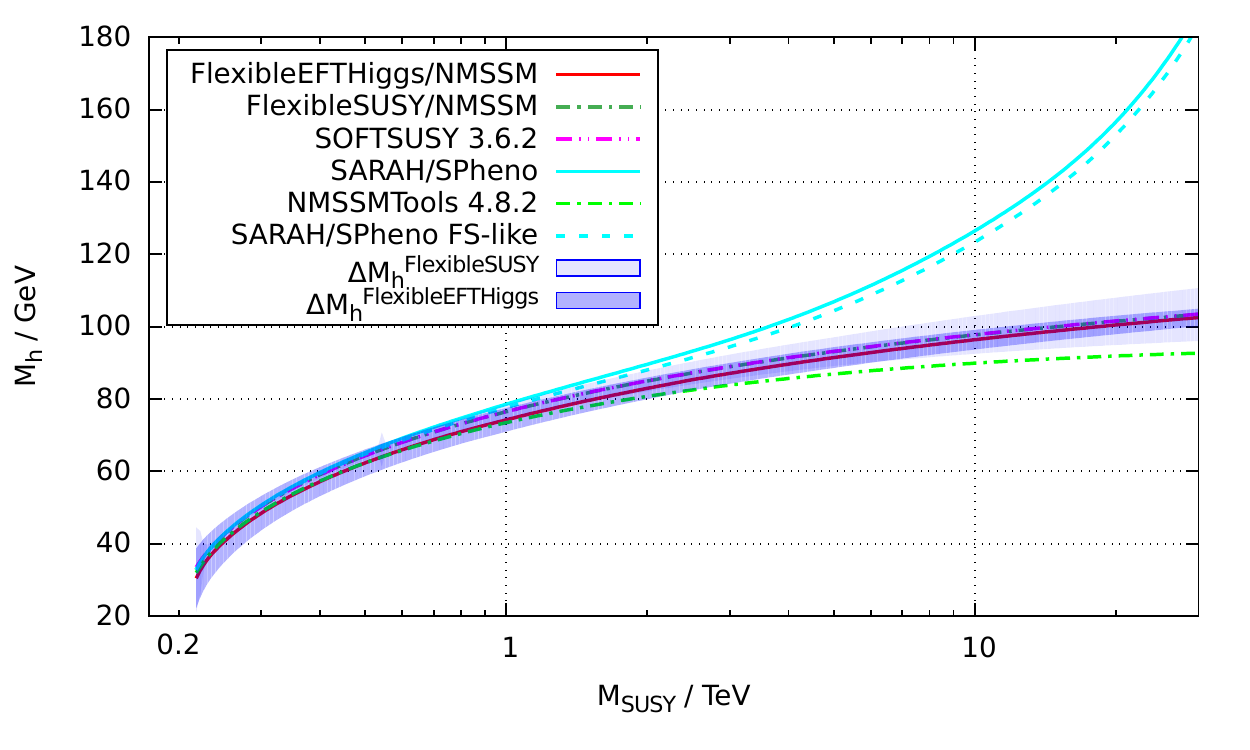}\\
  \caption{Predictions for $M_h$ and combined theoretical uncertainty estimates for
  \FS and  \FSTower   in the NMSSM, compared with results of other
  codes. We choose $X_t = 0$ and $\tan \beta = 5$ in all panels.
The top panel shows $\lambda = \kappa = 0.01$ (left), which is close to the MSSM limit and $\lambda = \kappa =0.2$ (right). In both cases the situation is quite similar to the \MSSM.  Row 2 shows $\lambda = \kappa = 0.4$ (left) and $\lambda = \kappa=0.6$ (right). Here we see significant deviation from the MSSM pattern with the \SPheno fixed-order calculation, which is due to the same infra-red divergences behind the known Goldstone boson catastrophe \cite{Martin:2013gka,Martin:2014bca, Elias-Miro:2014pca}.}
  \label{fig:Mh_NMSSM}
\end{figure}

If we combine these uncertainties in the manner described in
Section~\ref{sec:MSSM_uncertainty_summary}, we find that also in
the NMSSM \FSTower becomes more precise than the fixed-order
calculation for values of $\MS$ in the few-TeV region.  However, by
comparing the cases $\lambda = \kappa = 0.01$ and $\lambda = \kappa
=0.4$ (see Figure \ref{fig:Mh_NMSSM}) we find that the precise value
of $\MS$ at which this happens depends on the singlet Yukawa couplings.


We now turn to a comparison with the results of \FSTower and various
public NMSSM codes: the fixed-order \FS calculation, \nmspec
\cite{Ellwanger:2006rn} in the \nmssmtoolsv package, the
next-to-minimal extensions of \Softsusyv \cite{Allanach:2013kza} and
an NMSSM module generated with \SARAHv and compiled and run in
\SPhenov. We also include results from a modified version of
\SARAH/\SPheno, which calculates the top quark Yukawa coupling
$y_t^\NMSSM(M_Z)$ using Eq.~\eqref{eq:fixedmt_FlexibleSUSY} as is done
in \FS and \Softsusy, labeled as \SARAH/\SPheno FS-like. Here we omit the
calculation of \nmssmcalc \cite{Baglio:2013iia}, but one may see
comparisons between \nmssmcalc calculating in the \DRbar scheme and
the other fixed-order codes in Ref.~\cite{Staub:2015aea}. Note that the
\SARAH/\SPheno calculations take into account the full 2-loop
corrections in the gaugeless limit and effective potential
approximation, while the other fixed-order codes include 2-loop NMSSM
corrections of ${\mathcal O}((\alpha_t + \alpha_b) \alpha_s)$
from Ref.~\cite{Degrassi:2009yq} but include only MSSM-like 2-loop
corrections of ${\mathcal O}((\alpha_t+\alpha_b)^2 +
\alpha_\tau^2)$.

In Figure \ref{fig:Mh_NMSSM} we show the Higgs mass against \MS
for $\lambda(\MS) = \kappa(\MS) \in \{0.01, 0.2, 0.4, 0.6 \}$.  For
small $\lambda$ and $\kappa$ the results are like in the
\MSSM. \FSTower agrees very well with the fixed-order \FS calculation.
Among the fixed-order codes, \SARAH/\SPheno FS-like agrees very well with
\Softsusy and the fixed-order \FS. Due to the different definition of
the top Yukawa coupling, the Higgs mass calculated with \SPheno is
slightly higher than all other fixed-order codes; \nmssmtools is
slightly lower. The agreement between all these codes shows in
particular that the specific, non-MSSM-like 2-loop corrections that
are only included in \SPheno are small.

In contrast, for larger $\lambda=\kappa\gtrsim0.2$ both \SPheno results
(both in its original form and in the modified version with the \FS-like top
Yukawa coupling definition) deviate very strongly from all other results for
large $\MS\gtrsim2\unit{TeV}$. This effect has not been discussed in
Ref.~\cite{Staub:2015aea}, where only smaller $\MS$ were considered.
The discrepancy can be traced back to singularities in the 2-loop
effective potential calculation used in \SPheno,\footnote{We
  gratefully acknowledge clarifying discussions with the authors of
  Refs.~\cite{Goodsell:2014bna,Goodsell:2015ira} about these
  discrepancies and the expected range of validity of the
  \SPheno results.}  briefly described in Section 2.3 of
Ref.~\cite{Goodsell:2014bna}. As also mentioned in this reference,
these singularities are not present in the corresponding MSSM
calculations, and also not present in the other NMSSM codes, since
these codes do not take into account NMSSM-specific 2-loop
corrections involving $\alpha_\lambda$ and $\alpha_\kappa$.  These
singularities are similar to the ones related to Goldstone bosons and
discussed in Refs.~\cite{Martin:2013gka,Martin:2014bca}, but are
related to the smallness of the physical Higgs boson mass compared to
the renormalization scale. As explained in the mentioned references,
such singularities are spurious and appear only due to the truncation
of the perturbation series at fixed order.  For this reason we regard
the parameter region with large $\lambda,\kappa$ and large $\MS$ as
outside the range of validity of the \SPheno
calculation.\footnote{Note: our estimation of various sources of
  uncertainty in the fixed-order calculation cannot account for this
  kind of effect. So it is not surprising that the \SPheno result lies
  outside this band.}
\begin{figure}[tbh]
  \centering
  \includegraphics[width=0.49\textwidth]{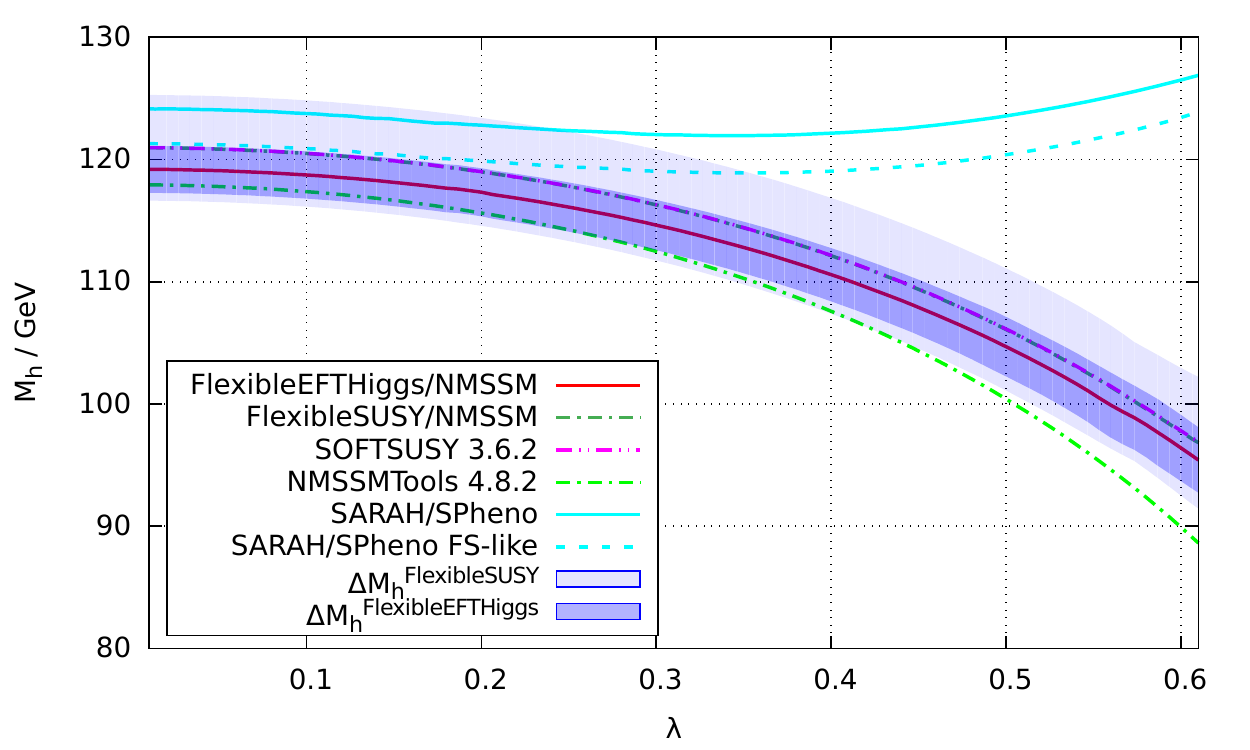}
  \includegraphics[width=0.49\textwidth]{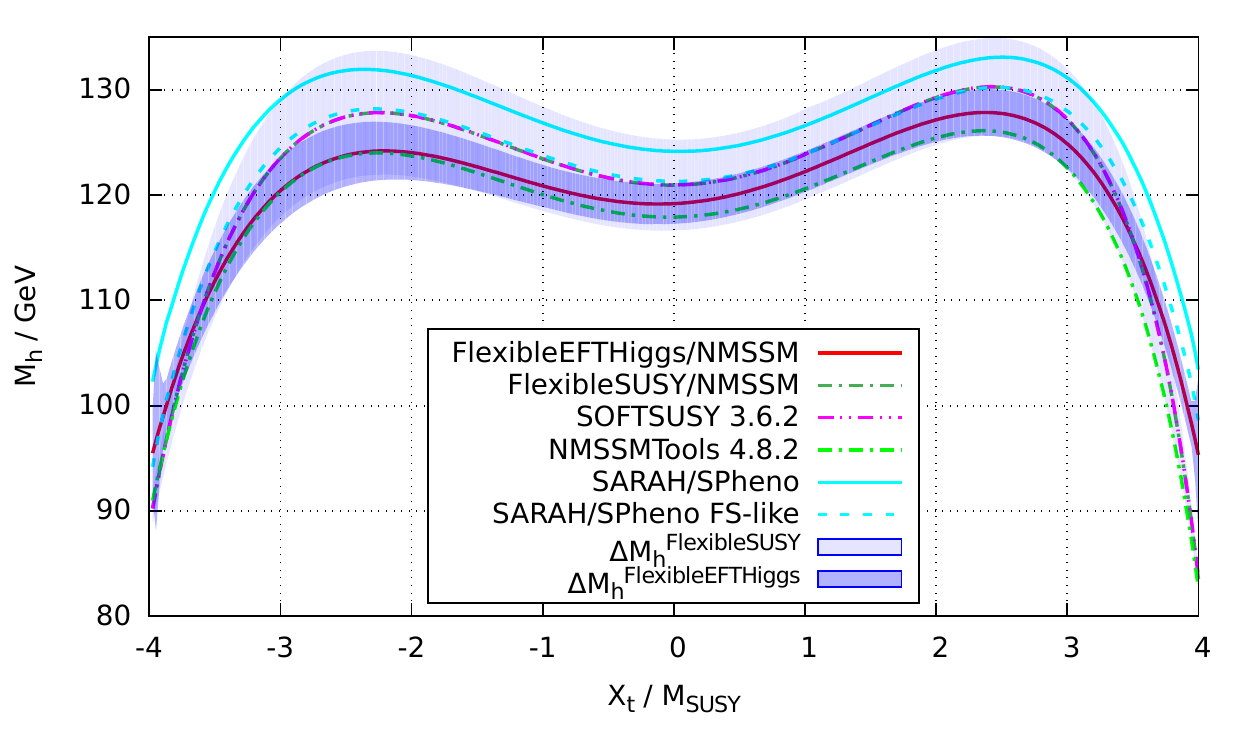}
  \caption{As Fig.~\ref{fig:Mh_NMSSM}, but for other parameter
    choices. We fix $\MS = 10$ TeV
    and $\tan\beta=5$ and plot $M_h$ against i) (left panel) $\lambda$,
    where $\kappa =\lambda$, $X_t = 0$ and ii) (right panel) $X_t$
    where $\kappa =\lambda = 0.01$.}
  \label{fig:Mh_NMSSM2}
\end{figure}

  The left panel of Figure \ref{fig:Mh_NMSSM2} confirms that \FSTower and
  the fixed-order calculations in \FS and \Softsusy agree remarkably
  well for all values of $\lambda$ and $\kappa$, indicating that these
  couplings do not disrupt this remarkable numerical cancellation
  amongst the 3-loop logs, which yield results close to the
  correct ones calculated using effective field theory techniques.  The same
  cancellation does not take place in the \nmssmtools calculation and
  the deviation between this result and fixed-order \FS gives an
  indication of the large uncertainty in these approaches.  By
  contrast the right panel of Figure \ref{fig:Mh_NMSSM2} shows that as
  with the MSSM, this cancellation depends on the value of $X_t$.  The
  results here are very similar to those of the MSSM, since we are in
  the MSSM limit.  Interestingly the fixed order calculation of \nmssmtools
  agrees very well with \FSTower when $X_t \approx -\sqrt{6}\MS$.

\section{Numerical results in the \ESSM}
\label{sec:E6SSM}
We now make a much bigger departure from minimality and consider an
$E_6$ inspired model, with an extra $U(1)$ gauge symmetry and matter
filling complete multiplets of the fundamental $\mathbf{27}$
representation of $E_6$.  Specifically we
consider the exceptional supersymmetric standard model (\ESSM)
\cite{King:2005jy,King:2005my,Athron:2010zz} which has previously been
shown to have very heavy sfermions \cite{Athron:2010zz,Athron:2009bs}
making effective field theory techniques very important for accurately
predicting the Higgs mass.  In the past 2-loop expressions for the Higgs
mass were obtained \cite{King:2005jy} by generalising MSSM \cite{Carena:1995wu}
and NMSSM \cite{Ellwanger:1999ji} results from effective field theory calculations that
had been expanded to fixed 2-loop order.  This was used in
determining the spectrum \cite{Athron:2010zz,Athron:2009bs} and
showing consistency with a $125$ GeV Higgs \cite{Athron:2012sq},
though the accuracy was strictly limited due to the very heavy
spectra.  A first attempt at improving precision of calculations in
the model was made when full 1-loop threshold corrections for the
gauge and Yukawa couplings were calculated \cite{Athron:2012pw}, with
the top Yukawa threshold corrections having a significant impact on
the Higgs mass.  With \SARAH the full 1-loop self energy can be
calculated for the first time in \FS and \SPheno and both
include the option to use NMSSM and MSSM 2-loop corrections, though
these will not be accurate when the exotic couplings are large and must be
used with care.  \SARAH/\SPheno can now calculate full
fixed-order 2-loop order corrections.\footnote{Calculated in the gaugeless
  limit with the effective potential approximation, where $p^2 =0$.}  Finally,
recently when studying the phenomenology of an $E_6$ inspired model
\cite{Athron:2015vxg}, \SUSYHD was used to resum the logs and
obtain the Higgs mass after matching to the MSSM at tree level.

Here we investigate the impact of \FSTower on the Higgs mass. We will
compare our results to the fixed-order calculations of \FS and also
compare with \SARAH/\SPheno.

The E$_6$SSM extends the matter content of the MSSM with the following superfields:
\begin{equation}
  \begin{aligned}
    \hat{H^d_\alpha}&:\textstyle (\mathbf{1}, \mathbf{2}, -\frac12, -3),
    &\hat{H^u_\alpha}&:\textstyle (\mathbf{1}, \mathbf{2}, \frac12, -2),
    &\hat{D^x_i}&:\textstyle (\mathbf{3}, \mathbf{1}, -\frac13, -2),
    &\hat{\overline{D^x_i}}&:\textstyle (\mathbf{3}, \mathbf{1}, \frac13, -3),\\
    \hat{S_i}&:\textstyle (\mathbf{1}, \mathbf{1}, 0, 5),
    &\hat{N_i^c}&:\textstyle (\mathbf{1}, \mathbf{1}, 0, 0),
    &\hat{H^\prime}&:(\mathbf{1}, \mathbf{2}, -\frac12, 2),
    &\hat{\overline{H^\prime}}&:(\mathbf{1}, \mathbf{2}, \frac12, -2),
  \end{aligned}
\end{equation}
where we include generation indices $i=1,2,3$ and $\alpha=1,2$ and we specify the $G_{\ESSM} = G_{\text{SM}}\times U(1)_N$ gauge group quantum numbers with the quantities in brackets specifically showing the $SU(3)$ representation, the $SU(2)$ representation, the $U(1)_Y$ charge without GUT normalization and the $U(1)_N$ charge also without GUT normalization.\footnote{The $E_6$ GUT normalization for the $U(1)_N$ charges is $\frac{1}{\sqrt{40}}$, while the $E_6$ GUT normalization for the $U(1)_Y$ charges is the same as the usual $SU(5)$ one, $\sqrt{3/5}$.}

The full $E_6$ superpotential is rather complicated, but with some simplifying assumptions including a $Z_2^H$ symmetry to forbid flavour changing neutral currents and a $Z_2^B$ symmetry to forbid proton decay, the superpotential can be written as \cite{Athron:2010zz},
\begin{align}
  \begin{split}
    \mathcal{W}_{\ESSM} &= \mathcal{W}_{\text{MSSM}}(\mu=0) +  \lambda \hat{S}_3 \hat{H}_u \hat{H}_d + \lambda_\alpha  \hat{S}_3 \hat{H}^u_\alpha \hat{H}^d_\alpha + \kappa_i \hat{S}_3 \hat{D}^x_i \hat{D}^x_i + \mu^\prime \hat{\overline{H^\prime}} \hat{H}^\prime.
  \end{split}
  \label{eq:E6SSM_superpotential}
\end{align}
The soft breaking Lagrangian then contains,
\begin{align}
  \LModelSoft{\ESSM} &= \LModelSoft{\MSSM} (B\mu = 0)  - m^2_{S_i}|S_i|^2 - m^2_{H^u_i}|H^u_i|^2 - m^2_{H^d_i}|H^d_i|^2 - m^2_{H^\prime}|H^\prime|^2 - m^2_{\overline{H^\prime}}|\overline{H^\prime}|^2 \nonumber \\
      &\phantom{={}}-   m^2_{D_i}|D_i|^2  - m^2_{\overline{D}_i}|\overline{D}_i|^2 - \frac12 M_1^\prime \bar{\tilde{B}}^\prime \tilde{B}^\prime \nonumber \\
      &\phantom{={}}-\left[B^\prime\mu^\prime \, {\overline{H^\prime}}\cdot{H^\prime} + T_{\lambda_i} S_3 H_i^d \cdot H_i^u + T_{\kappa_i} S_3 D_i\overline{D}_i +\textrm{h.c.}\right],
\end{align}
where $\tilde{B}^\prime$ is the gaugino superpartner of the $B^\prime$ gauge field, associated with the $U(1)_N$ gauge symmetry, and we have defined $H_3^d := H_d$ and $H_3^u:=H_u$ to write the soft trilinear couplings more compactly. The third generation singlet $S_3$ and the neutral components of $H_u$ and $H_d$ doublets are the Higgs fields which develop the VEVs,
$v_s/\sqrt{2}$, $v_u/\sqrt{2}$, and $v_d/\sqrt{2}$, respectively.
In our analysis here we set the soft-breaking scalar and gaugino mass
parameters and $\mu_{\text{eff}} = \lambda v_s / \sqrt{2}$ to $\MS$,
as defined in Eqs.~\eqref{eq:MSSM_MSUSY_scenario}.  In addition, we
fix
\begin{align}
  \begin{split}
    &(m_{s}^2)_{\alpha\beta}(\MS) = \delta_{\alpha\beta} \MS^2, \quad (s=S,H^u,H^d) \\
    &m^2_{H^\prime}(\MS) = m^2_{\overline{H^\prime}}(\MS) = \MS^2, \\
    &m^2_{D_i}(\MS) = m^2_{\overline{D}_i}(\MS) = \MS^2, \quad (i = 1,2,3) \\
    &B^\prime\mu^\prime(\MS) = \MS^2, \\
    &M_1^\prime(\MS) = \MS .
  \end{split}
\end{align}
To ensure that the exotic quarks, the inert Higgsinos and the
$Z^\prime$ boson all get \DRbar masses equal to \MS we set
\begin{equation}
    \kappa(\MS) = \lambda_{1,2}(\MS)  = g_1^\prime(\MS)\frac{5}{\sqrt{20}} = \lambda(\MS) .
  \end{equation}
We also require that $v_s$ is fixed so that $\mu_{\text{eff}} = \MS$, i.e.
\begin{equation}v_s(\MS) =  \frac{\sqrt{2} \MS}{\lambda(\MS)}.\end{equation}
The \ESSM-specific trilinear couplings are set to
\begin{align}
  \begin{split}
    & T_{\lambda_3} (\MS) = \frac{\sqrt{2}\MS^2 \sin\beta \cos\beta}{v_s}, \\
    & T_{\kappa_{1,2,3}}(\MS)  = T_{\lambda_{1,2}}(\MS) = 0.
\end{split}
\end{align}  
and the soft scalar Higgs masses $m_{H_u}^2, m_{H_d}^2,
(m_{S}^2)_{3,3}$ are fixed by the EWSB conditions.  For the scans we
use $\tan\beta(\MS) = 5$ and $\lambda(\MS) = 0.1$.

\begin{figure}[tbh]
  \centering
  \includegraphics[width=0.49\textwidth]{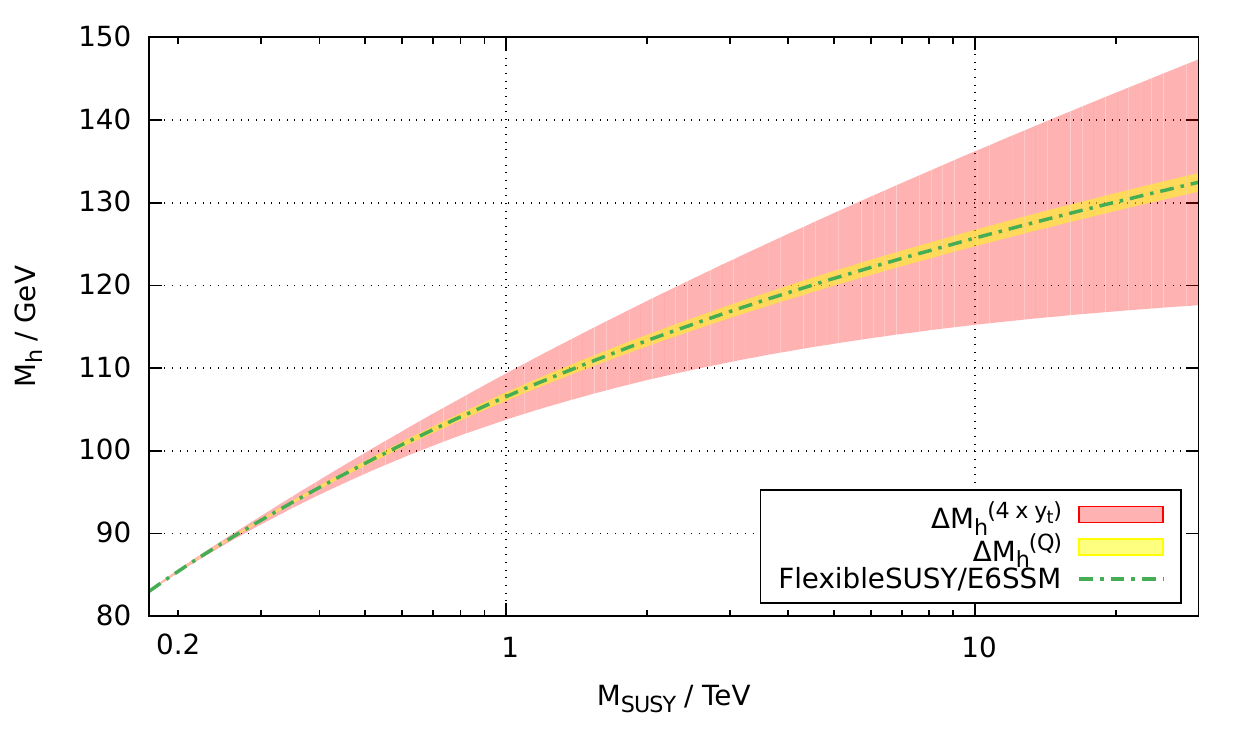}
  \includegraphics[width=0.49\textwidth]{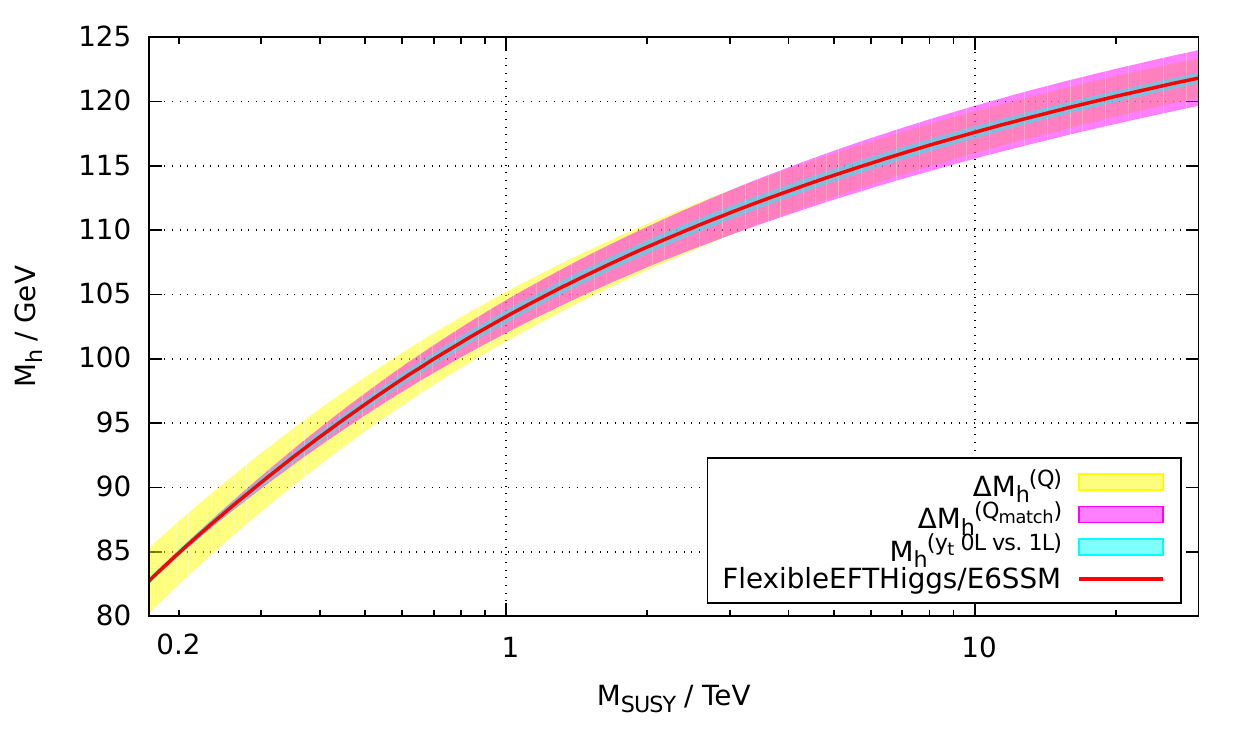} \\
  \includegraphics[width=0.49\textwidth]{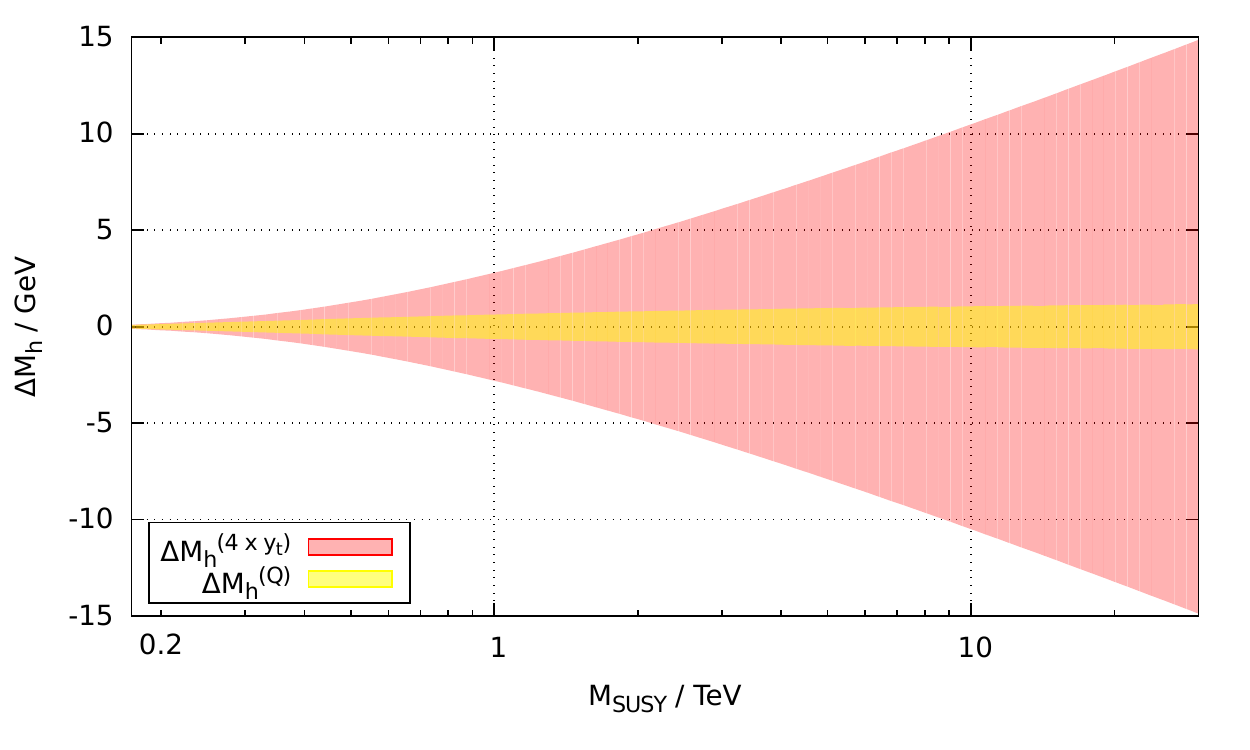}
  \includegraphics[width=0.49\textwidth]{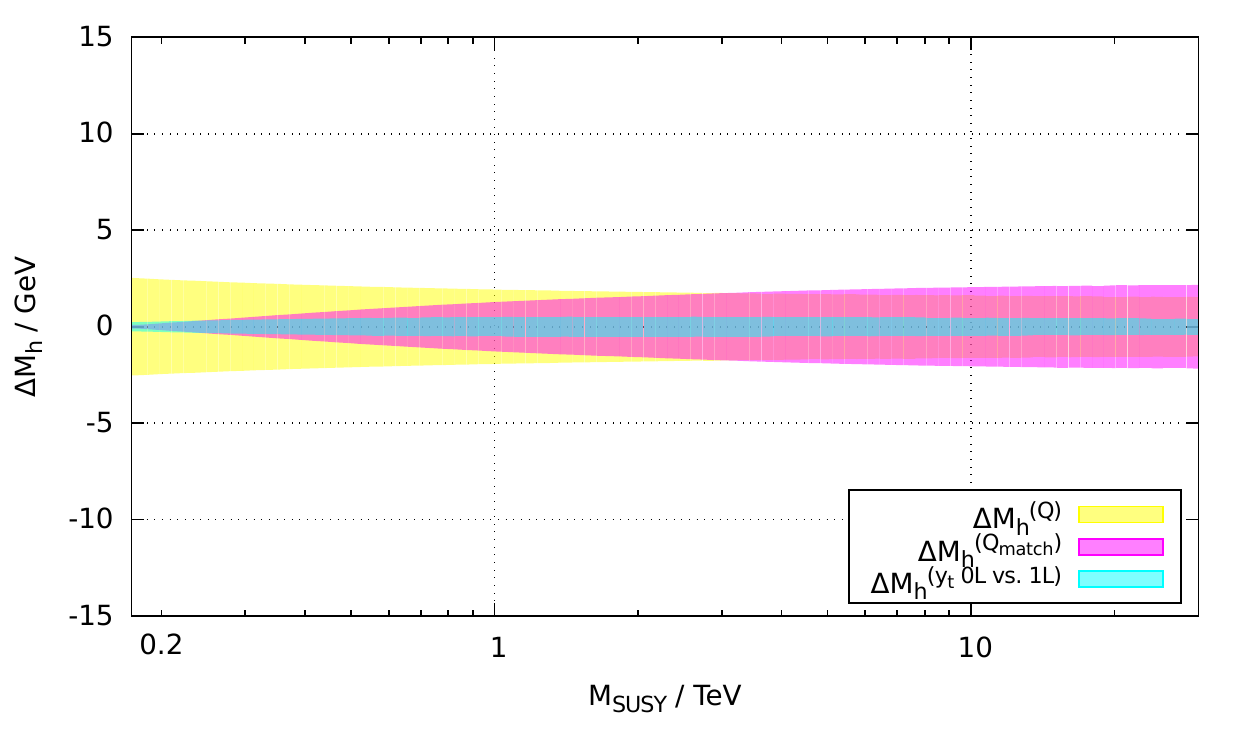}
  \caption{Higgs mass predictions and uncertainty estimates of
    Sec.~\ref{sec:MSSMtower_results} applied to the fixed-order
    calculations with \FS and to \FSTower in the \ESSM with fixed
    $X_t = 0$, $\tan\beta = 5$ and $\lambda = 0.1$. The
    top row shows the Higgs mass predictions and the uncertainty
    estimates, the lower row only the uncertainty estimates, as in
    Fig.~\ref{fig:MSSM_tower_uncertainties}.}
 \label{fig:MhFSErrsE6SSM}
 \end{figure}

In Figure \ref{fig:MhFSErrsE6SSM} one can see that the fixed-order \FS
result is quite different from the \FSTower result.  In this case it
seems that the cancellation between the logarithms is spoiled, due
to the substantially altered RGE running between the EW scale and \MS
caused by the additional colored matter which dramatically affect the
RGE trajectory of $\alpha_s$ and then indirectly $\alpha_t$ through
the gauge coupling contributions to the RGEs.  The fact that we are
shifted so far away from the cancellation is also reflected in the
enhancement of the fixed-order uncertainty estimate from extracting
$y_t$ in different ways, shown in red in the left panels. Figure
\ref{fig:MhFSErrsE6SSM} also shows our uncertainty estimates for
\FSTower in the right panels.  As with the MSSM and NMSSM we can see
that our estimation of the theory uncertainty
(shown on the right) is not increasing significantly with \MS, which is to be
expected from the construction of this approach.

\begin{figure}[tbh]
  \centering
  \includegraphics[width=0.7\textwidth]{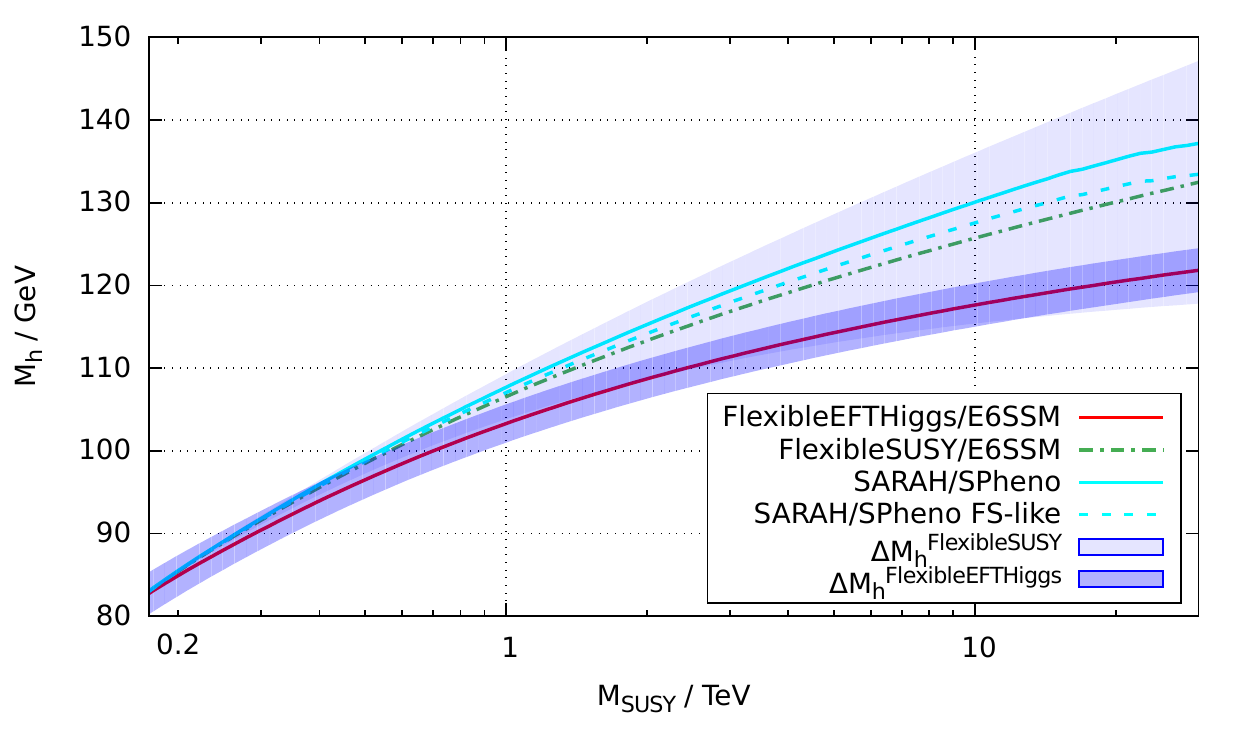}
  \caption{Predictions for $M_h$ and combined theoretical uncertainty
    estimates for \FS and \FSTower in the \ESSM, compared with results
    of other codes.  We fix $X_t = 0$, $\tan\beta = 5$
    and $\lambda = 0.1$.}
 \label{fig:MhFSvsSpheno}
\end{figure}

In Figure \ref{fig:MhFSvsSpheno} we show the \SARAH/\SPheno prediction, along
with combined uncertainty estimations for the fixed-order \FS and \FSTower
results.  Here the \SPheno prediction is close to the \FS fixed-order
prediction, particularly when the top Yukawa is extracted in the same
way.  This should be expected since the exotic couplings are all quite
small in this scenario, making the 2-loop corrections that are only
in \SPheno rather small.  Therefore the main difference between the
\SPheno and \FSTower results appears to be due to the resummed logs, which in this
case become important at much lower \MS values.  Notably at $\MS =1$
TeV there is already a $3$ GeV gap between the \FSTower prediction and
the fixed-order predictions, though the results are compatible within
estimated uncertainties. It is noteworthy that the estimated
uncertainty of \FSTower is in the range  $2$--$3\unit{GeV}$, like in
the MSSM, while the one of the fixed order results has significantly
increased. To improve the precision further, adding 2-loop matching
to the \FSTower calculation will be very important. It is also worth
noting that we see no evidence of the problems due to infra-red
divergences in the \ESSM-specific \SARAH/\SPheno 2-loop correction
here, which can be understood due to the small values of the exotic
couplings.
We do not investigate varying the exotic couplings here due to large
dimensionality of the parameter space, but leave this for dedicated
studies of this model.

\section{Numerical results in the MRSSM}
\label{sec:MRSSM}

As another example for a non-minimal model, we study the properties of
\FSTower in the MRSSM, a minimal supersymmetric model with unbroken
continuous R-symmetry \cite{Kribs:2007ac}. The model is motivated in a
number of ways, and particularly the mass of the SM-like Higgs boson
has been shown to be compatible with experiment in a variety of
parameter scenarios in
Refs.~\cite{Diessner:2014ksa,Diessner:2015yna,Diessner:2015iln}. In the following we employ
the conventions of these references.  The MRSSM has the same
field content as the MSSM, with the following additional superfields:
\begin{equation}
  \begin{aligned}
 \hat{R}_d&:\textstyle (\mathbf{     1} ,\mathbf{2},-\frac{1}{2}) ,
&\hat{R}_u&:\textstyle (\mathbf{     1} ,\mathbf{2}, \frac{1}{2}) ,
&\hat{S}  &:\textstyle (\mathbf{     1} ,\mathbf{1}, 0) ,
&\hat{T}  &:\textstyle (\mathbf{     1} ,\mathbf{3}, 0) ,
&\hat{O}  &:\textstyle (\mathbf{     8} ,\mathbf{1}, 0) .
  \end{aligned}
\end{equation}
The superpotential of the MRSSM is given by
\begin{align}
  \begin{split}
    \mathcal{W}_{\text{MRSSM}} &= \mathcal{W}_{\textrm{MSSM}}(\mu = 0)
    + \mu_d \hat{R}_d \cdot \hat{H}_d + \mu_u \hat{R}_u\cdot\hat{H}_u \\
    &\phantom{={}}+ \lambda_d \hat{S} \hat{R}_d \cdot \hat{H}_d +
    \lambda_u \hat{S} \hat{R}_u\cdot\hat{H}_u +\Lambda_d
    \hat{R}_d\cdot \hat{T} \hat{H}_d +\Lambda_u \hat{R}_u\cdot\hat{T}
    \hat{H}_u .
  \end{split}
  \label{eq:MRSSM_superpotential}
\end{align}
As in the \ESSM\ and $Z_3$-symmetric NMSSM, the $\mu$ term is
forbidden in the \MRSSM.  New $\mu_{u,d}$ terms and Yukawa-like
interactions between the $\hat{R}$ Higgs fields and $\hat{H}_{u,d}$
are allowed in general.  The soft-breaking trilinear couplings as well
as the Majorana mass terms for the gauginos are forbidden by the
R-symmetry.  The Lagrangian of the soft breaking scalar mass terms
reads
\begin{align}
  \begin{split}
    \mathcal{L}_{\MRSSM}^{\mathrm{soft},m^2}
    &= \LModelSoft{\MSSM} (B\mu = T_u = T_d = T_e = M_1 = M_2 = M_3 = 0) \\
    &\phantom{={}}
    - m_{R_u}^2 \left( |R_u^0|^2 + |R_u^-|^2 \right)
    - m_{R_d}^2 \left( |R_d^0|^2 + |R_d^+|^2 \right) \\
    &\phantom{={}}
    - m_S^2 |S|^2 - m_T^2 \left( |T^0|^2 + |T^+|^2 + |T^-|^2 \right) - m_O^2 |O|^2.
  \end{split}
\end{align}
The fermionic components of the $\hat{S}$, $\hat{T}$ and $\hat{O}$
fields mix with the gauginos $\tilde{B}$, $\tilde{W}$ and $\tilde{g}$
into Dirac fermions.  The Dirac mass terms can be interpreted as being
generated by the soft breaking of a supersymmetric hidden sector model
via spurions.  The resulting part of the soft-breaking MRSSM
Langrangian reads
\begin{align}
  \begin{split}
    \mathcal{L}_{\MRSSM}^{\mathrm{soft,M}}
    &= - M_B^D (\tilde{B}\tilde{S} - \sqrt{2} \mathcal{D}_B S)
    - M_W^D(\tilde{W}^a\tilde{T}^a-\sqrt{2}\mathcal{D}_W^a T^a) \\
    &\phantom{={}}- M_g^D(\tilde{g}^a\tilde{O}^a-\sqrt{2}\mathcal{D}_g^a O^a) +
    \text{h.c.} ,
  \end{split}
\end{align}
where the auxiliary $\mathcal{D}$ fields can be eliminated by their
equations of motion, giving rise to triple scalar interactions
governed by the Dirac mass parameters.  The Higgs fields $H_u$, $H_d$ and $S$ develop
VEVs as in Eqs.~\eqref{eq:NMSSM_Higgs_VEVs}.  In addition, the
electrically neutral linear combination of the Higgs triplet develops
a VEV as
\begin{align}
  \langle T^0 \rangle = \frac{v_T}{\sqrt{2}}.
\end{align}
For the MRSSM study presented in this section, we impose the boundary
conditions of Eqs.~\eqref{eq:MSSM_MSUSY_scenario} as well as
\begin{align}
  \begin{split}
    &m_S^2 = m_T^2 = m_O^2 = m_{R_d}^2 = m_{R_u}^2 = 10 \MS^2, \\
    &M_B^D = M_W^D = M_g^D = \MS,\\
    &\mu_u = \mu_d = 1 \unit{TeV}, \tan\beta = 5, \\
    &\Lambda_u = \Lambda_d = -0.5, \lambda_u = \lambda_d = -0.01.
  \end{split}
  \label{eq:MRSSM_parameterpoint}
\end{align}
at the scale $\MS$, which is inspired by BMP3$'$ from
Ref.~\cite{Diessner:2015yna}.
The parameters $m_{H_u}^2$, $m_{H_d}^2$, $v_S$, $v_T$ are fixed at the
scale $\MS$ by the four electroweak symmetry breaking conditions.  In
the fixed-order calculation, the MRSSM \DRbar gauge and Yukawa
couplings as well as the Standard Model-like vacuum expectation value
$v=\sqrt{v_u^2 + v_d^2}$ are calculated from the known values of
$\alphaEMMZ$, $\alphaSMZ$, $G_F$, $M_W$, $M_Z$ and from the known
Standard Model fermion masses at the 1-loop level at the low-energy
scale $M_Z$.\footnote{\label{ftn:MRSSM_gauge_couplings}In \SPheno,
  $\alphaEMMZ$ and the Fermi constant $G_F$ are used as input to calculate the
  \DRbar gauge couplings $g_1(M_Z)$ and $g_2(M_Z)$. This approach is a
  generalization of the one presented in \cite{Pierce:1996zz} for the
  \MSSM.  \FS, in contrast, uses the $W$ and $Z$ pole masses as input
  to calculate $g_1(M_Z)$ and $g_2(M_Z)$ in the MRSSM at the 1-loop level.}
In
particular, the top Yukawa coupling is calculated as described in
Section~\ref{sec:fixed_order_calculations}, where the 2-loop
SM-QCD corrections are taken into account.

\begin{figure}[tbh]
  \centering
  \includegraphics[width=0.49\textwidth]{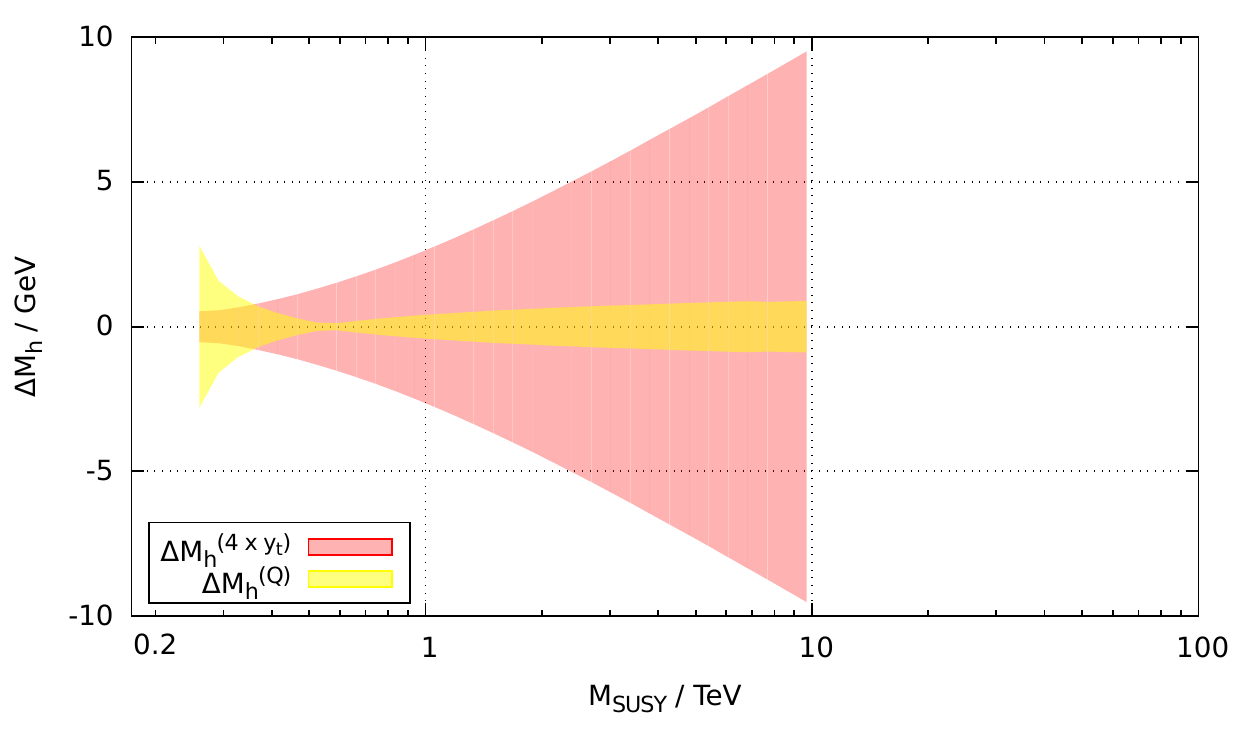}
  \includegraphics[width=0.49\textwidth]{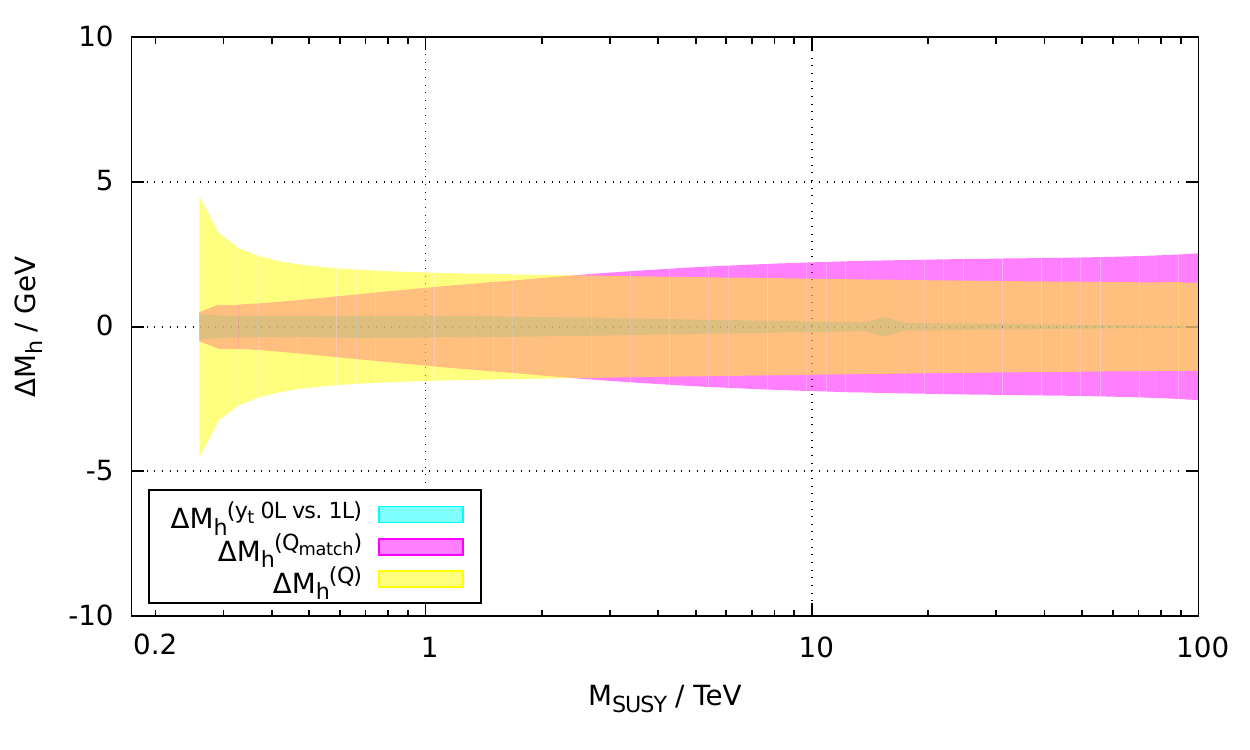}
  \caption{Uncertainty estimates of Sec.~\ref{sec:MSSMtower_results}
    for $M_h$, applied to the fixed-order approach with \SARAH/\SPheno
    (left panel) and to \FSTower/MRSSM (right panel) as a function of
    $\MS$ for the MRSSM parameter point
    \eqref{eq:MRSSM_parameterpoint}.}
  \label{fig:Mh_MRSSM_uncertainties}
\end{figure}
In the left panel of Figure~\ref{fig:Mh_MRSSM_uncertainties} we show
the two uncertainty estimate bands for the fixed-order 2-loop
calculation with \SARAH/\SPheno as described in
Section~\ref{sec:FStower_Xt=0_uncertainty_estimation}:
\begin{itemize}
\item The yellow uncertainty band shows $\DeltaMhQ$, i.e.\
  the variation of $M_h$ when
  the renormalization scale is varied, at which $M_h$ is calculated.
  We find that for \SPheno this estimation of part of the
  uncertainty is of the order $0.5$--$2\unit{GeV}$ for
  most of the displayed \MS range,
  which is relatively small, because of the 2-loop Higgs
  mass computation.  In contrast, this uncertainty is between
  $1$--$3\unit{GeV}$ for \FS due to the missing 2-loop Higgs mass
  contributions.
\item The red band shows $\DeltaMhFSYt$, i.e.\ the variation of $M_h$
  in the 1-loop fixed-order calculation with \FS when $y_t^\MRSSM$ is
  calculated in the four different ways presented in
  Section~\ref{sec:FStower_Xt=0_uncertainty_estimation}.  As
  discussed, it estimates a partial theory uncertainty of the
  fixed-order 2-loop calculation from missing 3-loop terms.  The red
  band would be a clear underestimation of the theory uncertainty of
  \FS's 1-loop calculation.  For both programs we find that this
  uncertainty estimate is dominant and can reach up to $17\unit{GeV}$
  for SUSY scales around $100\unit{TeV}$.
\end{itemize}
In the right panel of Figure~\ref{fig:Mh_MRSSM_uncertainties},
the three uncertainty bands introduced in
Section~\ref{sec:MSSM_for_Xt_ne_0} for \FSTower are shown:
\begin{itemize}
\item The turquoise uncertainty band shows $\DeltaMhFStowerYt$, which has
  been obtained by calculating $y_t^\MRSSM(\MS)$ using either a
  tree-level or a 1-loop top quark pole mass matching.
  For this scenario the resulting
  estimated uncertainty is below $0.5\unit{GeV}$ for all values of $\MS$.
  This uncertainty is smaller in the MRSSM than in
  the MSSM for maximal mixing, partially because $y_t^\MRSSM(\MS)$ is
  smaller than $y_t^\MSSM(\MS)$, for example
  $y_t^{\MRSSM,(1)}(100\unit{TeV}) = 0.81$,
  $y_t^{\MSSM,(1)}(100\unit{TeV}) = 0.87$.
\item The red band shows $\DeltaMhQmatch$, i.e.\ the variation of $M_h$ when the matching
  scale $\Qmatch$ is varied within the interval $[\MS/2, 2\MS]$.  We
  find that this uncertainty is between $0.5$--$2.5\unit{GeV}$ and thus
  dominates in the scenario considered here.
\item The yellow band shows $\DeltaMhQ$ in \FSTower.
  This estimated uncertainty is below $2\unit{GeV}$ for
  all SUSY scales above $1\unit{TeV}$, similarly to the results in the
  other non-minimal SUSY models.
\end{itemize}
Based on these estimated theoretical uncertainties, we conclude that
for the scenario studied here \FSTower leads to a more precise
prediction of $M_h$ than the fixed-order calculation for SUSY
scales above a few TeV\@.

\begin{figure}[tbh]
  \centering
  \includegraphics[width=0.7\textwidth]{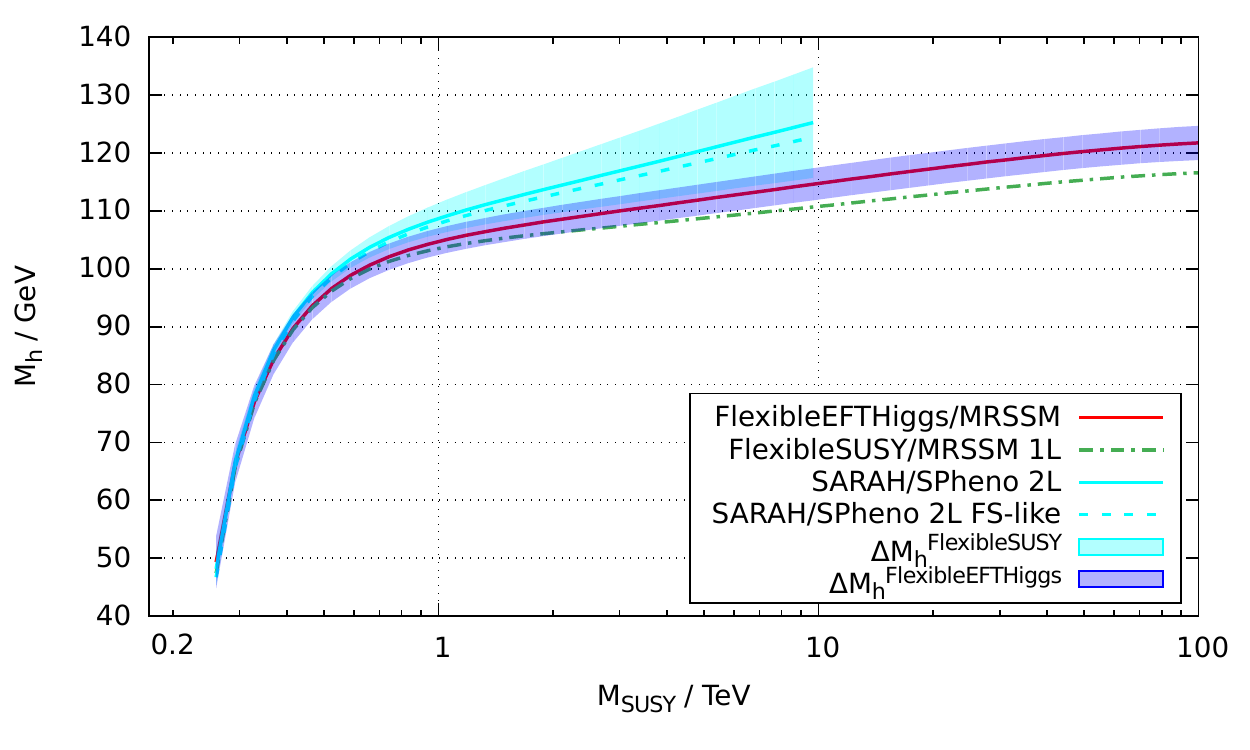}
  \caption{Predictions for $M_h$ and combined theoretical uncertainty estimates for
  \FS and  \FSTower   in the MRSSM, compared with results of other
  codes as a function of
    $\MS$ for the parameter point specified in
    Eqs.~\eqref{eq:MRSSM_parameterpoint}.}
  \label{fig:Mh_MRSSM}
\end{figure}

In Figure~\ref{fig:Mh_MRSSM}, the lightest CP-even Higgs pole mass in
the MRSSM is shown as a function of $\MS$ for the parameter point
\eqref{eq:MRSSM_parameterpoint}, together with the combined uncertainty
estimates. The green dash-dotted line shows the fixed-order 1-loop
calculation with \FS.  The difference between \FS's and \SARAH/\SPheno's
1-loop calculations originates again from the different definition of
the running top mass in the MRSSM at $M_Z$.
The turquoise solid line shows \SPheno's fixed-order 2-loop Higgs pole
mass calculation.  We find that the pure 2-loop corrections enhance
the Higgs mass significantly, up to $12\unit{GeV}$ for $\MS =
10\unit{TeV}$, compared to the 1-loop result.\footnote{We see no
  evidence for infra-red divergences in \SPheno's 2-loop corrections
  for this MRSSM scenario.  However, we find numerical instabilities
  for $\MS > 10\unit{TeV}$, which is why we do not draw the \SPheno 2-loop
  curve above this scale.}  Such large 2-loop corrections in the
\DRbar scheme have been found and studied already in
\cite{Diessner:2015yna} and have been compared to the on-shell scheme
in \cite{Braathen:2016mmb}.  The turquoise
dashed line again shows \SPheno's fixed-order 2-loop calculation, but
using the definition \eqref{eq:fixedmt_FlexibleSUSY} for the running
top Yukawa coupling.
We find that in this scenario all fixed-order curves become
linear for $\MS \gtrsim 1\unit{TeV}$, which indicates that in this
scenario $M_h$ is dominated by the leading logarithm for large SUSY
scales.
The red solid line shows $M_h$ as calculated with \FSTower in the
\MRSSM.  Since in \FSTower a 1-loop Higgs mass matching and 3-loop
renormalization group running is performed, \FSTower resums the
leading logarithmic contributions to all orders.
Since \FSTower resums large logarithms, and since these logarithmic
contributions dominate for SUSY scales above $1\unit{TeV}$ in this
scenario, we again expect \FSTower to give a
more precise Higgs mass prediction for SUSY scales above a few TeV\@.

In addition, we show in Figure~\ref{fig:Mh_MRSSM} the combined
uncertainty estimates introduced in
Section~\ref{sec:MSSM_uncertainty_summary}.  In the MRSSM, \SPheno is
the only publicly available program which can perform a (partial)
2-loop calculation.  Since $\Delta M_h^{\FS}$ is a partial estimation
of missing 3-loop corrections, we can reasonably draw it only around
\SPheno's 2-loop curve.  The corresponding combined uncertainty for
\FS's 1-loop calculation is expected to be significantly larger than
the shown size of $\Delta M_h^{\FS}$ and would require a separate
uncertainty estimation of the missing 2-loop contributions.  As
expected, $\Delta M_h^{\FS}$ grows logarithmically with $\MS$ and can
be as large as $10\unit{GeV}$ for $\MS = 10\unit{TeV}$.  In contrast,
the uncertainty estimate for \FSTower, $\Delta M_h^{\FSTower}$,
remains nearly constant and around $3\unit{GeV}$
for $\MS \gtrsim 2\unit{TeV}$.

For comparison to the study of the 2-loop Higgs pole mass
contributions in Ref.~\cite{Diessner:2015yna}, we show in
Table~\ref{tab:MRSSM-benchmark-points} the lightest CP-even Higgs pole
mass in the MRSSM for the benchmark points BM1$'$--BM3$'$ together
with combined uncertainties from
Eqs.~\eqref{eq:def_Delta_Mh_tower} and
\eqref{eq:def_Delta_Mh_fixed_order}.\footnote{The values of the
  lightest CP-even Higgs masses for BM1$'$--BM3$'$ presented in
  \cite{Diessner:2015yna} have been obtained with SARAH 4.5.3 and
  SPheno 3.3.6.  In addition, the authors have modified the generated
  SPheno code to predict the $W$ mass with higher precision, as
  described in \cite{Diessner:2014ksa}, Eqs.~(4.8)--(4.13).}
\begin{table}[tbh]
  \centering
  \begin{tabular}{rrrrrrr}
    \toprule
    Point & \SPheno & \SPheno & \SPheno & \SPheno & \FS & FlexibleEFT-\\
          & 1L      & 2L    & 1L, \eqref{eq:fixedmt_FlexibleSUSY}
          & 2L, \eqref{eq:fixedmt_FlexibleSUSY} & 1L & Higgs 1L\\
    \midrule
    BM1$'$ & $120.4$ & $125.6\pm 1.3$ & $120.0$ & $125.1\pm 1.3$ & $120.6$ & $122.1\pm 1.7$\\
    BM2$'$ & $120.8$ & $126.0\pm 1.1$ & $120.4$ & $125.6\pm 1.1$ & $120.2$ & $121.7\pm 1.8$\\
    BM3$'$ & $121.0$ & $125.7\pm 1.3$ & $120.5$ & $125.2\pm 1.3$ & $120.4$ & $121.9\pm 1.9$\\
    \bottomrule
  \end{tabular}
  \caption{Lightest CP-even Higgs pole mass in GeV for the MRSSM
    benchmark points BM1$'$--BM3$'$ of Ref.~\cite{Diessner:2015yna}.
    The given uncertainty estimates have been obtained using
    Eqs.~\eqref{eq:def_Delta_Mh_tower} and
    \eqref{eq:def_Delta_Mh_fixed_order}.}
  \label{tab:MRSSM-benchmark-points}
\end{table}
The first two data columns show $M_h$ calculated with \SARAH and
\SPheno at the 1- and 2-loop level, respectively.  These
computations use the definition \eqref{eq:fixedmt_SPheno} to calculate
$y_t^\MRSSM(M_Z)$.  In the third and fourth data columns $M_h$ has been
calculated with a modified version of \SARAH/\SPheno, where
Eq.~\eqref{eq:fixedmt_FlexibleSUSY} is used to calculate
$y_t^\MRSSM(M_Z)$.  These different Yukawa coupling definitions amount
to $0.4$--$0.5\unit{GeV}$ shift in $M_h$ for these benchmark points.
The fifth data column shows $M_h$ calculated with \FS at the 1-loop
level, where by default Eq.~\eqref{eq:fixedmt_FlexibleSUSY} is used to
calculate $y_t^\MRSSM(M_Z)$, and $g_{1,2}^\MRSSM(M_Z)$ are calculated
using $M_Z$ and $M_W$ as input (instead of $M_Z$ and $G_F$ as done in
\SPheno).  The difference between the fixed-order \FS and \SPheno
1-loop calculations using Eq.~\eqref{eq:fixedmt_FlexibleSUSY} (data
columns $3$ and $5$) originates from the different definitions of the
electroweak gauge couplings, which affect the Higgs pole mass already
at the tree-level.  The last column shows the calculation of $M_h$
with \FSTower, which resums the leading logarithms to
all orders.  The result of \FSTower lies between the 1- and 2-loop
calculations.

\section{Conclusions}
\label{sec:conculsion}
We have presented \FSTower, an EFT calculation of the SM-like Higgs
mass in any SUSY or non-SUSY model, that can make precise predictions for both
high and low new physics scales. A judicious choice of matching conditions,
equating pole masses, ensures that terms of
$\mathcal{O}(v^2/\MS^2)$ are included, which are missed by ``pure EFT''
calculations such as \SUSYHD and \FS/\HSSUSY. Thus large logarithms can be
resummed, while ensuring that the Higgs pole mass calculation is exact at
the 1-loop level. Since this choice of matching requires only self
energies and tadpoles, it is also very easy to automate and apply to
any SUSY (or even non-SUSY) model, where the Standard Model is the
valid low energy effective field theory. This method has been
implemented in \FS, and we have used this to obtain results in the
MSSM, NMSSM, E$_6$SSM and MRSSM\@.

We discussed several ways to estimate the theoretical uncertainty of
\FSTower and the fixed-order approaches of \FS/\Softsusy and \SPheno.
These estimates show the expected behaviour, i.e.\ the fixed-order
uncertainty rises with \MS while our \FSTower estimate
does not.  For example in the MSSM when
$\MS$ is larger than a few TeV our combined uncertainty estimate for
\FSTower is smaller than our combination of various fixed-order
uncertainty estimates, and similar results are obtained in the
NMSSM, \ESSM\ and MRSSM\@.  Moreover, even for SUSY scales close to the
EW scale we observe that the uncertainty of \FSTower is around
$2$--$3\unit{GeV}$ and is thus not much larger than the fixed-order
calculations, even for cases where the 2-loop contributions to the
Higgs mass are large.

We also compared \FSTower to other spectrum generators in all four of
these models. In the MSSM we demonstrated that we understand all the
differences between \FSTower and \SUSYHD and showed that the two codes
agree well at high \MS in scenarios where the 2-loop threshold
corrections are negligible.  In regions where the 2-loop threshold
corrections are non-negligible, the codes disagree by non-logarithmic
2-loop terms, which do not increase with \MS.  However, in these
regions the deviation of the codes lies within our uncertainty
estimate for \FSTower.  We also found that the fixed-order
calculations of \FS and \Softsusy agree surprisingly well with the EFT
results of \FSTower, \SUSYHD and \FS/HSSUSY even at very large \MS,
owing to an accidental cancellation among the 3-loop leading
logarithms.  This cancellation however depends on which partial 3-loop
corrections are included and \SPheno does not have the same tendency
despite being accurate to the same formal 2-loop order as \FS and
\Softsusy.  This cancellation also occurs in the NMSSM, even for
rather large values of the new singlet Yukawa couplings.  There we
see on the other hand that the full fixed-order 2-loop calculation in
\SARAH/\SPheno is not reliable when both \MS and the exotic couplings
are large, due to infrared divergences which appear in the 2-loop
functions, which are not present in the other fixed order codes as
they neglect these contributions.

In the \ESSM\ we studied cases where all exotic couplings are
rather small.  Nonetheless, there is already a large impact of the
exotic couplings on the fixed-order calculations at large \MS and thus
we find that there is no longer good agreement between the available
fixed-order calculations.  The very
different renormalization group flow in this model, where the
$\beta$-function of $\alpha_s$ vanishes at 1-loop level, spoils the
accidental cancellation between the 3-loop logarithms observed in the
MSSM and NMSSM\@.  We see that some of the
sources of uncertainty in the fixed-order calculation, specifically
the uncertainty that is estimated from different definitions of
the Yukawa couplings, rises with \MS much more rapidly than in the MSSM
or NMSSM\@.  Therefore, an effective field theory calculation
in this model is important presumably at lighter values of \MS than in the MSSM
and NMSSM\@.

Finally, we applied \FSTower to the MRSSM and compared the results
with the available 1- and 2-loop fixed-order calculations in a
scenario with sizable triplet couplings as well as with benchmark
points from the literature.  Similar to the \ESSM, we find that the
fixed-order programs no longer agree well with \FSTower for SUSY
scales above a few TeV\@.  One of the reasons is the different running
of $\alpha_s$ in the MRSSM, which again spoils the accidental
cancellation of higher-order logarithms.  We also find that the
uncertainty of the fixed-order calculations, estimated by the
different definitions of the top Yukawa coupling, increases more
rapidly with $\MS$ than in the MSSM or NMSSM\@.  In contrast, the
combined uncertainty estimate for \FSTower is independent of $\MS$,
making the \FSTower calculation more reliable already above a few TeV\@.

There are limitations of \FSTower motivating further
developments. The most obvious is the use of  1-loop matching
at the SUSY scale instead of 2-loop matching. As a result,
the implementation still misses 2-loop power-suppressed as well
as non-power-suppressed (but non-logarithmic) terms in the Higgs pole
mass.  This is complementary to \SARAH/\SPheno, the only
publicly available code 
that can include these  2-loop terms for all models, which
can however
become unreliable for large SUSY scales due to the lack of
large higher-order
logarithms. It is planned 
to extend \FSTower by using the Higgs pole mass matching condition at
the 2-loop level for a higher accuracy. Another possible extension
of \FSTower is to allow for more diverse mass hierarchies leading to additional
intermediate scales at which subsets of particles are sequentially integrated
out. In this way further types of potentially large logarithms can be
resummed.

\section*{Acknowledgements}

We thank Javier Pardo Vega and Giovanni Villadoro for providing us
with threshold corrections as \Mma expressions.  We also thank Mark
Goodsell, Kilian Nickel and Florian Staub, for answering questions
about the fixed-order 2-loop calculation in the SARAH/SPheno code.
AV thanks Philip Dießner for helpful comments and validation of the
studied MRSSM scenarios.  PA thanks Roman Nevzorov and Dylan Harries
for helpful comments regarding the \ESSM\ corrections.  This work has
been supported by the German Research Foundation DFG through grant
No.~STO876/2-1.  JP acknowledges support from the MEC and FEDER (EC)
Grants FPA2011--23596 and the Generalitat Valenciana under grant
PROMETEOII/2013/017.  The work of PA has in part been supported by the
ARC Centre of Excellence for Particle Physics at the Tera-scale.

\appendix
\section{\boldmath Equivalence of $M_h$-matching to $\Delta\lambda^{(1)}$}
\label{sec:appendix-equivalence-proof}

We show the equivalence of \FSTower' matching procedure,
set forth in Section \ref{sec:tower-approach},
to the 1-loop threshold corrections to $\lambda$
at ${\mathcal O}(\alpha_t)$ from the
MSSM presented in the literature such as Eq.~(10) in \cite{Bagnaschi:2014rsa}.
For this, we apply
the matching condition~\eqref{eq:polemassmatching} at the 1-loop level
which
requires that the Higgs pole mass calculated in the SM
be equal to the lightest Higgs pole mass in the MSSM\@.
In this appendix, the matching scale $\Qmatch$ shall be abbreviated to $Q$.
In the Standard Model Higgs pole mass in Eq.~\eqref{eq:SM polemass},
both $\Sigma^{\MSbar,\SM}_h$ and $t_h^{\MSbar,\SM}$ are evaluated
at the 1-loop level.  The lightest MSSM Higgs pole mass
$M_h^\MSSM$ is calculated at the renormalization scale $Q = \MS$
iteratively by Eq.\ \eqref{eq:fixedmh} as
\begin{align}
  (M_h^{\MSSM})^2 &= (m_h^\MSSM)^2
  - \Sigma_h^{\DRbar,\MSSM} + t_h^{\DRbar,\MSSM}/v ,
\end{align}
where $m_h^\MSSM$ is the running \DRbar Higgs mass in the MSSM,
$\Sigma_h^{\DRbar,\MSSM}$ is the \DRbar-re\-nor\-ma\-li\-zed 1-loop
self-energy of the SM-like Higgs in the MSSM, and $t_h^{\DRbar,\MSSM}$
is the corresponding tadpole, given by
\begin{align}
  \Sigma_h^{\DRbar,\MSSM} &=
  c_\alpha^2 \Sigma_{h_u h_u}^{\DRbar,\MSSM}
  + s_\alpha^2 \Sigma_{h_d h_d}^{\DRbar,\MSSM}
  - 2 s_\alpha c_\alpha \Sigma_{h_u h_d}^{\DRbar,\MSSM},\\
  \frac{t_h^{\DRbar,\MSSM}}{v} &=
  c_\alpha^2 \frac{t_{h_u}^{\DRbar,\MSSM}}{v_u}
  + s_\alpha^2 \frac{t_{h_d}^{\DRbar,\MSSM}}{v_d} .
\end{align}
In the SM coupling limit, the Higgs mixing angle $\alpha$ is given by
$\alpha = \beta - \frac{\pi}{2}$, and the SM-like Higgs self-energy
and the tadpole become
\begin{align}
  \Sigma_h^{\DRbar,\MSSM} &=
  s_\beta^2 \Sigma_{h_u h_u}^{\DRbar,\MSSM}
  + c_\beta^2 \Sigma_{h_d h_d}^{\DRbar,\MSSM}
  + 2 s_\beta c_\beta \Sigma_{h_u h_d}^{\DRbar,\MSSM},\\
  \frac{t_h^{\DRbar,\MSSM}}{v} &=
  s_\beta^2 \frac{t_{h_u}^{\DRbar,\MSSM}}{v_u}
  + c_\beta^2 \frac{t_{h_d}^{\DRbar,\MSSM}}{v_d} .
\end{align}
Keeping only the ${\mathcal O}(y_t^n)$ terms, the 1-loop corrections to the
SM-like Higgs in the MSSM is given by
\begin{align}
  &-\Sigma_h^{\DRbar,\MSSM} + t_h^{\DRbar,\MSSM}/v =
  -\Sigma_h^{\MSbar,\SM} + t_h^{\MSbar,\SM}/v \nonumber \\
  &\qquad\qquad
  -\frac{3 (y_t^{\SM})^2}{(4\pi)^2}
  \Big\{X_t^2 c_{2\theta}^2 B_0(p^2,\mstL^2,\mstR^2) \nonumber \\
  &\qquad\qquad\qquad
        + 2 \left(m_t + s_{2\theta} X_t / 2\right)^2
        \left[B_0(p^2,\mstL^2,\mstL^2) + B_0(p^2,\mstR^2,\mstR^2)\right]
        \nonumber\\
  &\qquad\qquad\qquad
      - \frac{X_t s_{2\theta}}{2 m_t} \left[
        A_0(\mstL^2) + A_0(\mstR^2)
        \right]
    \Big\} ,
    \label{eq:relation_SE_Xt_nonzero}
\end{align}
where $\theta$ is the stop mixing angle as defined in Eq.\ (19) of
Ref.\ \cite{Degrassi:2001yf} and $s_{2\theta} = \sin 2\theta$, $c_{2\theta} =
\cos 2\theta$, $X_t = A_t - \mu/\tan\beta$, $m_t = y_t^\SM v / \sqrt{2}$.
The MSSM top quark Yukawa
coupling has been replaced by the corresponding SM Yukawa coupling
using the tree-level relation $y_t^\MSSM = y_t^\SM / s_\beta$.  By
making use of the relation
\begin{align}
  \sin 2\theta = \frac{2 m_t X_t}{\mstL^2 - \mstR^2},
\end{align}
and inserting Eq.~\eqref{eq:relation_SE_Xt_nonzero} into
\eqref{eq:polemassmatching}, one obtains the running Higgs mass in
the Standard Model as
\begin{align}
  (m_h^\SM)^2 &= (m_h^\MSSM)^2 + \Delta m_h^2 ,
  \label{eq:mhSM-mhMSSM-1L}
\end{align}
with the 1-loop correction
\begin{align}
  (4\pi)^2 \Delta m_h^2 &=
  -3 X_t^2 (y_t^\SM)^2 \Bigg\{
  \frac{2 (y_t^\SM)^2 v^2 \left[
      B_0(p^2,\mstL^2,\mstL^2) - B_0(p^2,\mstR^2,\mstR^2)
    \right]}{\mstL^2-\mstR^2} \nonumber \\
&\phantom{={}}\qquad\qquad\qquad
  + B_0(p^2,\mstL^2,\mstR^2)
  + \frac{A_0(\mstR^2) - A_0(\mstL^2)}{\mstL^2-\mstR^2}
 \Bigg\} \nonumber \\
&\phantom{={}}
 -\frac{3 X_t^4 (y_t^\SM)^4 v^2}{(\mstL^2-\mstR^2)^2} \Big[
  B_0(p^2,\mstL^2,\mstL^2) + B_0(p^2,\mstR^2,\mstR^2) - 2 B_0(p^2,\mstL^2,\mstR^2)\Big] \nonumber \\
&\phantom{={}}
-3 (y_t^\SM)^4 v^2 \Big[
   B_0(p^2,\mstL^2,\mstL^2) + B_0(p^2,\mstR^2,\mstR^2)
\Big].
\end{align}
By inserting the stop masses in terms of the soft-breaking parameters
and $X_t$,
\begin{align}
  m_{\tilde{t}_{1,2}}^2 &= m_t^2 + \frac{1}{2}\left(
    m_{Q_3}^2 + m_{U_3}^2 \mp \sqrt{\left(m_{Q_3}^2 - m_{U_3}^2\right)^2 + 4 (m_t X_t)^2}\right) ,
\end{align}
evaluating the $B_0$ functions at the momentum $p^2 = \lambda v^2$,
and expanding in powers of $v^2/\MS^2$ up to ${\mathcal O}(v^2/\MS^2)$,
one obtains at ${\mathcal O}(y_t^4)$
\begin{align}
  &\frac{(4\pi)^2}{(y_t^\SM)^4 v^2} \Delta m_h^2 =
  3\ln\frac{m_{Q_3}^2}{Q^2} + 3\ln\frac{m_{U_3}^2}{Q^2}
  - \frac{p^2\left(m_{Q_3}^2+m_{U_3}^2\right)}{2 m_{Q_3}^2 m_{U_3}^2}
  \nonumber \\
&\quad\quad
  + X_t^2 \Bigg[
  \frac{6 \ln \frac{m_{Q_3}^2}{m_{U_3}^2}}{m_{Q_3}^2-m_{U_3}^2} \nonumber \\
&\quad\quad\quad\quad\quad\quad
  +\frac{p^2}{m_{Q_3}^2 m_{U_3}^2 \left(m_{Q_3}^2-m_{U_3}^2\right)^3}
  \Big(m_{Q_3}^6 - 6 m_{Q_3}^4 m_{U_3}^2+6 m_{Q_3}^2 m_{U_3}^4 \nonumber \\
&\quad\quad\quad\quad\quad\quad\quad\quad
  + \frac{3}{2} m_{Q_3}^2 m_{U_3}^2 \left(m_{Q_3}^2+m_{U_3}^2\right) \ln \frac{m_{Q_3}^2}{m_{U_3}^2}
  - m_{U_3}^6\Big)
 \Bigg] \nonumber \\
&\quad\quad
  + X_t^4 \Bigg\{
  -\frac{3 \left[
      \left(m_{Q_3}^2 + m_{U_3}^2\right) \ln\frac{m_{Q_3}^2}{m_{U_3}^2}
      -2 m_{Q_3}^2 + 2 m_{U_3}^2\right]}{\left(m_{Q_3}^2-m_{U_3}^2\right)^3} \nonumber \\
&\quad\quad\quad\quad\quad\quad
  +\frac{p^2}{2 m_{Q_3}^2 m_{U_3}^2 \left(m_{Q_3}^2-m_{U_3}^2\right)^5}
  \Bigg[
  -m_{Q_3}^8+17 m_{Q_3}^6 m_{U_3}^2-17 m_{Q_3}^2 m_{U_3}^6 \nonumber \\
&\quad\quad\quad\quad\quad\quad\quad\quad
  +3 m_{Q_3}^2 m_{U_3}^2 \left(m_{Q_3}^4+8 m_{Q_3}^2
   m_{U_3}^2+m_{U_3}^4\right) \ln
   \left(\frac{m_{U_3}^2}{m_{Q_3}^2}\right)+m_{U_3}^8
   \Bigg]
 \Bigg\}
 . \label{eq:delta_lambda_nonzero_Xt_expanded}
\end{align}
Using the relations
\begin{align}
  (m_h^\SM)^2 &= \lambda v^2, \\
  (m_h^\MSSM)^2 &= \frac{1}{4} (g_Y^2 + g_2^2)(v_u^2 + v_d^2) c_{2\beta}^2,
\end{align}
and exploiting that $\Delta m_h^2 = v^2 \Delta \lambda$ at the
1-loop ${\mathcal O}(y_t^4)$, one obtains from Eq.~\eqref{eq:mhSM-mhMSSM-1L} in
the limit $p^2 \rightarrow 0$
\begin{align}
  \lambda &= \frac{1}{4} (g_Y^2 + g_2^2) c_{2\beta}^2 + \Delta\lambda^{(1)}
\end{align}
with
\begin{align}
  (4\pi)^2 \Delta\lambda^{(1)} &=
  3 (y_t^\SM)^4 \left(\ln\frac{m_{Q_3}^2}{Q^2} + \ln\frac{m_{U_3}^2}{Q^2}\right)
  +\frac{6 (y_t^\SM)^4 X_t^2 \ln \frac{m_{Q_3}^2}{m_{U_3}^2}}{m_{Q_3}^2-m_{U_3}^2} \nonumber \\
&\phantom{={}}
  -\frac{3 (y_t^\SM)^4 X_t^4 \left[
      \left(m_{Q_3}^2 + m_{U_3}^2\right) \ln\frac{m_{Q_3}^2}{m_{U_3}^2}
      -2 m_{Q_3}^2 + 2 m_{U_3}^2\right]}{\left(m_{Q_3}^2-m_{U_3}^2\right)^3} ,
  \label{eq:delta_lambda_nonzero_Xt}
\end{align}
which is equivalent to the ${\mathcal O}(y_t^4)$ terms of Eq.~(10) of Ref.\
\cite{Bagnaschi:2014rsa}.  We conclude that in the MSSM \FSTower'
approach is equivalent to the 1-loop threshold corrections to
$\lambda$ from Ref.~\cite{Bagnaschi:2014rsa} in the SM coupling limit
$\alpha = \beta - \frac{\pi}{2}$ and $p^2 \ll \MS^2$.

\section{Leading logarithms in the EFT, \FS, and \SPheno-calculations}
\label{sec:appendix-leading-logs}
\newcommand{\bgsgs}{\beta_{\gshigh,\gshigh^2}}
\newcommand{\bytgs}{\beta_{\ythigh,\gshigh^2}}
\newcommand{\bytyt}{\beta_{\ythigh,\ythigh^2}}
\newcommand{\bvyt}{\beta_{\vhigh,\ythigh^2}}
\newcommand{\blambdaytyt}{\beta_{\lambdahigh,\ythigh^4}}
\newcommand{\blambdayt}{\beta_{\lambdahigh,\ythigh^2\lambdahigh}}
\newcommand{\blambdalambda}{\beta_{\lambdahigh,\lambdahigh^2}}
\newcommand{\btildegsgs}{\tilde\beta_{\gshigh,\gshigh^2}}
\newcommand{\btildeytgs}{\tilde\beta_{\ythigh,\gshigh^2}}
\newcommand{\btildeytyt}{\tilde\beta_{\ythigh,\ythigh^2}}
\newcommand{\btildevyt}{\tilde\beta_{\vhigh,\ythigh^2}}

Here we derive the leading $L$-loop logarithms of the form
$t_S\equiv\ln(\MS/M_t)$  governed by the two most important couplings
$\alpha_s=g_3^2/4\pi$ and 
$y_t$, which are contained in the   MSSM Higgs mass calculations in
\FS\ and \SPheno. We compare them to the  correct leading logarithms,
which are contained in the EFT calculation. For simplicity we work in
the approximation of large $\tan\beta$ and identify
$\sin\beta=1$, $v_u =v$ in the definition of the MSSM top Yukawa
coupling. In 
the present section, we 
use the following notation for the required running couplings in the
SM and MSSM at scale $t\equiv\ln(Q/M_t)$:
\begin{align}
\text{SM:}&&\gshigh(t),&& \ythigh(t),&& \vhigh(t),&& \lambda(t);\\
\text{MSSM:}&&\gsMSSMhigh(t),&& \ytMSSMhigh(t),&& \vMSSMhigh(t).
\end{align}
As an abbreviation, we write the quantities at the top-mass scale and
the SUSY scale as
\begin{align}
\text{SM, low:}&&\gs&= \gshigh(0),& \ytlow&=\ythigh(0),&
 \vlow &=\vhigh(0),& \lambdalow&=\lambdahigh(0);\\
\text{SM, high:}&&\gshigh&= \gshigh(t_S),& \ythigh&=\ythigh(t_S),&
 \vhigh &=\vhigh(t_S),& \lambdahigh&=\lambdahigh(t_S);\\
\text{MSSM, low:}&&\gsMSSM&=\gsMSSMhigh(0),& \ytMSSMlow&=\ytMSSMhigh(0),&
 \vMSSM&=\vMSSMhigh(0);\\
\text{MSSM, high:}&&\gsMSSMhigh&=\gsMSSMhigh(t_S),& \ytMSSMhigh&=\ytMSSMhigh(t_S),&
 \vMSSMhigh&=\vMSSMhigh(t_S).
\end{align}
The relevant $\beta$ functions are the 1-loop $\beta$ functions for
these parameters, $\beta_{X}(t)\equiv\frac{dX(t)}{dt}$. The relevant terms
can be written as
\begin{equation}
\begin{aligned}
\beta_{\gshigh}(t) &=
\bgsgs \gshigh^3(t),&
\beta_{\ythigh}(t) &=
\ythigh(t)\left(\bytgs \gshigh^2(t)+\bytyt \ythigh^2(t)\right),
\\
\beta_{\vhigh}(t) &=
\bvyt \vhigh(t) \ythigh^2(t),&
\beta_{\lambdahigh}(t) &=
\blambdaytyt \ythigh^4(t)+\blambdayt
\ythigh^2(t)\lambdahigh(t)+\blambdalambda \lambdahigh^2(t)
\,.
\end{aligned}
\end{equation}
The values of the appearing coefficients depend on the model. In the
SM, they read
\begin{equation}
\begin{aligned}
\bgsgs &=  -7\kappaL,&
\bytgs &= -8\kappaL,&
\bytyt &=  \frac{9}{2}\kappaL,&
\bvyt &= -{3}\kappaL  ,
\\
\blambdaytyt &=  -12\kappaL,&
\blambdayt &= 12\kappaL,&
\blambdalambda &=  12\kappaL.
\end{aligned}
\end{equation}
Here we use the common loop factor constant $\kappaL=1/(16\pi^2)$.
In the
MSSM, we denote the corresponding coefficients with a tilde; their
values are
\begin{align}
\btildegsgs &= -{3}\kappaL ,&
\btildeytgs &= -\frac{16}{3}\kappaL,&
\btildeytyt &= 6\kappaL,&
\btildevyt &=  -{3}\kappaL  .
\end{align}
As is well known, the leading logarithms can be obtained in the EFT
approach by integrating the RGEs in the SM\@. In a first step this
yields the running couplings
\begin{align}
\gshigh(t) &=\sqrt{\frac{1}{\frac{1}{\gs^2}-2 \bgsgs t}}
=\gs-7 t \kappa _L \gs^3+\frac{147}{2} t^2 \kappaL^2 \gs^5
-\frac{1715}{2} t^3  \kappaL^3 \gs^7+\ldots
,
\label{eq:gshighrunning SM}
\\
\ythigh(t) &= \ytlow +t \left(\bytgs\gs^2 \ytlow+\bytyt
   \ytlow^3\right) 
\nonumber\\
&\quad{}
+\frac{1}{2} t^2 \ytlow \left(2 \bgsgs \bytgs\gs^4+\bytgs^2\gs^4+4
   \bytgs \bytyt\gs^2 \ytlow^2+3 \bytyt^2 \ytlow^4\right)
+\ldots
\label{eq:ythighrunning}
\\
&=\ytlow +
t \kappaL \left(\frac{9 \ytlow^3}{2}-8\gs^2 \ytlow \right)
+
t^2\kappaL^2 \ytlow \left(88\gs^4 -72\gs^2
   \ytlow^2 + \frac{243}{8} \ytlow^4 \right)
+\ldots
,
\label{eq:ythighrunning SM}
\\
\vhigh(t) &=\vlow \left(1
+
\bvyt t \ytlow^2
+
t^2 \left(\frac{\bvyt^2 \ytlow^4}{2}+\bvyt \bytgs\gs^2 \ytlow^2+\bvyt \bytyt
   \ytlow^4\right)\right)
\\
&=
\vlow \left(1 -3 t \kappaL \ytlow^2
+
t^2 \kappaL^2\left(24\gs^2 \ytlow^2
-9 \ytlow^4 \right)\right) 
.
\end{align}
As indicated, the running couplings on the left-hand
side are taken at 
scale $t$, while the couplings without argument on the right-hand side 
are running couplings at the fixed low scale $t=0$, i.e.\ at $Q=M_t$. 
In a second step these results can be used to integrate the RGE for
$\lambda$, to express $\lambdalow$ as a function of
$\lambdahigh(t)$,
\begin{align}
\lambdalow &=
\lambdahigh(t)
-\blambdaytyt t
   \ytlow^4
+\frac{1}{2} \blambdaytyt t^2 \ytlow^4 \left(
   (\blambdayt-4 \bytyt)\ytlow^2
-4 \bytgs  \gs^2\right)
\nonumber\\
&\quad{}
-\frac{1}{6} \blambdaytyt t^3 \ytlow^4 \Big[
(8 \bgsgs \bytgs+16 \bytgs^2)\gs^4-10 \bytgs
   (\blambdayt-4 \bytyt)\gs^2 \ytlow^2
\nonumber\\
&\quad\quad{}
+ \left(2\blambdalambda
   \blambdaytyt+\blambdayt^2-10\blambdayt \bytyt+24 \bytyt^2\right)\ytlow^4\Big]
\end{align}
In this equation, terms of higher-order in $\lambdahigh(t)$ have been
neglected. In leading-logarithmic approximation, the high-scale
coupling $\lambdahigh(t_S)$ is given by matching the SM to the
tree-level MSSM Higgs boson mass $m_h^2$ at the SUSY scale.  The EFT
prediction for the Higgs boson mass is then, in this approximation,
\begin{align}
M_h^2 &=
\vlow^2 \lambdalow
\\
&=m_h^2
+ \vlow^2 
\ytlow^4\Big[12 t_S \kappaL
-12 t_S^2 \kappaL^2 
   \left(16 \gs^2 -3 \ytlow^2 \right)
\nonumber\\&\quad{}
+4 t_S^3 \kappaL^3  \left(736 \gs^4 -240
\gs^2 \ytlow^2 
-99 \ytlow^4 \right)+\ldots\Big].
\end{align}
The previous equations agree with
Eq.~(11) from Ref.\ \cite{Degrassi:2002fi} and Eq.~(A.17) from
Ref.\ \cite{Martin:2007pg}.

Now we compare these results with the leading logarithms contained in
the ``fixed-order'' calculations. The most important difference is the
definition of the Yukawa coupling. In the original \FS\ (and
\Softsusy) or \SPheno, the low-scale MSSM Yukawa coupling is defined
by Eq.~(\ref{eq:fixedmt_FlexibleSUSY}) or 
Eq.~(\ref{eq:fixedmt_SPheno}), respectively.
These equations contain leading logarithms within the self energy
parts 
$\widetilde{\Sigma}_t^{(1),L,R}$, and the coefficients of these
logarithms is given by the difference of the SM and MSSM $\beta$
functions for the Yukawa coupling. Hence, to the leading logarithmic
level, these equations imply
\begin{align}
\ytMSSMlow^{\text{\FS}} &= \ytlow \Big[1 + 
\left((\bytgs-\btildeytgs)
\gsMSSM^2+ (\bytyt-\btildeytyt)(\ytMSSMlow^{\text{\FS}})^2\right)
 t_S\Big],
\\
\ytMSSMlow^{\text{\SPheno}} &= \frac{\ytlow}{1 -
\left((\bytgs-\btildeytgs)
\gsMSSM^2 - (\bytyt-\btildeytyt)(\ytMSSMlow^{\text{\SPheno}})^2\right)
 t_S
},
\end{align}
which has to be iterated to find the solutions for the low-scale
Yukawa couplings in terms of $\ytlow$. The strong gauge coupling is
determined by low-scale matching to the SM, but the matching
condition is such that, at the leading logarithmic level, we obtain
$\gsMSSMhigh=\gshigh$, and
\begin{align}
{\gsMSSM^2} &=\left[{\frac{1}{\gs^2}-2 (\bgsgs-\btildegsgs) t_S}\right]^{-1}.
\end{align}
The low-scale Yukawa couplings are then run up to the SUSY scale with
the MSSM $\beta$ function. For the running
Eq.~(\ref{eq:ythighrunning}) applies, with the replacement of SM by
MSSM quantities. Plugging in the values of all coefficients, the final
result for the SUSY-scale MSSM Yukawa couplings used in the
fixed-order calculations is therefore 
\begin{align}
\ytMSSMhigh^{\text{\FS}} &=
\ytlow +
 t_S \kappaL \left(\frac{9 \ytlow^3}{2}-8 \gs^2
 \ytlow\right)
+
t_S^2\kappaL^2 \left(\frac{976 \gs^4 \ytlow}{9}-96
\gs^2 
\ytlow^3+\frac{63
   \ytlow^5}{2}\right) +\ldots,
\\
\ytMSSMhigh^{\text{\SPheno}} &= 
\ytlow +
t_S\kappaL \left(\frac{9 \ytlow^3}{2}-8
\gs^2 \ytlow\right) 
+
t_S^2 \kappaL^2\left(\frac{1040 \gs^4 \ytlow}{9}-88 \gs^2 \ytlow^3+\frac{135
   \ytlow^5}{4}\right)+\ldots ,
\end{align}
which agrees at the 1-loop level with the EFT result but disagrees
at the 2-loop level.
The fixed-order calculations of the MSSM Higgs boson mass then plug
the SUSY-scale MSSM parameters into the  \DRbar-Higgs self energy. At
the leading logarithmic level, this gives
\begin{align}
M_h^2 &=
m_h^2
+
\vMSSMhigh^2 \ytMSSMhigh^4\Big(
t_S c_1 
+
t_S^2 ( c_{21} \gsMSSMhigh^2 + c_{22} \ytMSSMhigh^2)
\Big),
\\
c_1 &=-\blambdaytyt =12\kappaL ,
\\
c_{21} &=   2 \blambdaytyt \bytgs =192\kappaL^2,
\\
c_{22} &= 
 \blambdaytyt \left( \blambdayt/2  +2  \bvyt
+2    \bytyt \right) 
= -108\kappaL^2,
\end{align}
where $m_h$ is the running tree-level Higgs mass and where the values
of the coefficients $c$ follow from the agreement  
of the 2-loop leading logarithms with the correct EFT result.
Hence, plugging in number, we obtain the leading logarithms up to the
3-loop level, contained in the fixed-order calculations,
\begin{align}
(M_h^2)^{\text{\FS}} \nonumber
&=m_h^2
+ \vlow^2 
\ytlow^4\Big[12 t_S \kappaL
-12 t_S^2 \kappaL^2
   \left(16 \gs^2 -3 \ytlow^2 \right)
\\&\quad{}
+4 t_S^3 \kappaL^3 \left(\frac{736}{3} \gs^4 +144
   \gs^2 \ytlow^2 -\frac{351}{2} \ytlow^4 \right)\Big],
\\
(M_h^2)^{\text{\SPheno}} \nonumber
&=m_h^2
+ \vlow^2 
\ytlow^4\Big[12 t_S \kappaL
-12 t_S^2 \kappaL^2
   \left(16 \gs^2 -3 \ytlow^2 \right)
\\&\quad{}
+4 t_S^3 \kappaL^3  \left(\frac{992}{3} \gs^4 +240
   \gs^2 \ytlow^2 -\frac{297}{2}
   \ytlow^4\right)
\Big].
\end{align}

So far, all results are expressed in terms of low-scale SM
couplings, which are connected to low-energy observables without large
logarithms. It is useful to record here the equivalent results, in
which these low-scale SM couplings are replaced by SUSY-scale running
SM parameters, which are connected to the fundamental high-scale SUSY
parameters without large logarithms. For this purpose,
Eqs.~\eqref{eq:gshighrunning SM}, \eqref{eq:ythighrunning SM}
can be inverted. The results for the high-scale Yukawa couplings used
in \FS\ and \SPheno\ are then
\begin{align}
\ytMSSMhigh^{\text{\FS}} &= 
\ythigh +
 t_S^2\kappaL^2 \left(\frac{184}{9} \gshigh^4 \ythigh -24 \gshigh^2
 \ythigh^3 
+\frac{9}{8} \ythigh^5 \right) 
+\ldots , \\
\ytMSSMhigh^{\text{\SPheno}} &= 
\ythigh +
 t_S^2 \kappaL^2 \left(\frac{248}{9} \gshigh^4 \ythigh-16 \gshigh^2
 \ythigh^3
+\frac{27}{8} \ythigh^5\right) 
+\ldots ,
\end{align}
and the results for the Higgs boson mass in the EFT, \FS, and
\SPheno\ are
\begin{align}
(M_h^2)^{\text{EFT}}&=
m_h^2
+ \vhigh^2 
\ythigh^4\Big[12 t_S \kappaL
+12 t_S^2 \kappaL^2 
\left(16 \gshigh^2 - 9\ythigh^2 \right)
\nonumber\\&\quad{}
+
4 t_S^3\kappaL^3  \left(736 \gshigh^4-672 \gshigh^2 \ythigh^2+90
   \ythigh^4\right) 
+\ldots\Big],
\\
(M_h^2)^{\text{\FS}}&=
m_h^2
+ \vhigh^2 
\ythigh^4\Big[12 t_S \kappaL
+12 t_S^2 \kappaL^2 
\left(16 \gshigh^2 - 9\ythigh^2 \right)
\nonumber\\&\quad{}
+4 t_S^3\kappaL^3 \left(\frac{736 \gshigh^4}{3}-288 \gshigh^2
   \ythigh^2+\frac{27 \ythigh^4}{2}\right)
+\ldots\Big],
\\
(M_h^2)^{\text{\SPheno}}&=
m_h^2
+ \vhigh^2 
\ythigh^4\Big[12 t_S \kappaL
+12 t_S^2 \kappaL^2 
\left(16 \gshigh^2 - 9\ythigh^2 \right)
\nonumber\\&\quad{}
+4 t_S^3\kappaL^3 \left(\frac{992 \gshigh^4}{3}-192 \gshigh^2
   \ythigh^2+\frac{81 \ythigh^4}{2}\right)
+\ldots\Big].
\end{align}
The EFT result here agrees with Ref.\ \cite{Martin:2007pg}, Eq.~(A.22).

\clearpage
\pagestyle{plain}

\bibliographystyle{JHEP}
\bibliography{bibliography}

\end{document}